\documentclass[english,aps,showpacs,showkeys,preprintnumbers,prd,floatfix,nofootinbib,superscriptaddress,notitlepage,twocolumn]{revtex4-1}
\usepackage{graphicx}
\usepackage{amsmath}
\usepackage{amsfonts}
\usepackage{hyperref}
\usepackage{color}
\usepackage{placeins}
\usepackage{tabu}
\usepackage{nicefrac}
\usepackage{ulem}

\newcommand{\chidof}{\chi^2/N_{dof}}
\newcommand{\chieff}{\chi^2_{\rm eff}}
\definecolor{darkgreen}{rgb}{.2,0.6,.2}

\newcommand{\orcid}[1]{\,\href{https://orcid.org/#1}{\includegraphics[width=9pt]{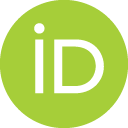}}\,}

\newcommand{\orcidTJ}{0000-0002-1334-7607} %
\newcommand{\orcidPD}{0000-0001-7960-7953} %
\newcommand{\orcidMK}{0000-0002-4665-3088} %
\newcommand{\orcidKK}{0000-0003-1412-447X} %
\newcommand{\orcidAK}{0000-0002-4090-0084} %
\newcommand{\orcidFM}{0000-0002-3888-1697} %
\newcommand{\orcidFO}{0000-0001-6799-2436} %
\newcommand{\orcidIS}{0000-0003-0373-474X} %
\newcommand{\orcidJY}{0000-0001-8366-0968} %
\begin{document}
\preprint{
\vbox{
\hbox{SMU-HEP-21-05,\ \ P3H-21-033}
\hbox{KA-TP-10-2021,\ \  MS-TP-21-11}
\hbox{IFJPAN-IV-2021-9}
\vspace{0.3in}
}}

\title{Impact of inclusive hadron production data on nuclear gluon PDFs}

\author{P.~Duwent\"aster\orcid{\orcidPD}}
\email{pit.duw@uni-muenster.de}
\affiliation{Institut f{ü}r Theoretische Physik, Westf{ä}lische Wilhelms-Universit{ä}t
M{ü}nster, Wilhelm-Klemm-Stra{ß}e 9, D-48149 M{ü}nster, Germany }

\author{L.~A.~Husov\'a}
\affiliation{Institut f{ü}r Kernphysik, Westf{ä}lische Wilhelms-Universit{ä}t
M{ü}nster, Wilhelm-Klemm-Stra{ß}e 9, D-48149 M{ü}nster, Germany }

\author{T.~Je\v{z}o\orcid{\orcidTJ}}
\affiliation{Institute for Theoretical Physics, KIT,  D-76131 Karlsruhe, Germany}

\author{M.~Klasen\orcid{\orcidMK}}
\affiliation{Institut f{ü}r Theoretische Physik, Westf{ä}lische Wilhelms-Universit{ä}t
M{ü}nster, Wilhelm-Klemm-Stra{ß}e 9, D-48149 M{ü}nster, Germany }

\author{K.~Kova\v{r}\'{\i}k\orcid{\orcidKK}}
\affiliation{Institut f{ü}r Theoretische Physik, Westf{ä}lische Wilhelms-Universit{ä}t
M{ü}nster, Wilhelm-Klemm-Stra{ß}e 9, D-48149 M{ü}nster, Germany }

\author{A.~Kusina\orcid{\orcidAK}}
\affiliation{Institute of Nuclear Physics Polish Academy of Sciences, PL-31342 Krakow, Poland}

\author{K.~F.~Muzakka\orcid{\orcidFM}}
\affiliation{Institut f{ü}r Theoretische Physik, Westf{ä}lische Wilhelms-Universit{ä}t
M{ü}nster, Wilhelm-Klemm-Stra{ß}e 9, D-48149 M{ü}nster, Germany }

\author{F.~I.~Olness\orcid{\orcidFO}}
\affiliation{Southern Methodist University, Dallas, TX 75275, USA }

\author{I.~Schienbein\orcid{\orcidIS}}
\affiliation{Laboratoire de Physique Subatomique et de Cosmologie, Université
Grenoble-Alpes, CNRS/IN2P3, 53 avenue des Martyrs, 38026 Grenoble,
France }

\author{J.~Y.~Yu\orcid{\orcidJY}}
\affiliation{Southern Methodist University, Dallas, TX 75275, USA }
\date{\today}

\begin{abstract}
\vspace*{0.5cm}

A precise knowledge of nuclear parton distribution functions (nPDFs) is --- among other things --- important
for the unambiguous interpretation of hard process data taken in $pA$ and $AA$ collisions at the Relativistic Heavy Ion Collider (RHIC)
and the Large Hadron Collider (LHC). The available fixed target data for deep inelastic scattering (DIS) and Drell-Yan (DY) 
lepton pair production mainly constrain the light quark distributions. 
It is hence crucial to include more and more collider data in global analyses of nPDFs in order to better pin down
the different parton flavors, in particular the gluon distribution at small $x$.
To help constrain the nuclear gluon PDF, we extend the nCTEQ15 analysis by
including single inclusive hadron (SIH) production data from RHIC
(PHENIX and STAR) and LHC (ALICE).
In addition to the DIS, DY and SIH data sets, we will also include LHC
$W/Z$ production data.
As the SIH calculation is dependent on hadronic fragmentation
functions (FFs), we use a variety of FFs available in the literature
to properly estimate this source of uncertainty.
We study the impact of these data on the PDFs, and compare with both the
nCTEQ15 and nCTEQ15WZ sets.
The calculations are performed using a new implementation of the
nCTEQ code (\texttt{nCTEQ++}) including a modified version of INCNLO
which allows faster calculations using pre-computed grids.
The extension of the nCTEQ15 analysis to include the SIH data
represents an important step toward the next generation of PDFs.

\vspace*{2.5cm}
\end{abstract}

\maketitle
\clearpage  %
\tableofcontents{}

\clearpage  %
\section{Introduction}\label{sec:intro}

Parton distribution functions (PDFs) are fundamental quantities required to calculate predictions for any process involving hadrons in the initial state. The  
QCD parton model  has been used successfully to make predictions for a variety of experiments at SLAC, HERA, TeVatron, RHIC and LHC. This theoretical framework  will also be essential  for both the physics program of the EIC, and proposed future experiments such as the FCC. 
While precise constraints have been imposed on the proton PDFs, 
for the case of nuclear PDFs (nPDFs), there is still much room for improvement of the uncertainties~\cite{Hou:2019efy,Ball:2017nwa,Kovarik:2015cma,Eskola:2016oht,AbdulKhalek:2019mzd,AbdulKhalek:2020yuc,Ethier:2020way,Khalek:2018mdn,Gao:2017yyd,Kovarik:2019xvh,Alekhin:2017olj,Nadolsky:2008zw,Sato:2019yez,Harland-Lang:2014zoa,Thorne:2019mpt,Ball:2009mk,Lin:2017snn,Lin:2020rut, Guzey:2019kik, Klasen:2017kwb, Klasen:2018gtb, Kovarik:2019xvh, Armesto:2015lrg}. The gluon PDFs are particularly problematic because the cross sections for the deep inelastic scattering (DIS) and the Drell-Yan  (DY) processes, which represent the bulk of the precision data in nPDF fits like nCTEQ15~\cite{Kovarik:2015cma}, are not directly sensitive to the gluon PDF at leading order.

 While many different microscopic models for nuclear effects on PDFs exist,
 no unambiguous picture has yet emerged for either the shadowing region
 \cite{Armesto:2006ph,Frankfurt:2011cs,Kopeliovich:2012kw,Kulagin:2004ie}, antishadowing
 region \cite{Brodsky:1989qz,Brodsky:2004qa,Kulagin:2004ie}, or the EMC effect
 \cite{Geesaman:1995yd,Norton:2003cb,Hen:2013oha,Malace:2014uea,Hen:2016kwk,Kulagin:2004ie}. 
 A particularly promising unified approach is provided by the Color Glass
 Condensate \cite{Iancu:2000hn,Gelis:2010nm}.
 On the other hand, unbiased fits to the experimental data provide important
 global constraints on these theoretical ideas and are an indispensable
 ingredient for many current and future experimental (i.e. at LHC, but also
 RHIC and EIC) and theoretical analyses (e.g., for the very successful
 Statistical Hadronization Model describing the freeze-out of the QGP
 \cite{Andronic:2017pug}). This is the approach we take in the following.
 Note that there are currently ongoing studies at the LHC of medium, i.e.
 final state effects also in small systems created in pA and even pp collisions
 \cite{ALICE:2016fzo,Ortiz:2019osu}. In our analysis below, we will
 demonstrate that our results are largely independent of the final state
 hadron fragmentation and thus that our interpretation of the nuclear effects
 as modifications of a cold initial state is currently totally consistent with
 the available experimental data.

\begin{figure}[tb]
	\centering
	\includegraphics[width=0.48\textwidth]{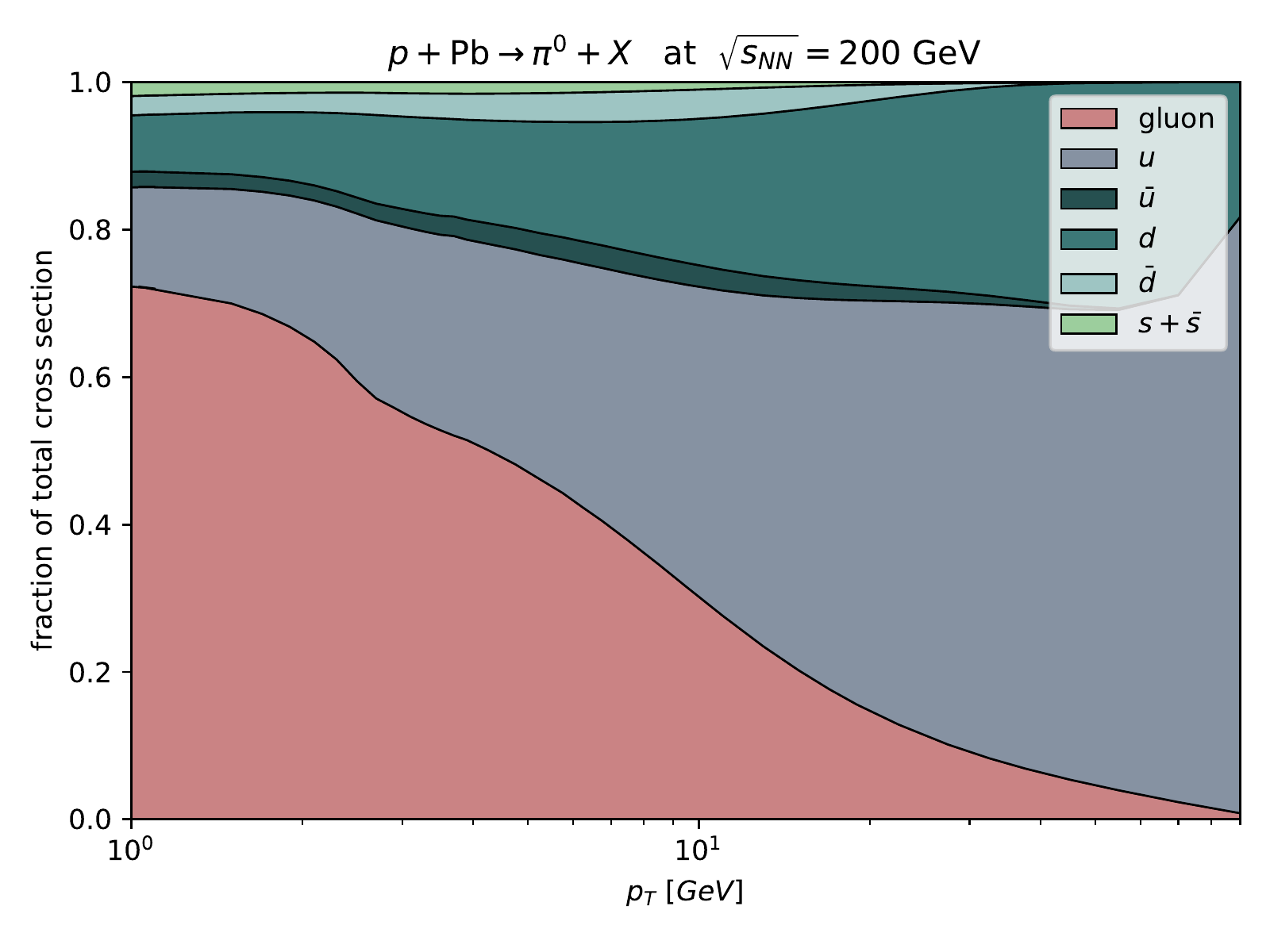}
	\centering
	\includegraphics[width=0.48\textwidth]{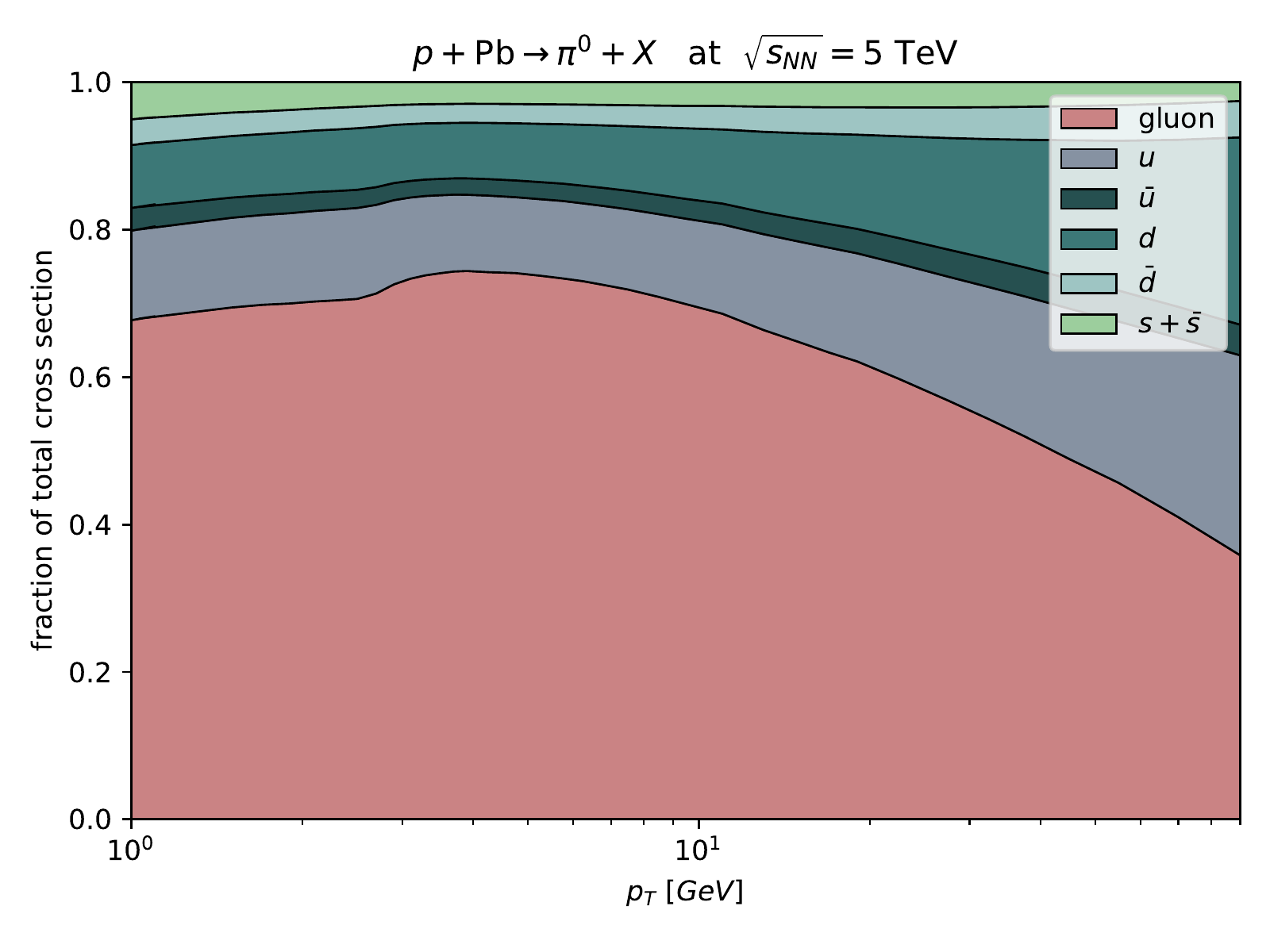}
    \caption{Fractional contributions of the total 
    \mbox{$p{+}\mathrm{Pb} \rightarrow \pi^0{+}X$} 
    cross section initiated by each PDF flavor $f_i^{\mathrm{Pb}}(x,Q)$ of the lead nucleus at $\sqrt{s_{NN}}=200$\,GeV (upper panel) and 5\,TeV (lower panel) for $i\in \{g, u, d, \bar{u}, \bar{d}, s+\bar{s}\}$.}
    \label{fig:xsec_per_flavour}
\end{figure}
\begin{figure}[tb]
	\centering
	\includegraphics[width=0.48\textwidth]{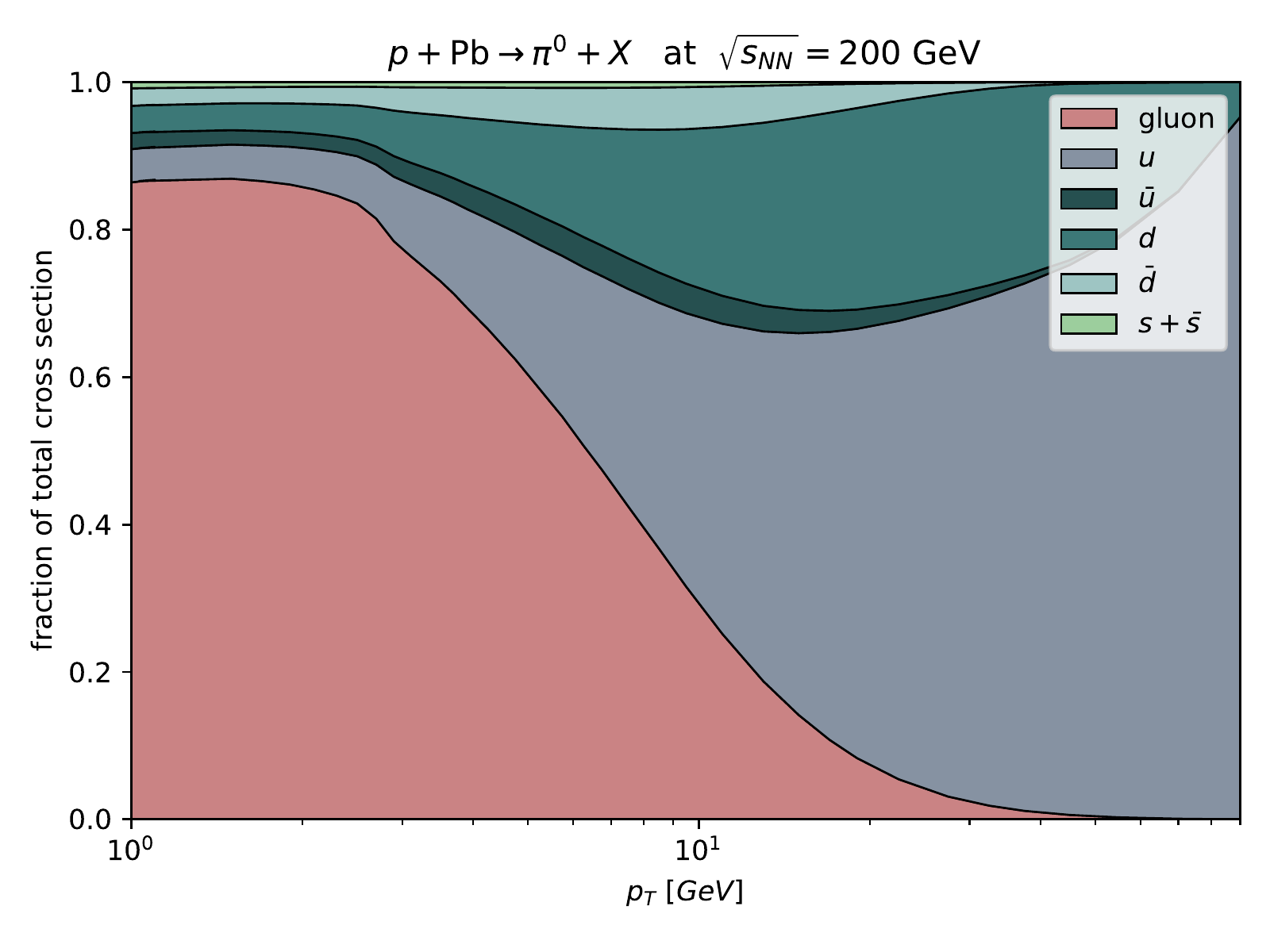}
	\centering
	\includegraphics[width=0.48\textwidth]{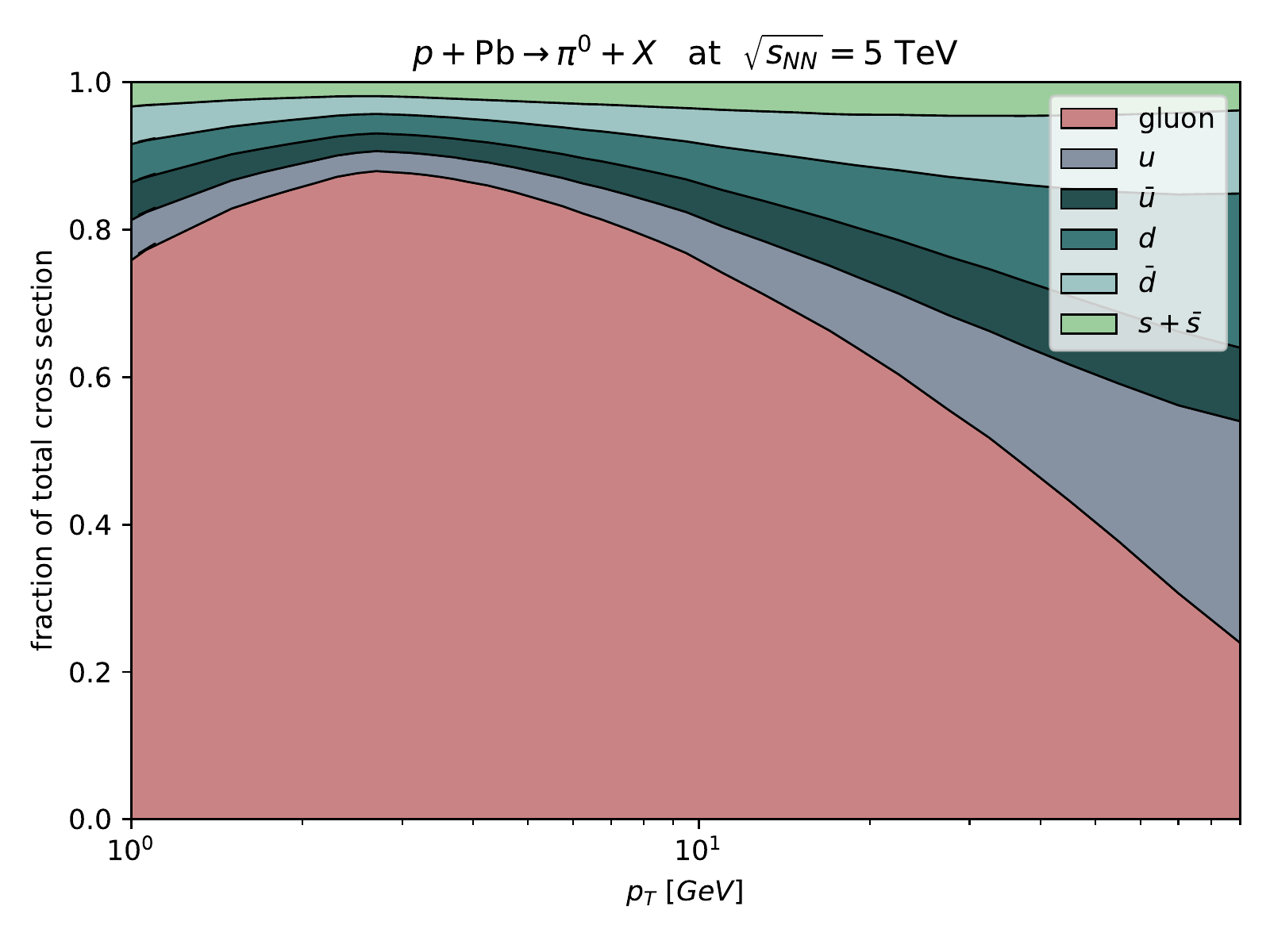}
	\caption{Fractional contribution  of the total 
	\mbox{$p{+}\mathrm{Pb} \rightarrow \pi^0{+}X$} cross section contributed by each fragmentation function, \mbox{$D^{\pi^0}_i(z,Q)$}, at $\sqrt{s_{NN}}=200$\,GeV (upper panel) and 5\,TeV (lower panel) for $i\in \{g, u, d, \bar{u}, \bar{d}, s+\bar{s}\}$.}
	\label{fig:FF_flavour}
\end{figure}
\subsection{The gluon PDF}\label{sec:overview}

Single Inclusive Hadron (SIH) production data has the potential to yield new constraints on the gluon PDF because the gluon contributes a significant part to the overall cross section of this process.
The importance of the gluon contribution   can be seen in Fig.~\ref{fig:xsec_per_flavour}, which shows the fractional contribution
to the process \mbox{$p{+}\mathrm{Pb} \rightarrow \pi^0 {+} X$} as a function of the transverse momentum $p_T$ 
for the various subprocesses initiated by gluons, up, down, and strange partons inside a lead nucleus.
In particular, the red 
shaded area shows the fraction where a parton from the proton interacts with a gluon from the lead nucleus to produce a neutral pion. The gluon contribution dominates in the low to mid~$p_T$ region at a center of mass energy per nucleon of $\sqrt{s_{NN}}=200$\,GeV.
At 5\,TeV, the gluon  is the dominant contribution even in the mid- to high-$p_T$ region. 
The remaining contribution is shared roughly evenly between the up and down quarks, while the antiquarks (including up and down) contribute a minor fraction.  Charm, bottom and top are omitted in this figure due to their negligible contributions, but they are fully incorporated in the calculation. 
The partonic fractions for kaons and eta mesons are similar to those of pions, so we do not present a separate figure. 

Figure~\ref{fig:FF_flavour} shows the relative contributions to the cross section of 
\mbox{$p{+}\mathrm{Pb} \rightarrow \pi^0{+}X$} of each parton's fragmentation function (FF). 
For instance, the red   
area shows the contribution from processes where the initial scattering event produces a gluon which then fragments into a neutral pion. These contributions are very similar to  those of the PDF flavors (Fig.~\ref{fig:xsec_per_flavour}), but with slightly larger contributions from the antiquarks. Both figures are computed with nCTEQ15WZ PDFs~\cite{Kusina:2020lyz} and DSS FFs~\cite{deFlorian:2014xna}, but there are no qualitative differences  when other nPDFs or FFs are used.

In this investigation we will study single inclusive hadron production in proton-lead and deuterium-gold collisions. The  focus will be to incorporate this process into the global 
analysis, including the dependence of the fragmentation function, and 
to determine the resulting  impact on the nuclear gluon PDF. 
The remainder of this section provides an overview of the nCTEQ framework and the available data sets. 
In Sec.~\ref{sec:theoryFF} we investigate the fragmentation function dependence, along with other theory considerations like the scale dependence. 
In Sec.~\ref{sec:fits} we present the fits obtained using the  SIH data,
and compare with the theoretical predictions. 
The main conclusions are summarized in Sec.~\ref{sec:conclusions}.

\subsection{The nCTEQ++ framework}

The nCTEQ project expands upon the foundation of the proton PDF global fitting analysis by including the nuclear dimension. In early proton PDF analyses~(e.g., Ref.~\cite{Olness:2003wz}), the nuclear data was used to calculate correction factors which were then applied to the proton PDF fit without any uncertainties. In contrast, the nCTEQ framework enables full communication between nuclear and proton data, which means that observed tensions between data sets can be investigated through the lens of nuclear corrections. 

The details of the nCTEQ15 nPDFs are presented in Ref.~\cite{Kovarik:2015cma}. The current analysis, along with the other recent nCTEQ analyses, such as  nCTEQ15WZ~\cite{Kusina:2020lyz} and nCTEQ15HIX~\cite{Segarra:2020gtj}, 
are performed with a new \texttt{C++} based code \texttt{nCTEQ++}. This allows us to easily interface external programs such as HOPPET~\cite{Salam:2008qg}, \mbox{APPLgrid}~\cite{Carli:2010rw}, and INCNLO~\cite{INCNLO}. In particular, we work at leading twist and next-to-leading order (NLO) of
 QCD for both the PDF and FF evolution equations as well as the hard
 scattering coefficients.

For the fits in this investigation, we use the same 19 parameters as for the nCTEQ15WZ set. These 19 parameters include the 16 free parameters of the nCTEQ15 analysis, with an additional 3 open parameters for the strange distribution. Recall that for the nCTEQ15 set, the strange PDF was constrained by the relation $s = \bar{s} = {(\kappa/2)(\bar{u} {+} \bar{d})}$ at the initial scale $Q_0 = 1.3$\,GeV so that it had the same form as the other sea quarks. 

Our PDFs are parameterized at the initial scale
$Q_0 = 1.3$ GeV as
\begin{align}
    xf_i^{p/A}(x,Q_0)=c_0x^{c_1}(1-x)^{c_2}e^{c_3x}(1+e^{c_4}x)^{c_5} \ ,
\end{align}
and the nuclear $A$ dependence is encoded in the coefficients as
\begin{align}
    c_k \longrightarrow c_k(A) \equiv p_{k}+ a_{k}(1-A^{-b_{k}}) \ ,
\end{align}
where $k = \{1, ..., 5\}$.
The 16 free parameters used for the nCTEQ15 set describe the $x$-dependence of the 
\{$g, u_v, d_v, \bar{d} {+} \bar{u}$\} PDF combinations, and we do not vary the $\bar{d}/\bar{u}$ parameters; see Ref.~\cite{Kovarik:2015cma} for details. 
As in the nCTEQ15WZ analysis, we have added three strange PDF parameters: $\{a_{0}^{s+\bar{s}}, a_{1}^{s+\bar{s}}, a_{2}^{s+\bar{s}}\}$; these parameters correspond to the nuclear modification of the overall normalization, the low~$x$ exponent and the large-$x$ exponent of the strange distribution,
respectively. 
In total,  the 19 open parameters are:
\begin{align*}
    \{a^{u_v}_{1},\ a^{u_v}_{2},\ a^{u_v}_{4},\ a^{u_v}_{5},\ a^{d_v}_{1},\ a^{d_v}_{2},\ a^{d_v}_{5},\ a^{\bar{u}+\bar{d}}_{1},\  a^{\bar{u}+\bar{d}}_{5},\\ a^{g}_{1},\ a^{g}_{4},\ a^{g}_{5},\
    b^{g}_{0},\ b^{g}_{1},\ b^{g}_{4},\  b^{g}_{5},\ \bm{a^{s+\bar{s}}_{0},\ a^{s+\bar{s}}_{1},\ a^{s+\bar{s}}_{2}}\} .
\end{align*}

To obtain the cross section for single inclusive hadron production, 
the  PDFs of the two initial state particles are convoluted with the cross section of the partonic subprocess and the final state fragmentation function:
\begin{align}
\sigma_{p{+}\mathrm{Pb}\to h+X} = f_{a}^{P} \otimes f_{b}^{\mathrm{Pb}} \otimes \hat{\sigma}_{ab\to c}  \otimes D_c^{h} 
\ ,
\label{eq:xsec}
\end{align}
where $h$ is the produced light hadron and a sum over all possible subprocesses $ab\to c+X$ is understood.
The twist-2 factorization formula has an error which
is suppressed by a power of the ratio $\Lambda/Q$ where
$\Lambda$ is a hadronic scale and $Q$ the hard scale of the process (for example the $p_T$ of the light hadron).
This factorization formula is the result of a rigorous factorization theorem (see \cite{Collins:1989gx,Albino:2008gy} and references therein) originally devised for $pp$ collisions. 
It is supposed to hold true also for $pA$ collisions; however, the error (higher twist terms)
is possibly enhanced by the nuclear $A$ and one has to assess phenomenologically which minimum value for the hard scale is necessary for the twist-2 factorization formula
to be a good approximation.\footnote{%
For additional details regarding target mass corrections, see Refs.~\cite{Schienbein:2007gr,tmc}.}
A detailed overview is given in Ref.~\cite{Aurenche:1999nz}.
Performing all convolutions for each data point in each iteration of a fit 
is however too computationally expensive. A solution is to perform the convolution of the proton PDF
(or deuteron PDF in case of RHIC data) and the pion FF ahead of time and store the results to a grid; thus, the cross section evaluation can be reduced to a single convolution during the fitting process.
In order to perform the corresponding calculations and produce such grids we have
modified the INCNLO~\cite{INCNLO} program. 
The obtained grids have been validated to reproduce the full calculation within a margin
significantly smaller than the data uncertainty.

\begin{table}[tb]
\renewcommand{\arraystretch}{1.2}  %
	\caption{Overview of the available data sets, including their center of mass energy, observable, and number of data points.
	\\
	}
	\centering	
	\begin{tabular}{|c|c|c|c|c|c|c|c|}
		\hline 
		Data set & Ref. & ID & $\sqrt{s_{NN}}$\,[GeV] & Observ. & No. points \\ 
		\hline
		\hline
		PHENIX $\pi^0$ &~\cite{Adler:2006wg}  & 4003 & 200 & $R_{dAu}$ & 21  \\ 
		\hline 
		PHENIX $\eta$ &~\cite{Adler:2006wg} & 4403 & 200 & $R_{dAu}$ & 12  \\ 
		\hline 
		PHENIX $\pi^\pm$ &~\cite{Adare:2013esx}   & 4103 & 200 & $R_{dAu}$ & 20  \\ 
		\hline 
		PHENIX $K^\pm$ &~\cite{Adare:2013esx} & 4203 & 200 & $R_{dAu}$ & 15  \\ 
		\hline 
		STAR$ \pi^0$ &~\cite{Abelev:2009hx} & 4002 & 200 & $R_{dAu}$ & 13 \\ 
		\hline 
		STAR $\eta$ &~\cite{Abelev:2009hx} & 4402 & 200 & $R_{dAu}$ & 7 \\ 
		\hline 
		STAR $\pi^\pm$ &~\cite{Adams:2006nd} & 4102 & 200 & $R_{dAu}$ & 23 \\ 
		\hline 
		ALICE 5\,TeV $\pi^0$ &~\cite{Acharya:2018hzf} & 4001 & 5020 & $R_{pPb}$ & 31 \\ 
		\hline 
		ALICE 5\,TeV $\eta$ &~\cite{Acharya:2018hzf} & 4401 & 5020 & $R_{pPb}$ & 16 \\ 
		\hline 
		ALICE 5\,TeV $\pi^\pm$ &~\cite{Adam:2016dau} & 4101 & 5020 & $R_{pPb}$ & 58 \\ 
		\hline 
		ALICE 5\,TeV $K^\pm$ &~\cite{Adam:2016dau} & 4201 & 5020 & $R_{pPb}$ & 58 \\ 
		\hline 
		ALICE 8\,TeV $\pi^0$ &~\cite{Acharya:2021yrj} & 4004 & 8160 & $R_{pPb}$ & 30 \\ 
		\hline 
		ALICE 8\,TeV $\eta$ &~\cite{Acharya:2021yrj} & 4404 & 8160 & $R_{pPb}$ & 14 \\ 
		\hline 
	\end{tabular}     	
	\label{tab:data_table}
\end{table}
\begin{figure*}[tb]
	\centering
	\includegraphics[width=1.6\columnwidth]{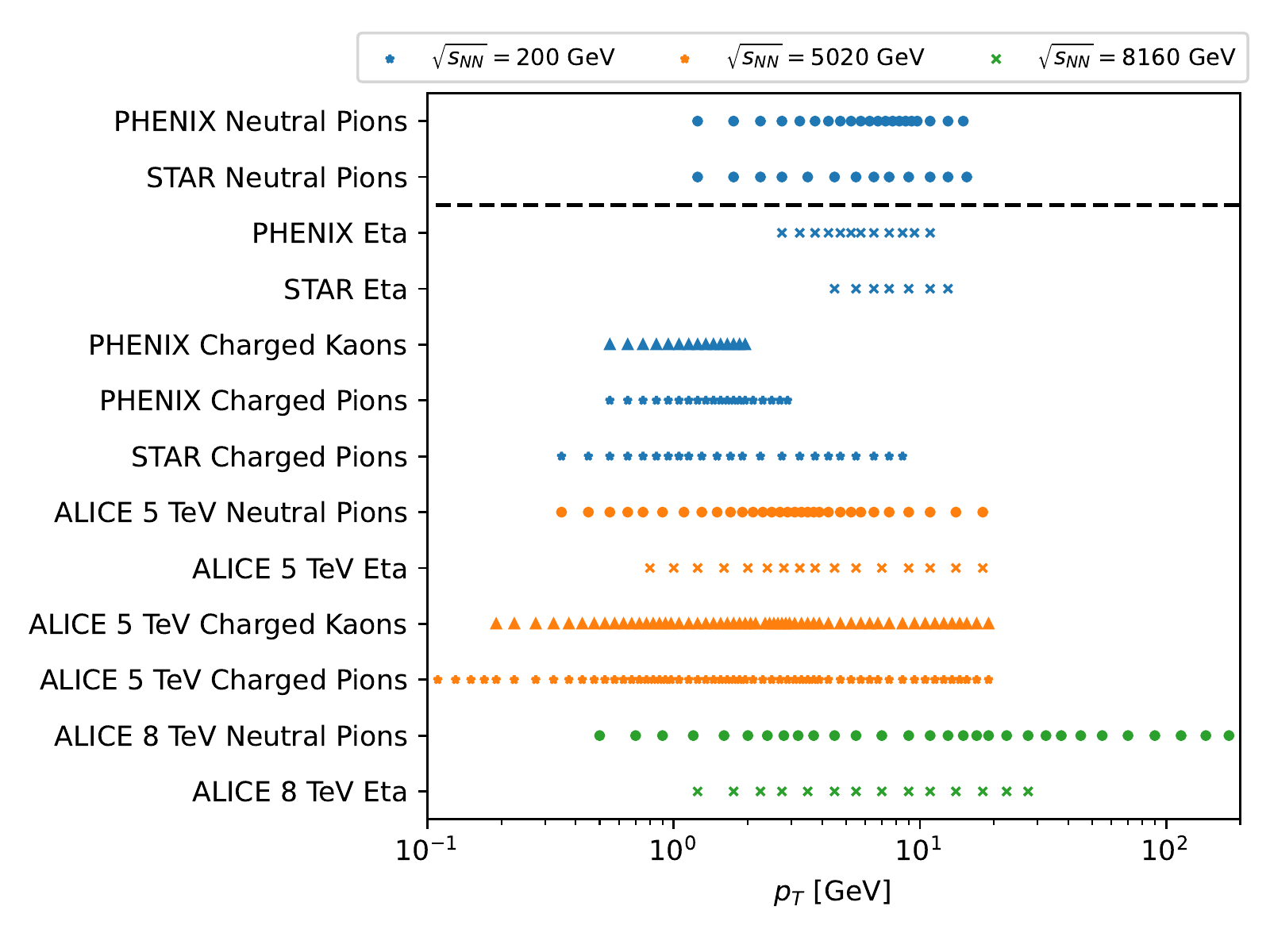}
	\caption{$p_T$ values of all data points, separated by set and colored by $\sqrt{s_{NN}}$. The data sets above the dashed line were included in the nCTEQ15 analysis.} 
	\label{fig:sihkin_pt}
\end{figure*}
\subsection{SIH data sets}
In this analysis we include the same deep inelastic scattering (DIS) and Drell-Yan (DY) lepton pair production data as in the nCTEQ15 analysis. 
The original nCTEQ15 analysis used also the RHIC pion data allowing to provide constraints on the gluon PDF. 
We now extend this analysis to include additional hadrons from the RHIC data, 
as well as   new data from recent ALICE analyses. 
We will study  four types of hadrons: neutral pions, charged pions, charged kaons and eta mesons. The charged mesons always appear as the average of their positively and negatively charged version, which is also how the neutral pions are calculated for their fragmentation functions. The data is taken at center of mass energies per nucleon of 200\,GeV (PHENIX, STAR), 5020\,GeV and 8160\,GeV (ALICE). Table~\ref{tab:data_table} gives an overview of the available data sets, while Fig.~\ref{fig:sihkin_pt} shows the $p_T$ distribution of the available data points for each set.

As in other types of experiments, 
kinematic cuts  are applied to  remove data that cannot be adequately described by the theory. 
For example,  in the very low $p_T$ region, the SIH  process becomes non-perturbative,
so we will impose a lower $p_T$ cut on the data. 
Additional restrictions may come from the FFs $D_i^h(z,Q^2)$ required to compute the cross sections.
The available sets of FFs are typically only reliable for momentum fractions $z_{\rm min}< z < 1$
with some minimal $z_{\rm min} \sim 0.1$. This issue will be further discussed in Sec.~\ref{sec:data_selection}.

All the single inclusive hadron production data is given in terms of ratios
\begin{align}
    R_{d\mathrm{Au}}^h &=\frac{1}{A_dA_\mathrm{Au}}\ \frac{\sigma_{d+\mathrm{Au}\rightarrow h+X}}{\sigma_{p+p\rightarrow h+X}}\quad(\mathrm{RHIC})\\
    R_{p\mathrm{Pb}}^h &=\frac{1}{A_\mathrm{Pb}}\ \frac{\sigma_{p+\mathrm{Pb}\rightarrow h+X}}{\sigma_{p+p\rightarrow h+X}}\quad(\mathrm{ALICE})
    \quad . 
\end{align}
All fits are also repeated including the $W^\pm$ and $Z$ production data from our recent nCTEQ15WZ~\cite{Kovarik:2015cma} analysis because  this data has a significant impact on the gluon PDF. Since the impact of SIH production on the high-$x$ region is negligible, we do not include a further comparison with the nCTEQ15HIX~\cite{Segarra:2020gtj} analysis.

Since all SIH data sets have considerable normalization uncertainties, we need to account for those to avoid data sets pulling the fit in unphysical directions due to their normalization errors. This is done by including a normalization factor for each set as a parameter in the fit. By adding a $\chi^2$ penalty that increases as normalizations stray further from unity we ensure that they are kept at reasonable values. This is done according to the prescription outlined in the Appendix~\ref{sec:norm} to avoid the D'Agostini bias \cite{DAgostini:1993arp}.

\section{Fragmentation functions}\label{sec:theoryFF}
\begin{table}[tb]
\renewcommand{\arraystretch}{1.2}
	\caption{Overview of the  available sets of FFs $D_i^h(z,Q)$ and their available particles.
        \\
        }
	\centering	
	\begin{tabular}{|c|c|c|c|c|c|}
		\hline 
		FF & \ Ref. \ & \ Year \ & Available particles \\ 
		\hline
		\hline
		BKK &~\cite{Binnewies:1994ju}  & 1994 & $\pi_0, \pi^\pm, K^\pm$  \\ 
		\hline 
		KKP &~\cite{Kniehl:2000fe} & 2000 & $\pi_0, \pi^\pm, K^\pm$  \\ 
		\hline 
		KRETZER &~\cite{Kretzer:2000yf}   & 2000 & $\pi_0, \pi^\pm, K^\pm$  \\ 
		\hline 
		\textbf{HKNS07} &~\cite{Hirai:2007cx} & 2007 & $\pi_0, \pi^\pm, K^\pm$  \\ 
		\hline 
		AKK &~\cite{Albino:2008fy} & 2008 & $\pi_0, \pi^\pm, K^\pm$ \\ 
		\hline 
		\textbf{NNFF} &~\cite{Bertone:2017tyb} & 2017 & $\pi_0, \pi^\pm, K^\pm$ \\ 
		\hline 
		\textbf{JAM20} &~\cite{Moffat:2021dji} & 2021 & $\pi_0, \pi^\pm, K^\pm$ \\ 
		\hline 
		\textbf{DSS14} &~\cite{deFlorian:2014xna} & 2014 & $\pi_0, \pi^\pm$ \\ 
		\hline 
		\textbf{DSS17} &~\cite{deFlorian:2017lwf} & 2017 & $K^\pm$ \\ 
		\hline 
		AESSS &~\cite{Aidala:2010bn} & 2011 & $\eta$ \\ 
		\hline 
	\end{tabular}     	
	\label{tab:FF_table}
\end{table}

In this analysis we investigate a total of ten 
different fragmentation functions (FFs), as listed in Table~\ref{tab:FF_table}. We will give a brief overview of their properties, and then compare them both in terms of predictions for proton-proton and proton-nucleus collisions. 

\subsection{Available fragmentation functions}

In a manner complementary to the PDFs, the  FFs describe the hadronization 
of a partonic constituent into a final-state hadron. 
Both the PDFs and FFs are  non-perturbative objects, and hence must be obtained 
by fitting to data.
The pioneering fits of FFs used only single-inclusive hadron production in electron-positron annihilation. 
In more recent analyses, groups have added data from semi-inclusive deep inelastic lepton-nucleon scattering and other processes to improve the accuracy and kinematic range of their fits~\cite{dEnterria:2013sgr, Metz:2016swz}. 

A selection of FFs for various mesons\footnote{We include in our analysis only data for the inclusive production of pions, eta mesons and kaons. Note that the neutral pion FFs are always calculated as the average of $\pi^+$ and $\pi^-$ FFs.} is shown in  Table~\ref{tab:FF_table}.
The fragmentation function sets listed in bold  provide uncertainties. We will henceforth denote the combination of DSS14 for pions and DSS17 for kaons simply as ``DSS."
Note that HKNS16~\cite{Hirai:2016loo} exists as an updated version of HKNS07, but no code is available to use these updated fragmentation functions. 

Additional fragmentation functions exist for other final states like protons, antiprotons and unidentified charged hadrons (SGK18~\cite{Soleymaninia:2018uiv}, NNFF1.1h~\cite{Bertone:2018ecm}). Some of the aforementioned FFs (AKK, BKK, HKNS, KKP, KRETZER) also include those, but we exclude those from the analysis due to the comparatively large uncertainties both on the data and the fragmentation functions. There have also been studies on the effect of the nuclear medium on the fragmentation~\cite{Sassot:2009sh,Sievert:2019cwq}, but we exclude the fragmentation functions obtained there from our analysis in order to avoid double counting of the shared data points. Also, any possible medium modifications of the FFs are small compared to the uncertainties of the FFs.

\subsection{Comparison in proton-proton collisions}\label{sec:ppPi}

\begin{figure*}[tb]
		\centering
 		\includegraphics[width=0.9\textwidth]{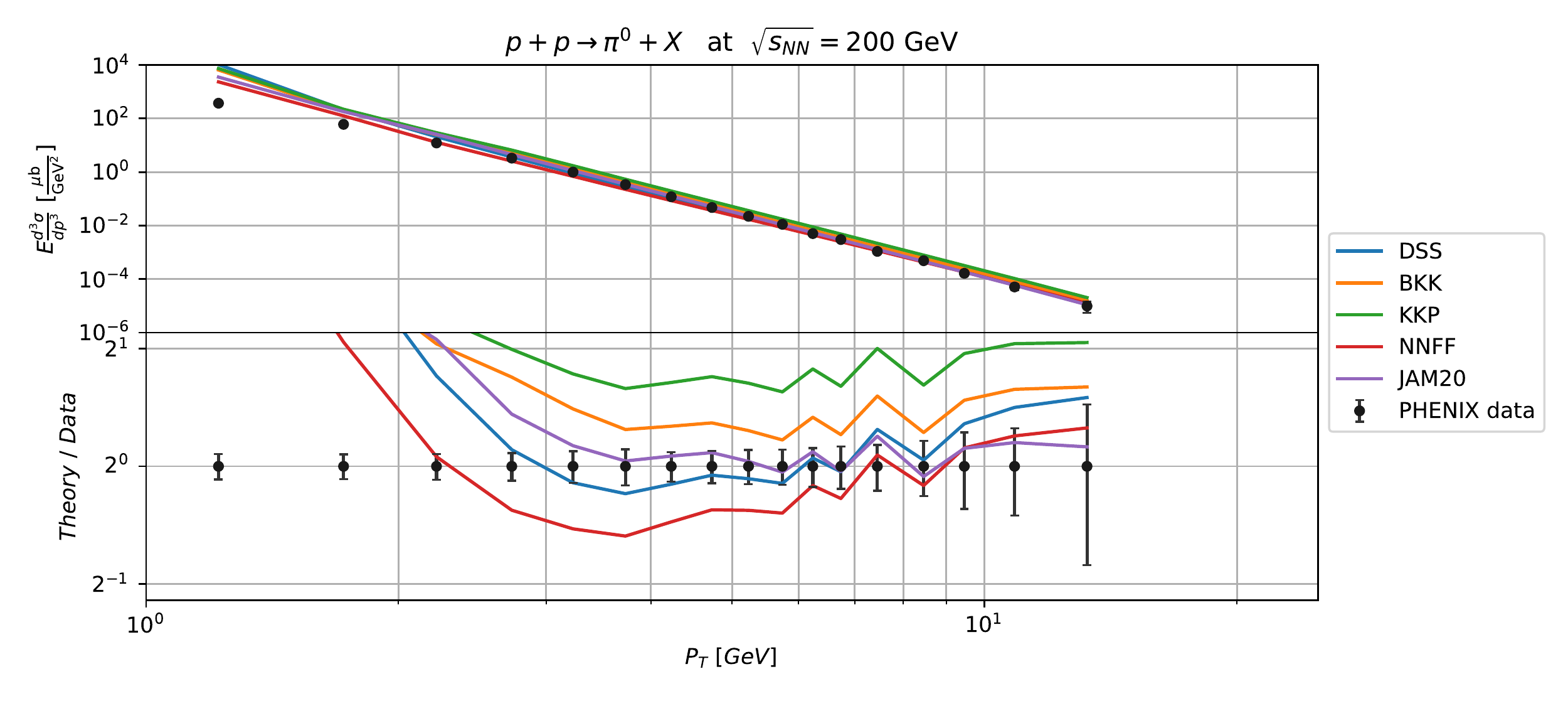}
		\centering
		\includegraphics[width=0.9\textwidth]{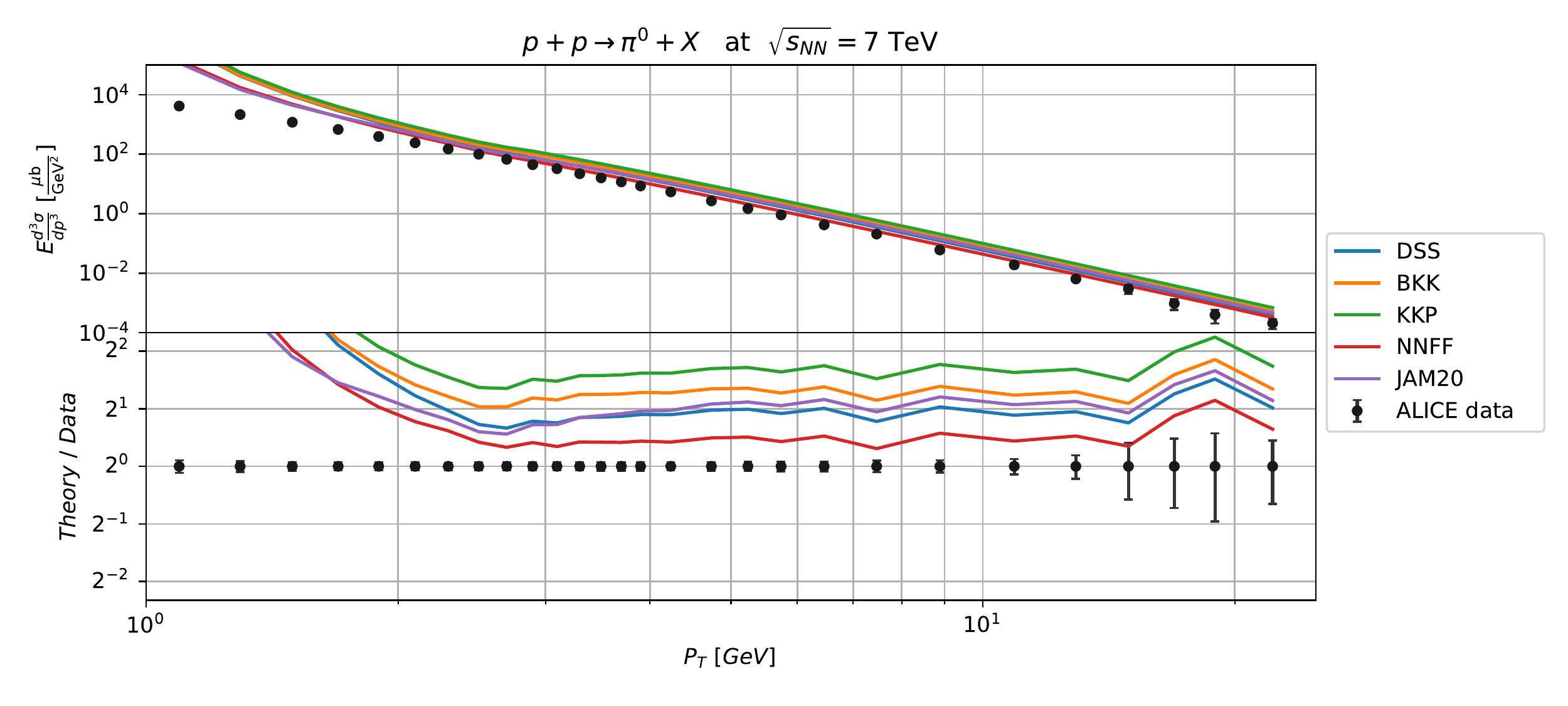}
	\caption{Comparison of predictions made with different fragmentation functions for ${p+p\rightarrow\pi^0+X}$. The calculations are done using nCTEQ15 proton PDFs. Both panels show data for neutral pions, with PHENIX data \cite{Adler:2003pb} in the upper and ALICE data \cite{Abelev:2012cn} in the lower one. 
	}
	\label{fig:ppPi0}
\end{figure*}

Before we examine the $d \mathrm{Au}$ and $p\mathrm{Pb}$ cases, 
let us first  look at the 
${p+p \rightarrow h+X}$ baseline process to help  understand the limitations of the theory prediction due to the uncertainties  of the FFs. 

Figure~\ref{fig:ppPi0} shows a comparison
of predictions from various FF sets with data taken by the PHENIX and ALICE experiments at $\sqrt{s_{NN}}=200$\,GeV and 7000\,GeV respectively.
The nCTEQ15 proton PDF is used in these calculations.
The fragmentation functions displayed are BKK, KKP, DSS, NNFF and JAM20. 
The KRETZER and HKNS FFS are not shown   
as their more strict kinematic restrictions preclude comparison with the ALICE data. 

At 200\,GeV all fragmentation functions are able to describe the data for 
$p_T \geq 3$\,GeV  with a satisfactory $\chidof<1$ if one allows for a normalization shift. 
Below $p_T$  of 3\,GeV,  all the curves display a significant upward slope in
Fig.~\ref{fig:ppPi0}  which points to a  qualitative disagreement. 
There is also a slight upwards slope towards higher $p_T$ for all fragmentation functions, 
but it is well within the data uncertainties, considering the allowed normalization shift. 
 
At ALICE energies, the data can be well described by BKK, KKP, DSS and NNFF down to $p_T$ values of 3\,GeV if a normalization is introduced. 
The JAM20 result is also relatively constant across the $p_T$ range, but it begins to decrease slightly
for lower $p_T$ values in the range of $p_T\approx5$\,GeV and below.\footnote{%
In principle, for the computation of Fig.~\ref{fig:ppPi0} 
the FFs should be combined with their matching PDFs, 
\textit{i.e.,} JAM20 FFs with JAM20 PDFs, and 
DSS FFs with MSTW2008~\cite{Martin:2009iq} PDFs. 
Since our focus is the impact on the nuclear PDFs however, 
we use our proton baseline instead.}
Again, the theory predictions increasingly overshoot the data the further one goes below 3\,GeV. Since this effect is not fragmentation function dependent, it is also independent of the produced final state.

\begin{figure*}[tb]
		\centering
		\includegraphics[width=0.9\textwidth]{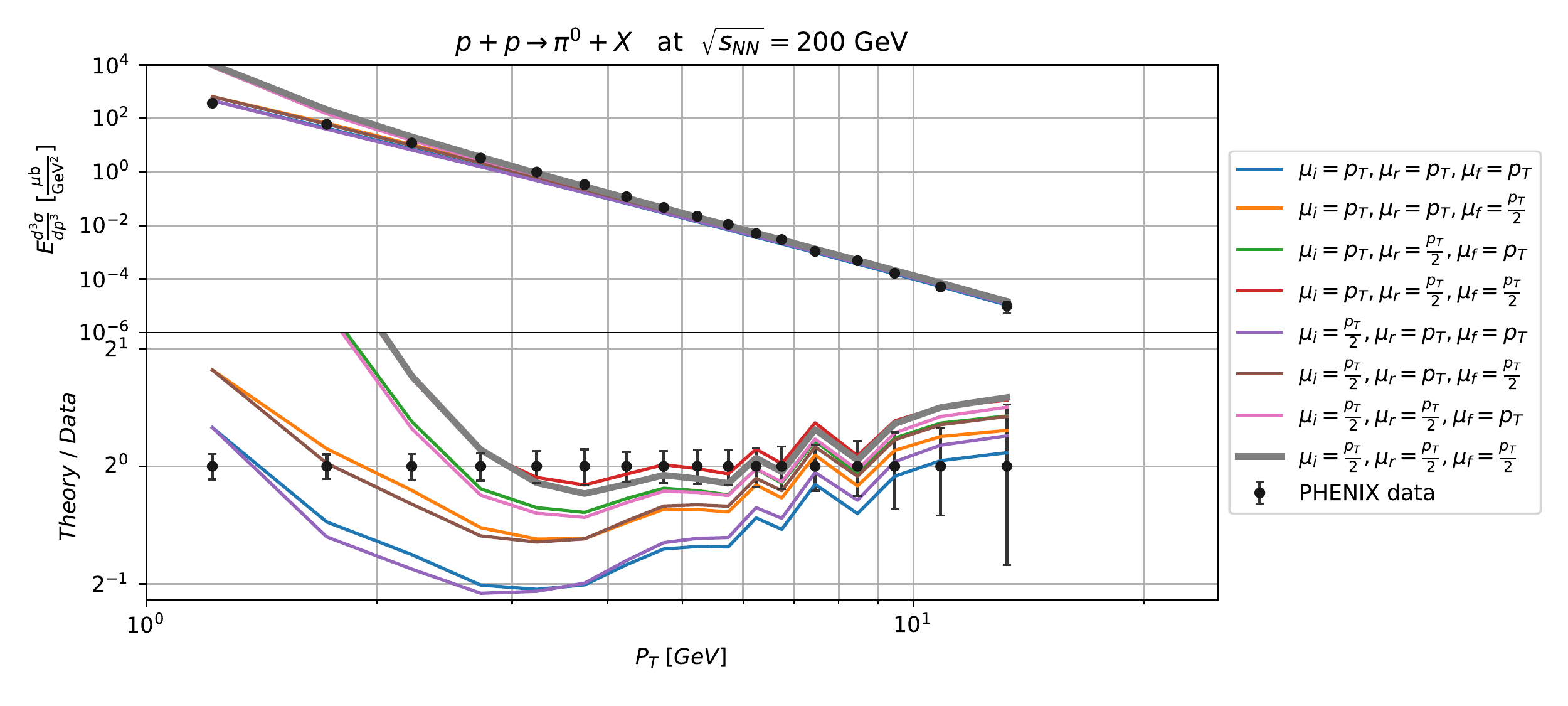}
		\centering
		\includegraphics[width=0.9\textwidth]{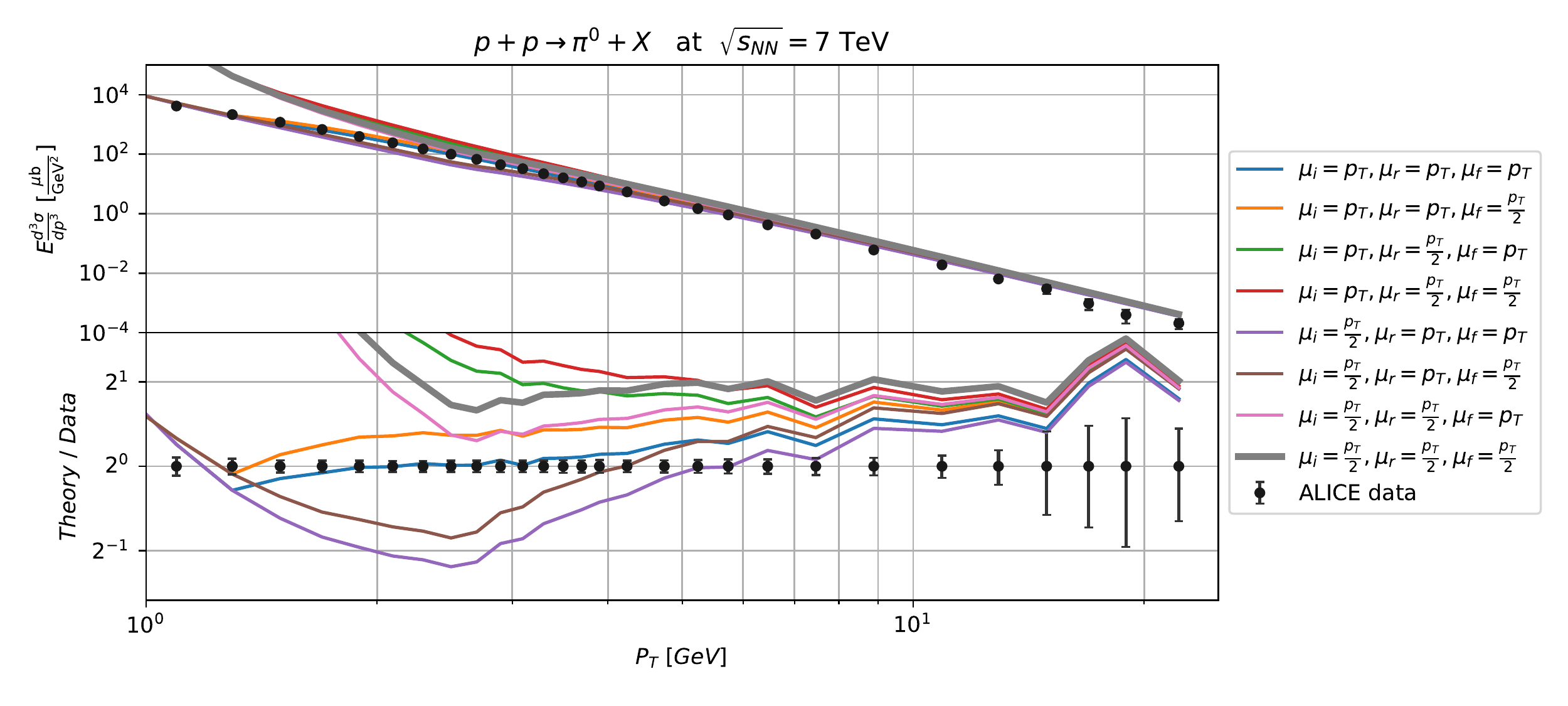}
	\caption{Comparison of predictions made with different scale choices for ${p+p\rightarrow\pi^0+X}$. The calculations are performed using nCTEQ15 proton PDFs with DSS fragmentation functions. Both panels show data for neutral pions, with PHENIX data in the upper panel and ALICE data in the lower one.}
	\label{fig:scale}
\end{figure*}
\subsection{Scale uncertainties}
The prediction for the SIH production cross section depends on three scale choices: the initial state factorization scale $\mu_i$, the final state factorization scale $\mu_f$ and the renormalization scale $\mu_r$. 
Frequently, they are taken to be $\mu_i = \mu_f = \mu_r = c\, p_T$, where $c$ is a constant that is commonly chosen as either $\nicefrac12$ or $1$, but there is no  unambiguous prescription for their choice. 

Figure~\ref{fig:scale} shows the prediction for pion production at 200\,GeV and 7\,TeV with each scale varied independently between the two common choices,  $c=\{\nicefrac12, 1\}$.
The case where all scales are equal to $\nicefrac12 \, p_T$ (bold, grey) gives the best description of the $200$\,GeV data; additionally, this  is also the only scale choice that 
yields  $\chidof<1$ for the 7\,TeV data with $p_T>3$\,GeV, if the normalization is chosen freely. 
Therefore, in the following comparisons we make the choice $c=\nicefrac{1}{2}$ going forward. 
This means that we need to freeze the initial state factorization scale ($\mu_i$) to the initial scale of our PDF evolution ($Q_0 =1.3$ GeV)
whenever $c \, p_T < Q_0$, i.e., for $p_T \le 2.6$ GeV:
\begin{align}
    \mu_{i} & =
    \begin{cases}
    1.3\ \text{GeV} \quad \text{for}\, \frac{1}{2} p_T < 1.3\ \text{GeV}\, ,
    \\
    \frac{1}{2} p_T \quad\text{otherwise}\, .
    \end{cases}
\end{align}
Otherwise we would have to interpolate the PDFs to scales
below the initial scale which is technically challenging.

Note that the 200\,GeV and 7\,TeV  data sets shown in Fig.~\ref{fig:scale} are actually included in the fit of the DSS FFs, where they impose a cut of $p_T\geq 5$\,GeV in their analysis which uses a scale choice of $\mu_i = \mu_f = \mu_r = p_T$ (blue curve).

\begin{figure*}[p]
		\centering
		\includegraphics[width=0.31\textwidth]{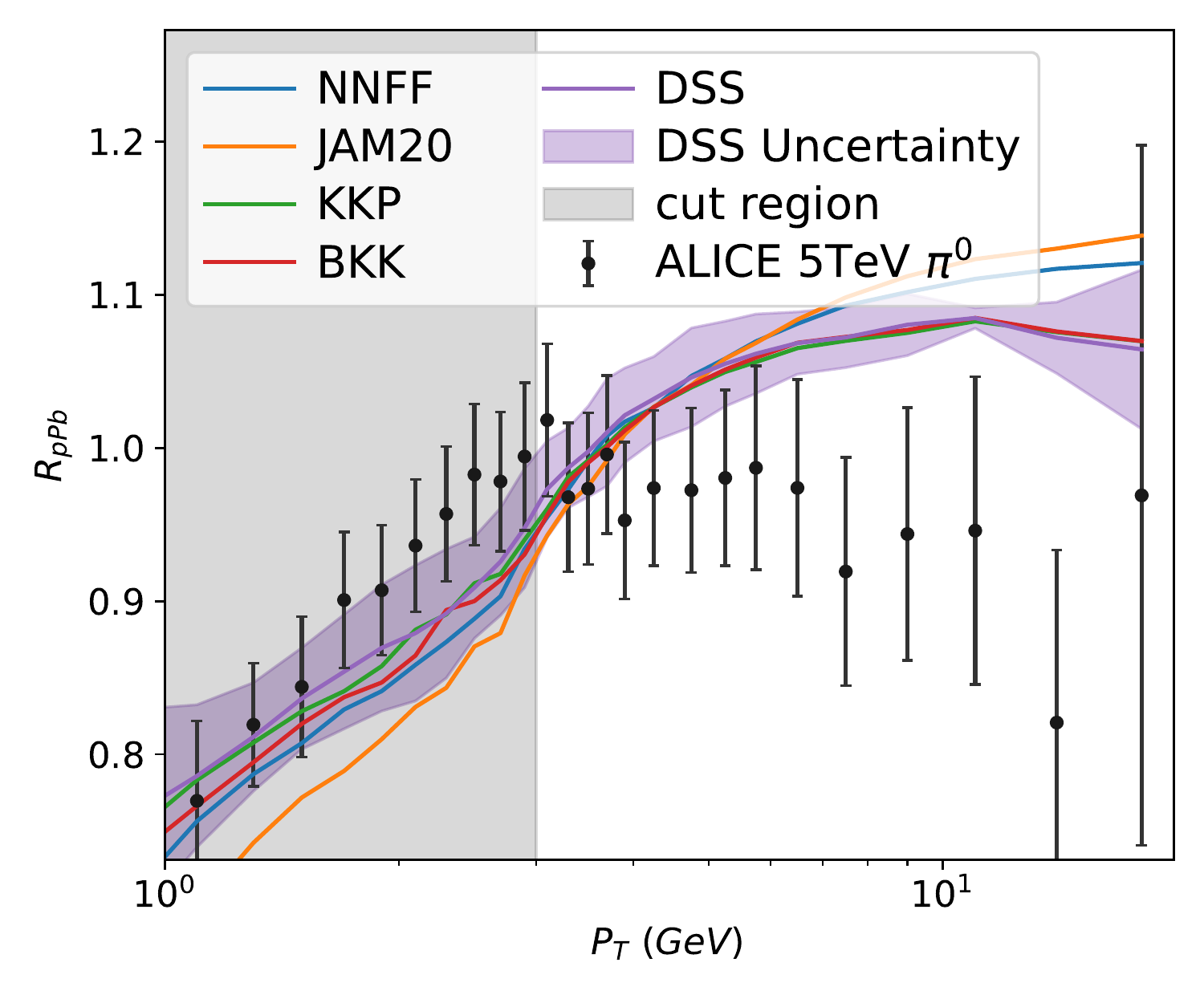}
		\includegraphics[width=0.31\textwidth]{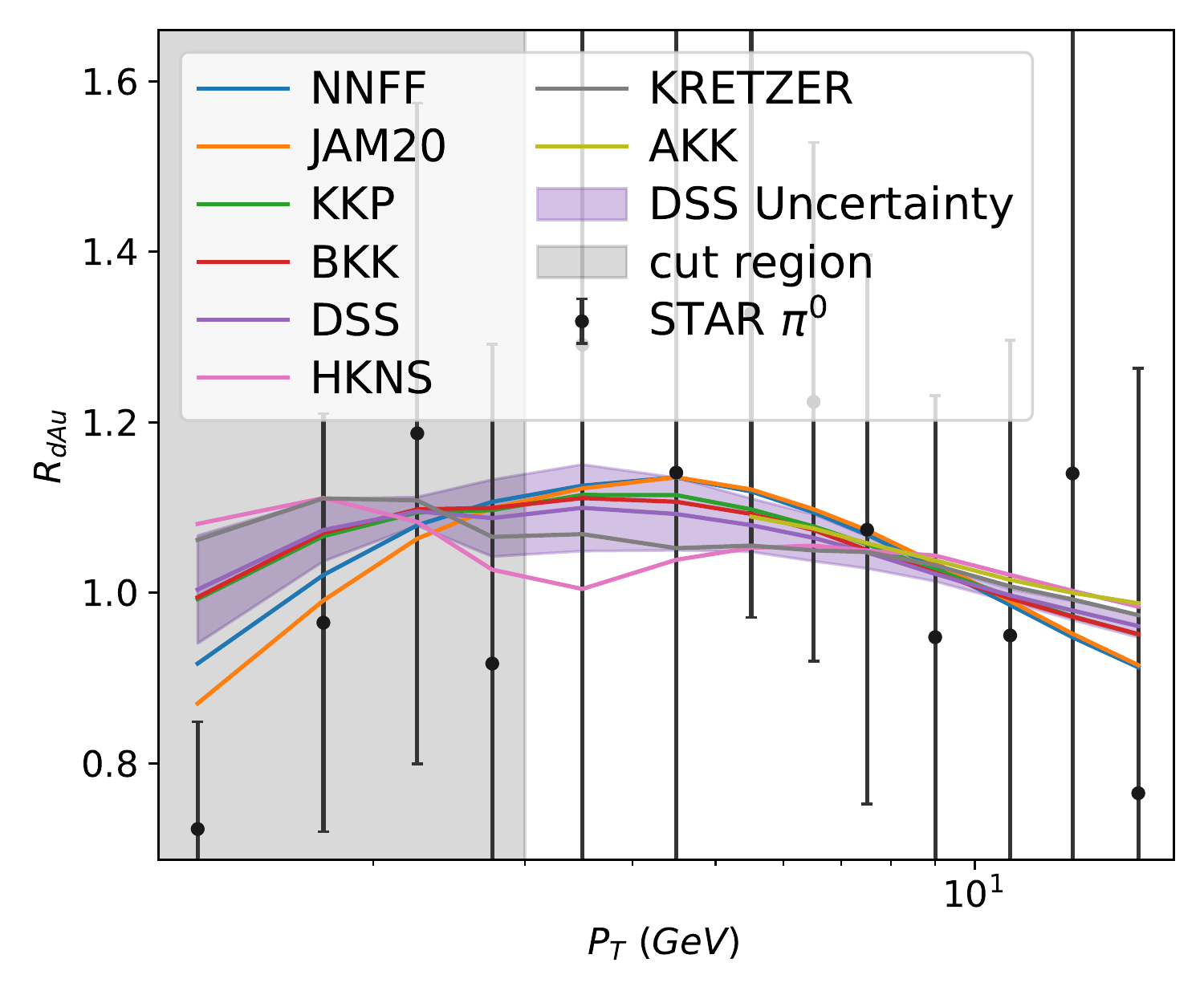}
		\includegraphics[width=0.31\textwidth]{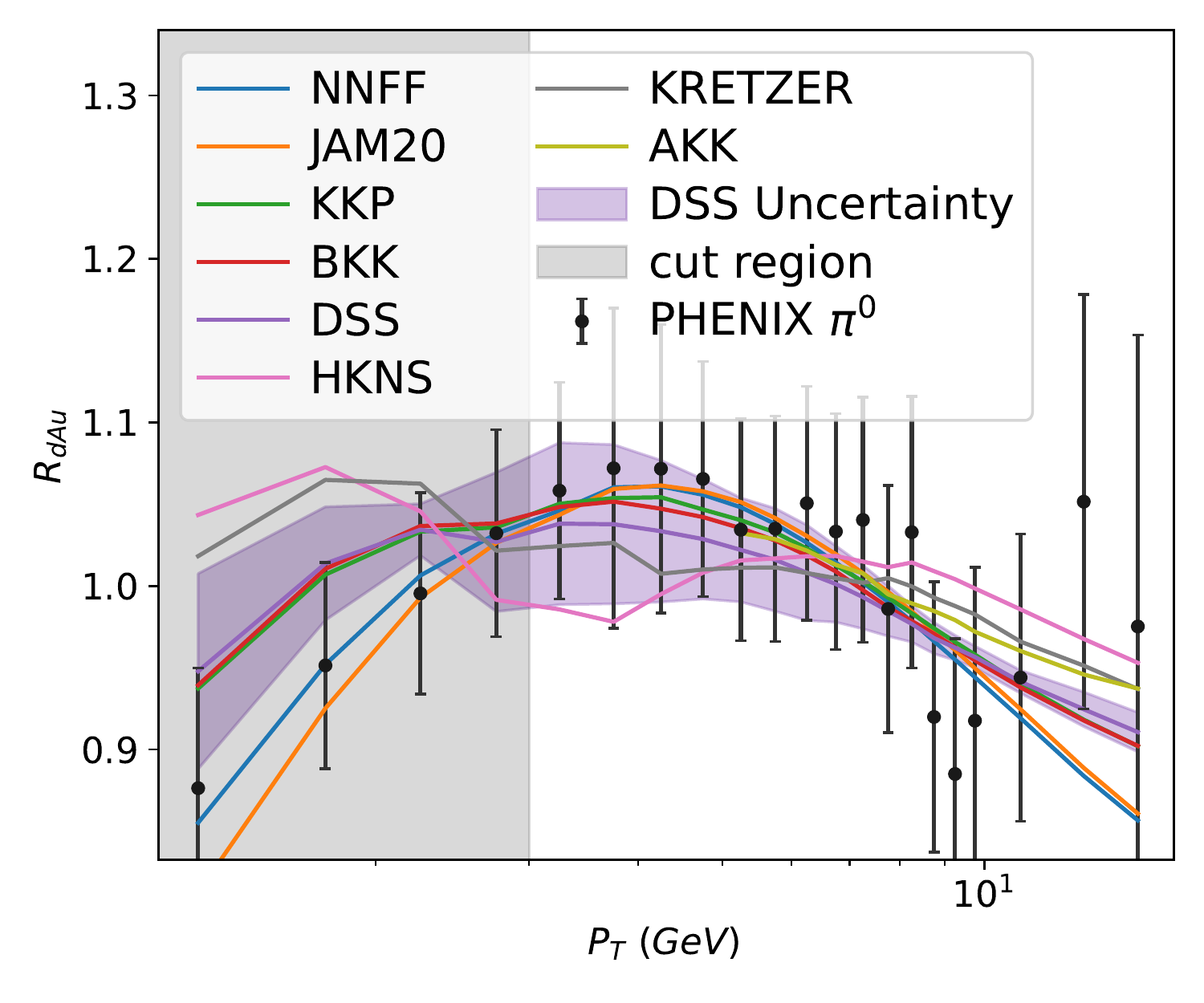}
		\includegraphics[width=0.31\textwidth]{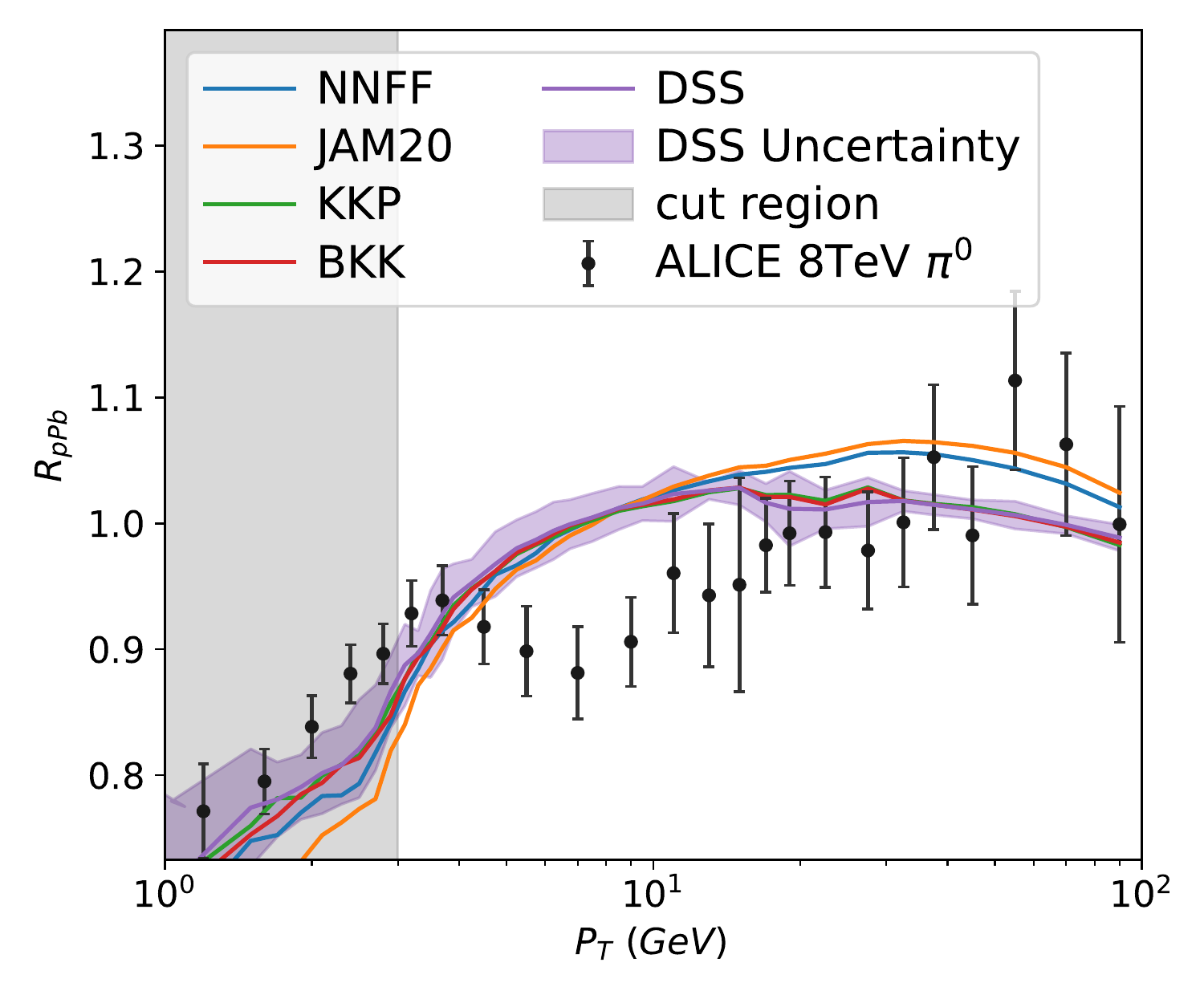}
		\includegraphics[width=0.31\textwidth]{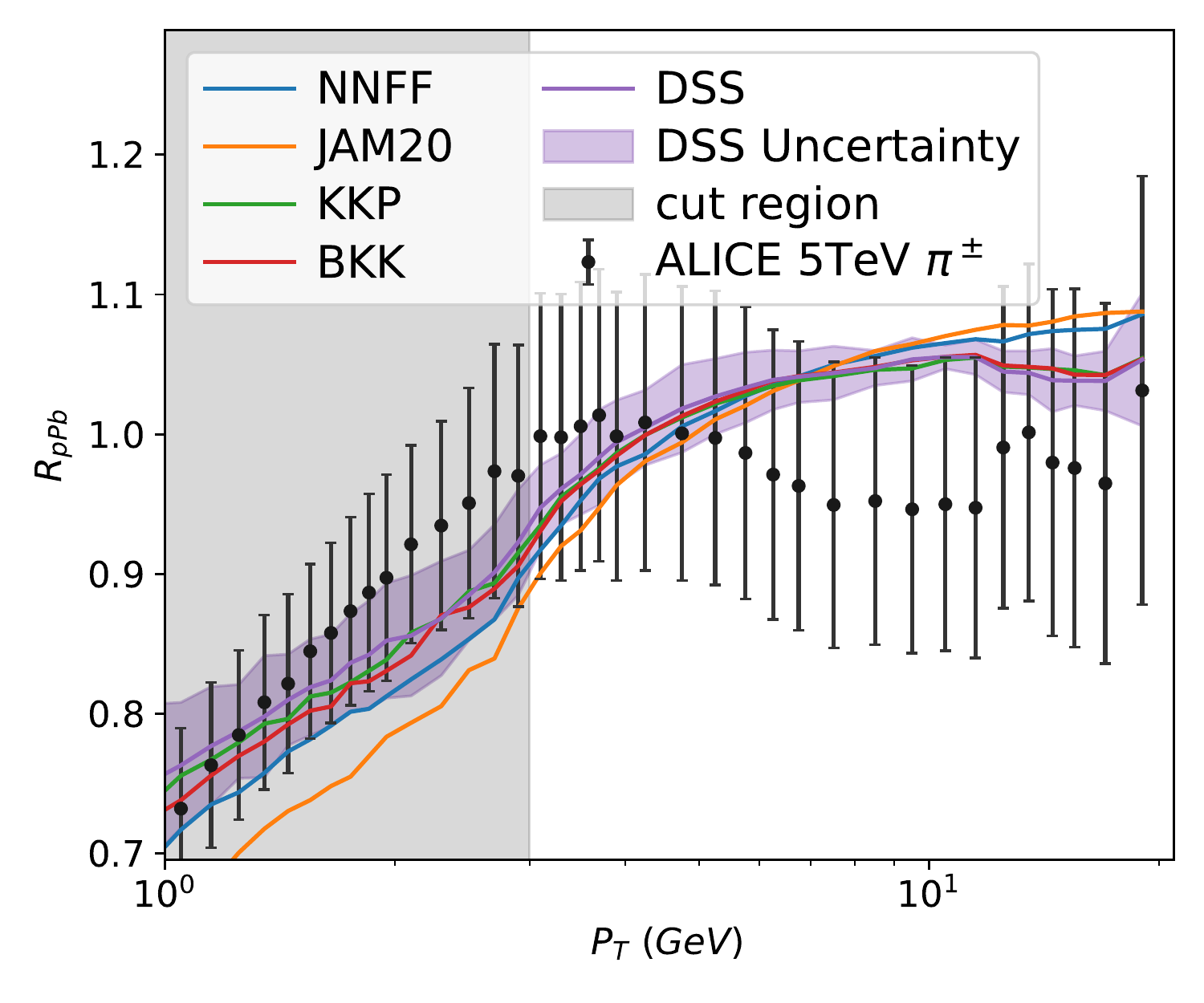}
		\includegraphics[width=0.31\textwidth]{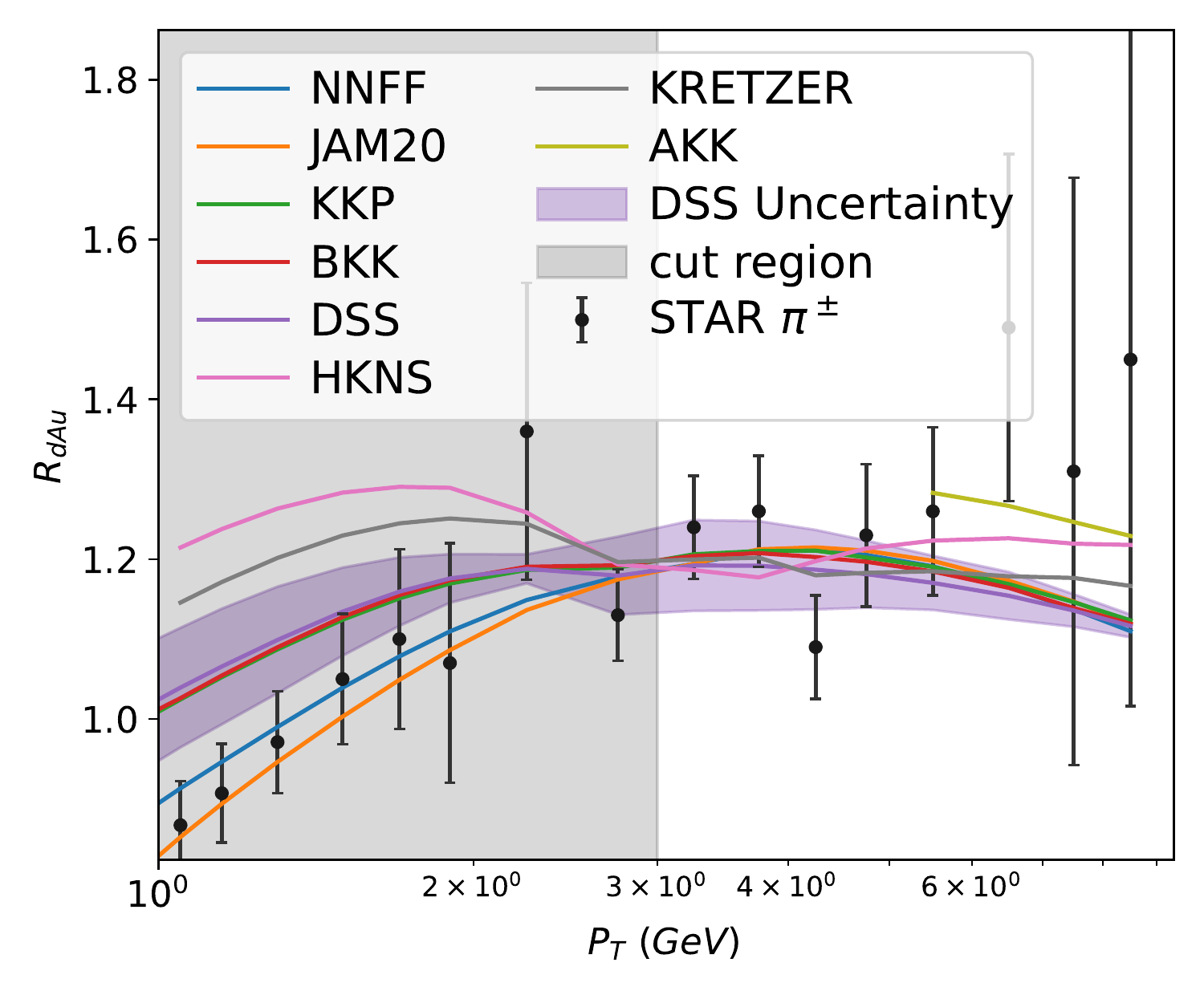}
		\includegraphics[width=0.31\textwidth]{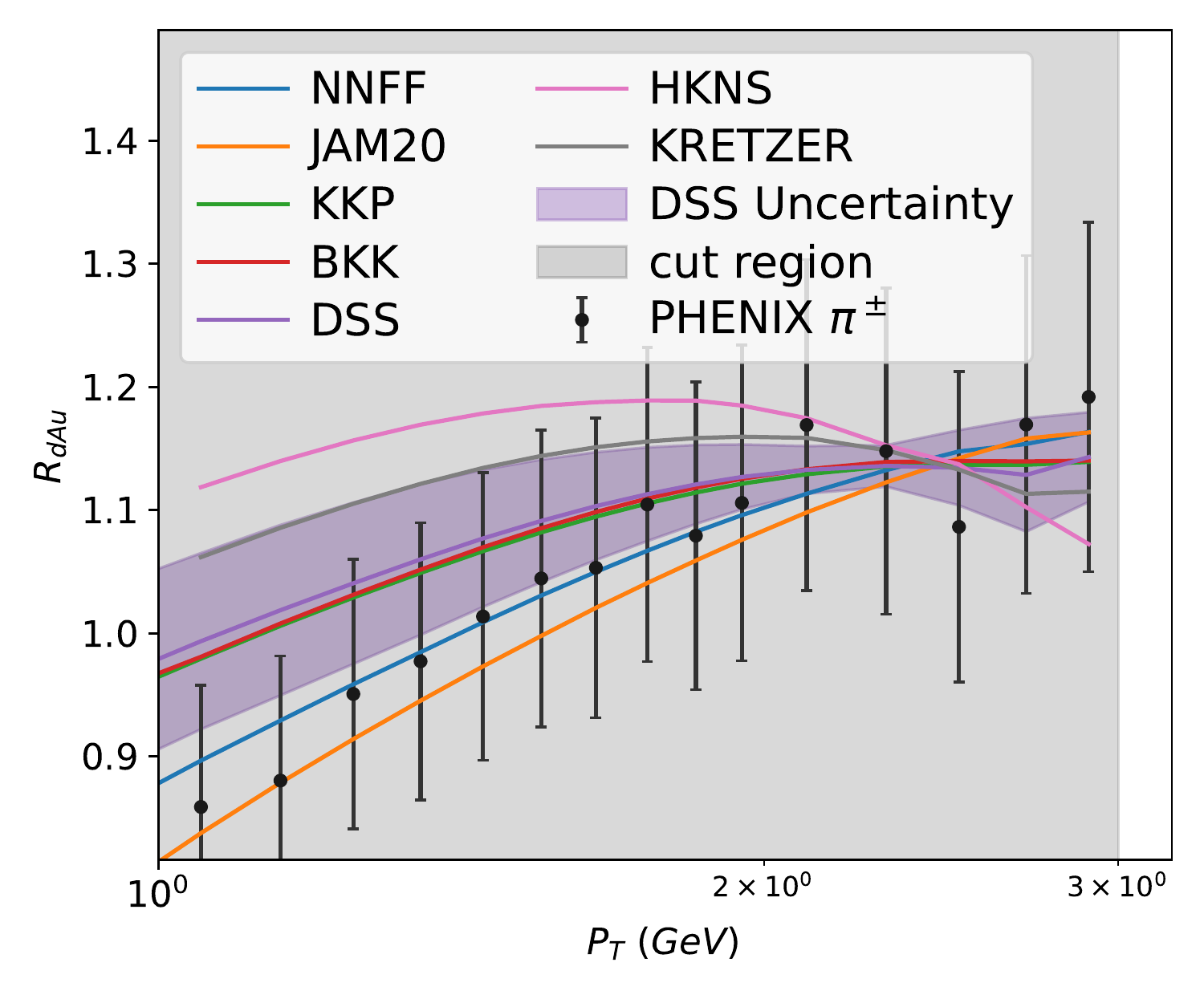}
		\includegraphics[width=0.31\textwidth]{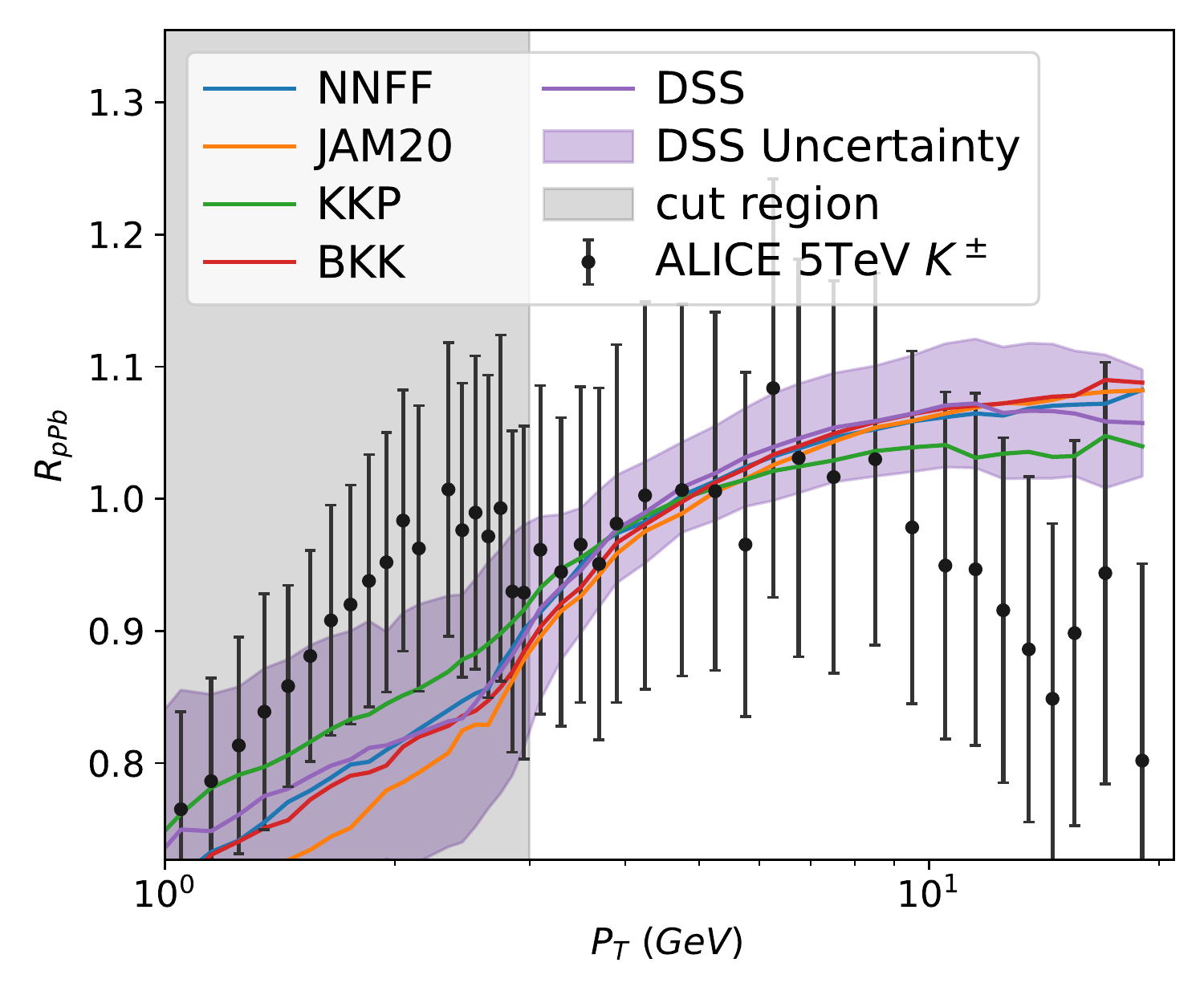}
		\includegraphics[width=0.31\textwidth]{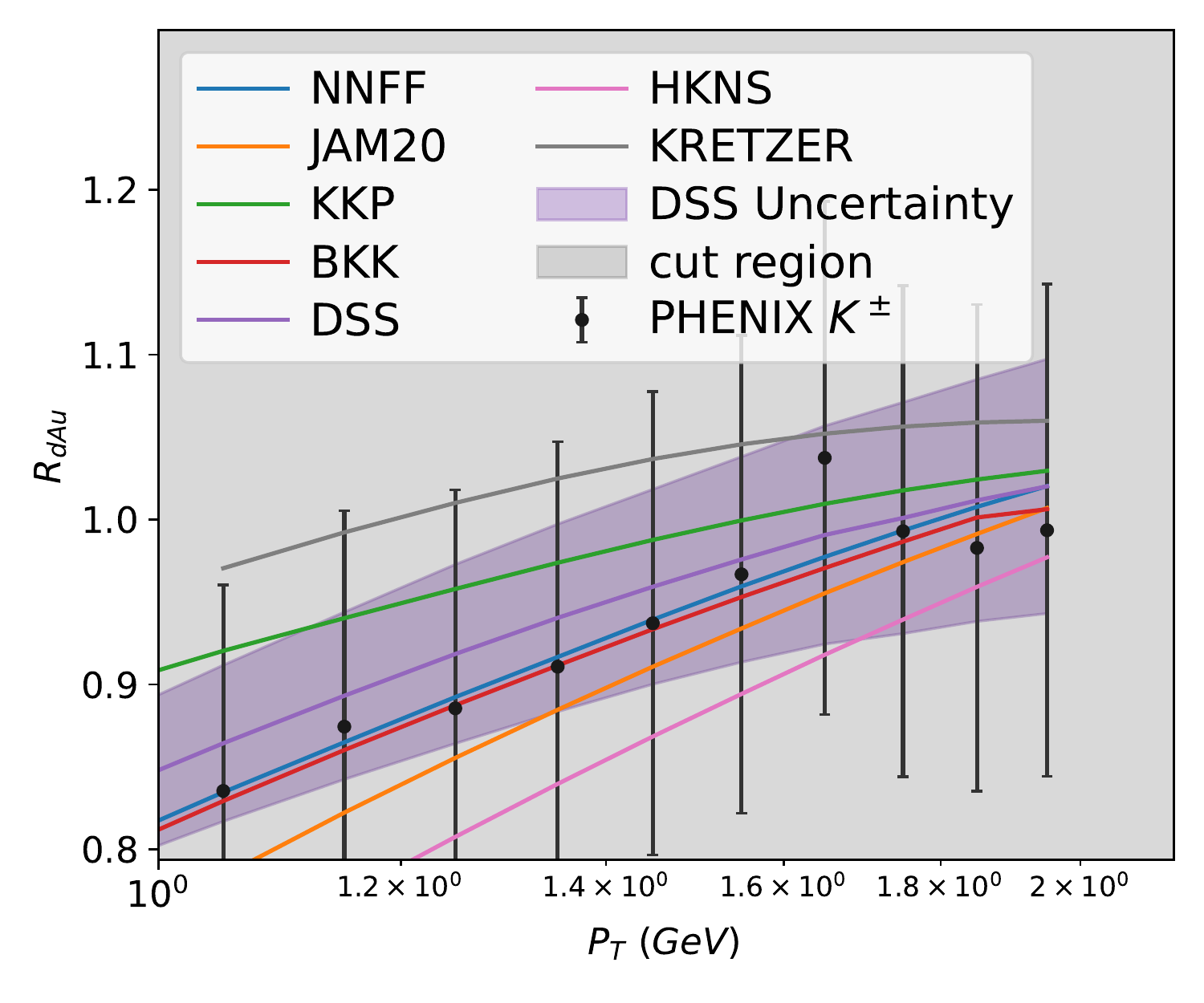}
		\includegraphics[width=0.31\textwidth]{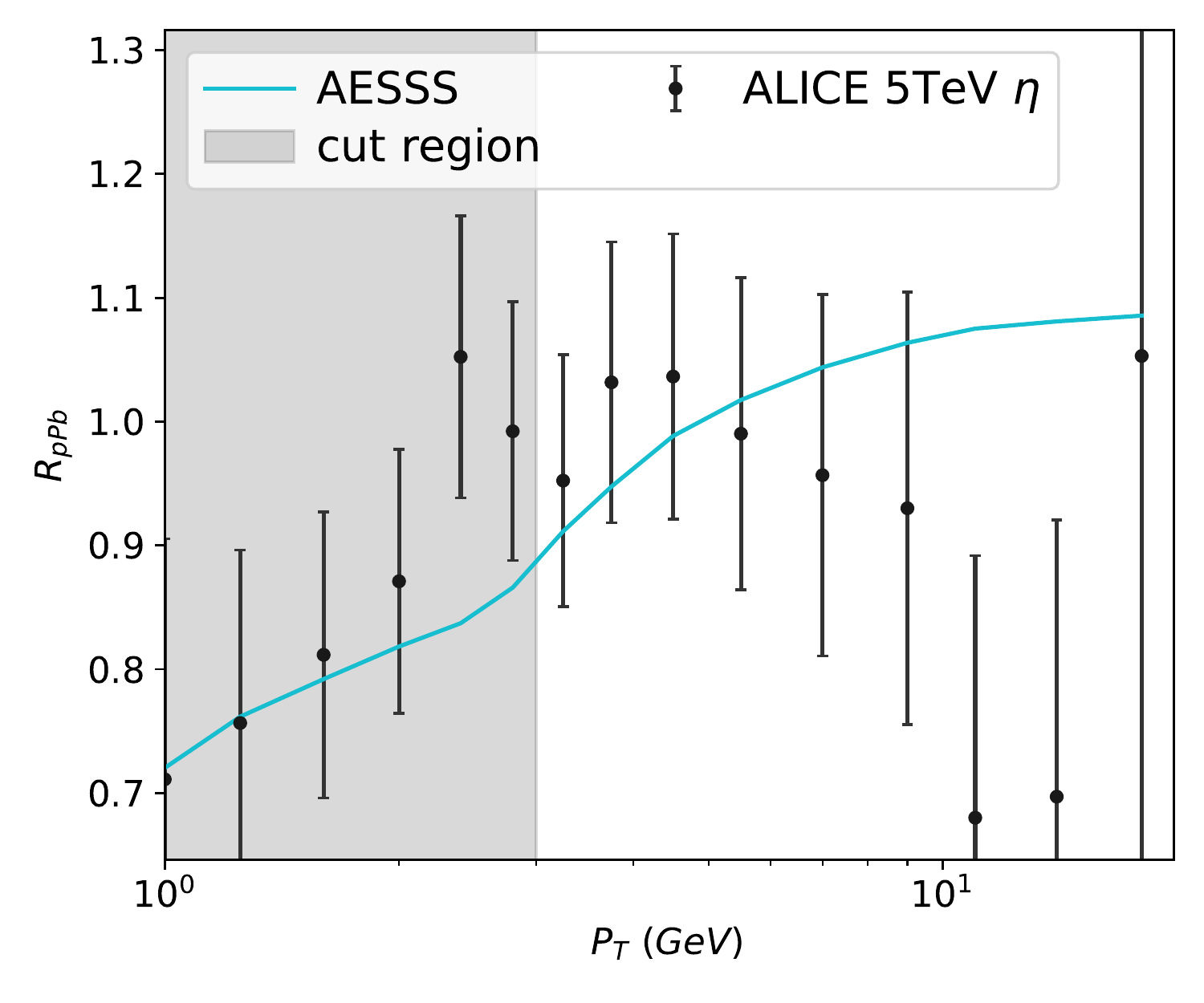}
		\includegraphics[width=0.31\textwidth]{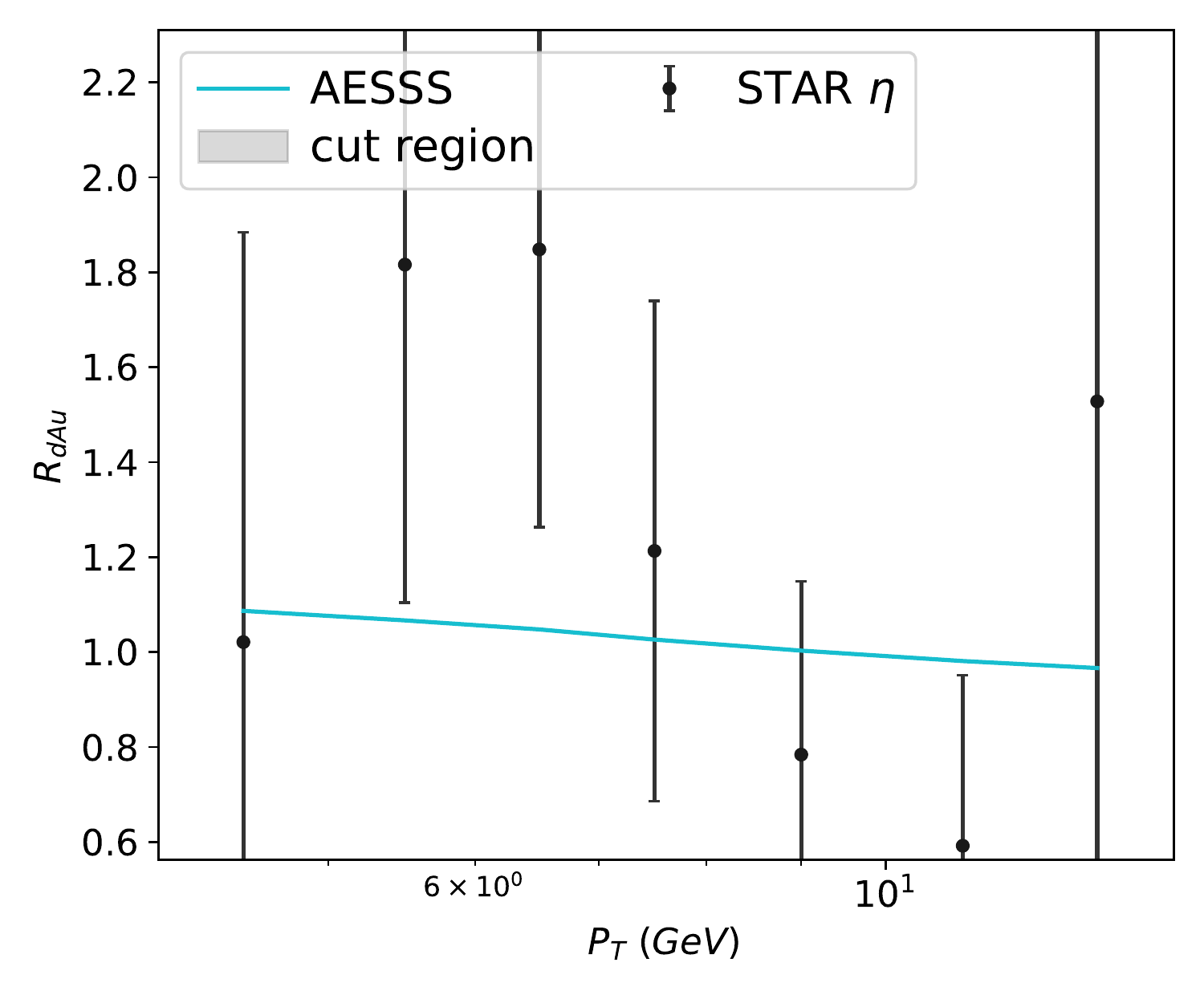}
		\includegraphics[width=0.31\textwidth]{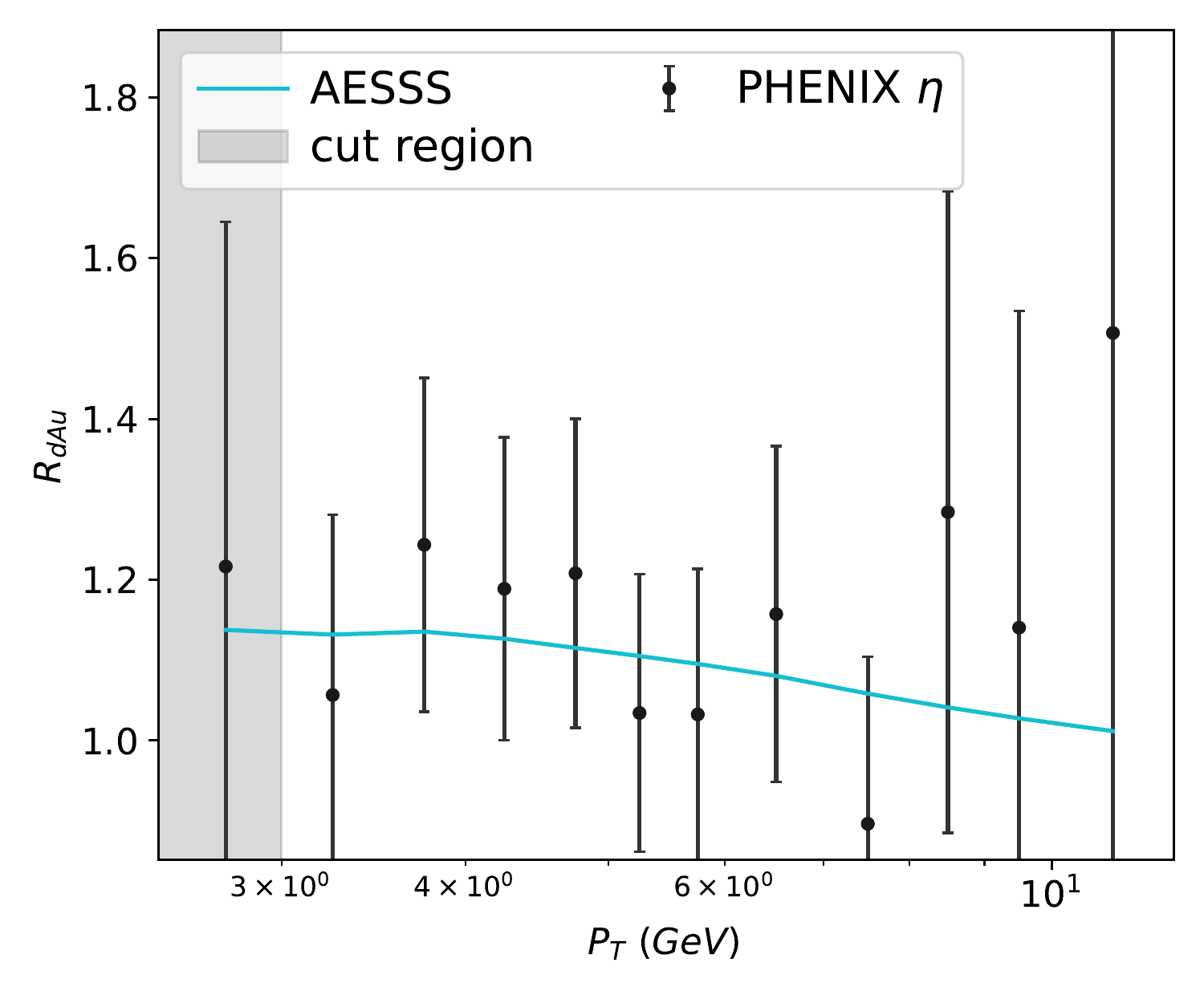}
		\includegraphics[width=0.31\textwidth]{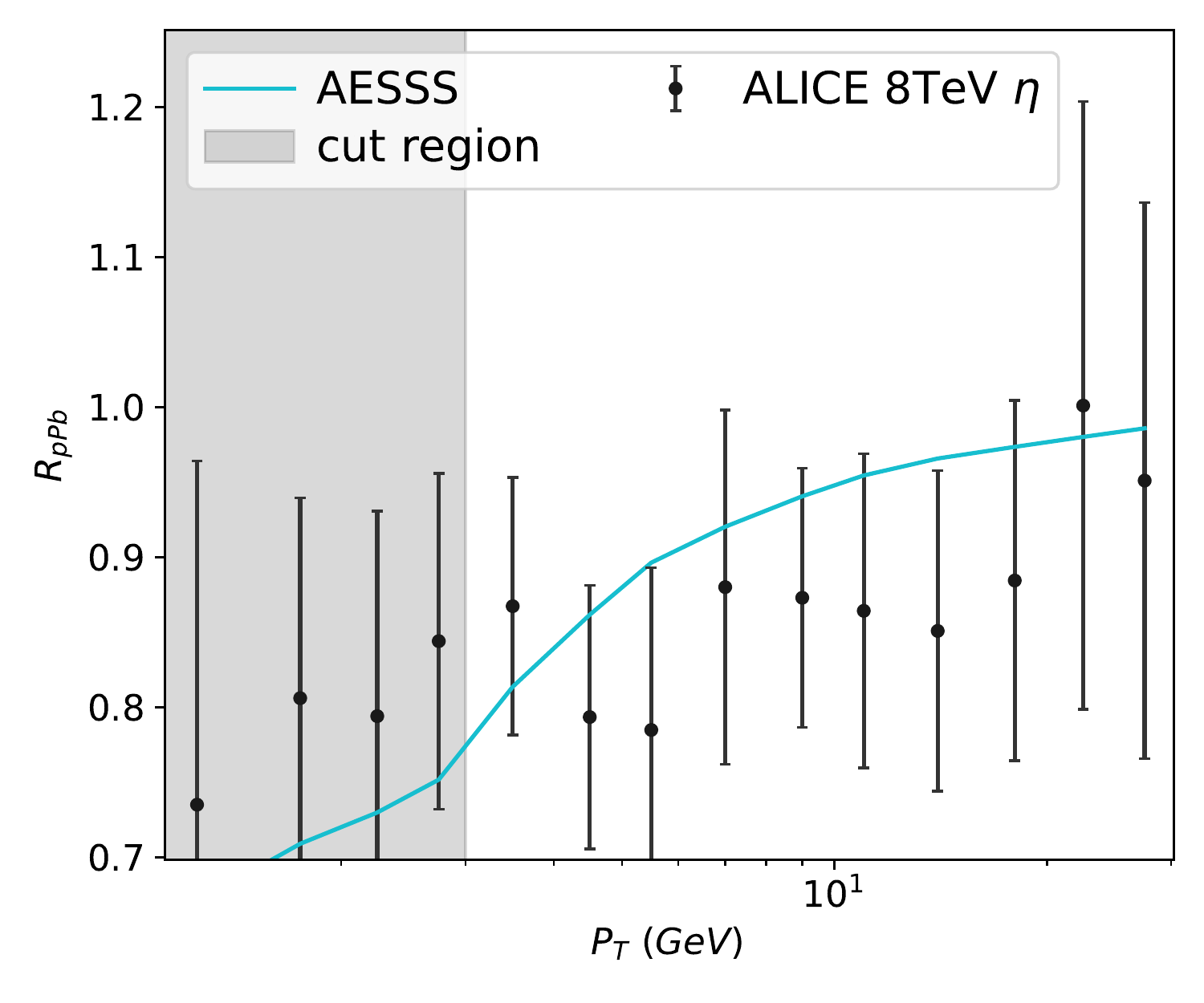}
	\caption{Comparison of data for nuclear ratios $R_{AA'}$ for pion, kaon, and eta production with theoretical predictions at NLO QCD using nCTEQ15WZ nuclear PDFs and different FFs. The predictions are scaled by the inverse of their fitted normalization. The uncertainties of the DSS fragmentation functions are shown as purple bands. The grey region shows the data that is cut from the fits.}
	\label{fig:nCTEQ15WZ_FFcomp}
\end{figure*}
\subsection{Comparison in dAu and pPb collisions}

We now examine the impact of the different FFs on the nuclear ratios $R_{p\mathrm{Pb}}$ and $R_{d\mathrm{Au}}$ 
for pion, kaon, and eta production. 
Figure~\ref{fig:nCTEQ15WZ_FFcomp} compares all the data sets with predictions using 
the nCTEQ15WZ nuclear PDFs for each set of FFs. Data taken at $p_T<1$\.GeV is not shown as the twist-2 formula for the theory is certainly not valid in that region.
We also display the uncertainty band for the DSS FFs to gauge the spread of the various FFs
as this represents  a typical FF uncertainty. 

We observe that the predictions with BKK, KKP and DSS show very close agreement. 
The most notable difference between BKK and KKP is seen in the charged kaon production, where KKP lies a bit lower 
for high $p_T$ values. 
NNFF and JAM20 also agree very well with each other across all data sets, and the only instance where they lie outside of the uncertainty given by DSS is for the high-$p_T$ ALICE pion data. Since the data uncertainties in this region is quite large, this should not have any significant impact on our fits.
In the  kinematic region where AKK allows predictions, they also agree with the previous FFs. The KRETZER FFs show some qualitative differences in the region just above the cut, but lie within the uncertainty of DSS. For HKNS the disagreement is slightly larger but still well below the data uncertainties. 
The predictions made with AESSS agree well with the eta meson production data, but since AESSS is the only available fragmentation function for eta mesons no comparisons can be made. 

We calculate  fragmentation function uncertainties from the DSS FFs (see below) for each data point and add these  as a systematic uncertainty in our fit. 
Although these uncertainties also depend on the PDF,  this dependence is very weak
and  can be neglected.
Note also that the predictions are already quite close to the data values. While this suggests that the data will not significantly change the central value of the PDFs, the data may well reduce their uncertainties.

\subsection{Uncertainties of fragmentation functions}

We now consider the FF uncertainties in further detail. 
Four of the available FF sets include uncertainties; HKNS and DSS provide their uncertainties in terms of Hessian eigenvectors, while JAM20 and NNFF provide Monte Carlo replicas. 
We show the uncertainties for NNFF and JAM20 
in Figs.~\ref{fig:NNFFunc} and~\ref{fig:JAM20unc} of  Appendix~\ref{sec:FFunc}.
The HKNS FFs yield uncertainties that are larger than the data uncertainty for $p_T$ values below 10\,GeV; hence, they will not help constrain the PDFs in this kinematic region.
The NNFF fragmentation functions yield slightly larger uncertainties than those of DSS shown in Fig.~\ref{fig:nCTEQ15WZ_FFcomp}. 
This may be due, in part, to the use of a parameterization-free neural network instead of a ``traditional" parameterization, and a slightly smaller data set. 
Lastly, the uncertainties of the JAM20 fragmentation are so small across the kinematic region with $p_T{>}1$\,GeV that they can be neglected when compared with the data uncertainty.

It is important to note that the displayed bands do not reflect the full uncertainty of the theory prediction, but rather represent a lower bound for the following reasons: 
Firstly, the theory predictions for low~$p_T$ points may depend on fragmentation functions extrapolated beyond their fitted kinematic region and the accuracy of the Hessian method outside of the region where data exists is heavily dependent on the specific parameterization of the FF.
Even more important are low~$p_T$ corrections. As we move to lower values of $p_T$, perturbation theory begins to  break down as contributions from non-perturbative sources increase. 
This can also make it difficult to disentangle initial from final state effects in hadron production processes. Medium suppression due to energy loss can be observed not only in $AA$, but even in $pA$ and $pp$ collisions~\cite{Aad:2016zif, Khachatryan:2016odn, Acharya:2018qsh}. These higher twist effects, however, are suppressed by powers of the hard scale $p_T$.
Thus, for our predictions in the lower $p_T$ range, we may reach the transition region between stable perturbative predictions and unreliable non-perturbative predictions. 
These  factors are the reason why we need to impose cuts on low $p_T$ values to ensure reliable predictions.

\begin{figure}[tb]
	\centering
	\includegraphics[width=0.48\textwidth]{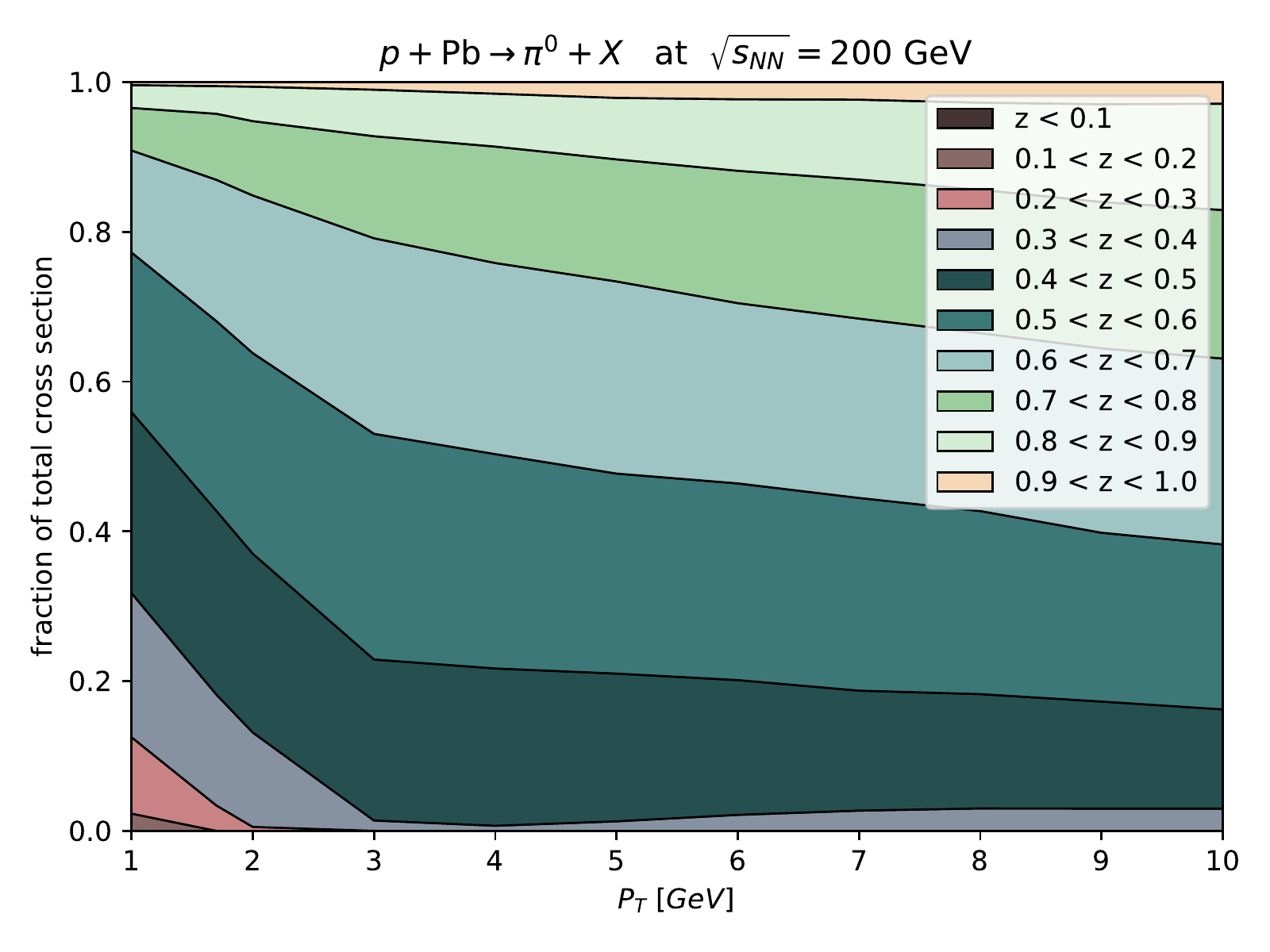}
	\includegraphics[width=0.48\textwidth]{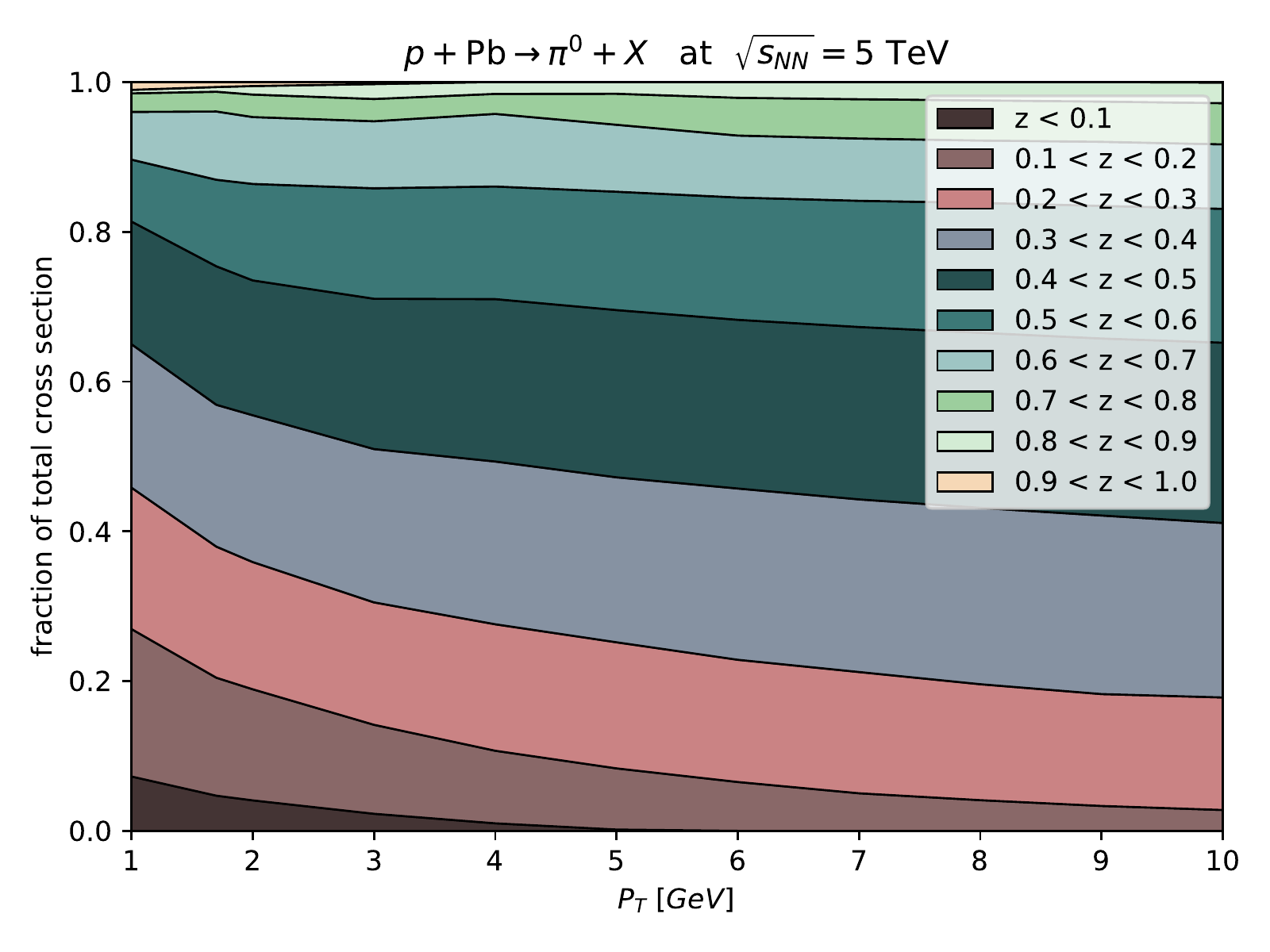}
	\caption{Contribution of different $z$ regions of the fragmentation functions, $D^{\pi^0}(z,Q)$, to the total ${p{+}\mathrm{Pb} \rightarrow \pi^0{+}X}$ cross section at $\sqrt{s_{NN}}=200$\,GeV (top) and 5\,TeV (bottom).
	}
	\label{fig:FF_z}
\end{figure}
\subsection{Fragmentation kinematics}
Finally, it is interesting to investigate the correspondence of the $p_T$ value of the data to the $z$ region of the fragmentation function. 

Figure~\ref{fig:FF_z} shows the contribution of different $z$ regions to the total 
${p{+}\mathrm{Pb} \rightarrow \pi^0{+}X}$ cross section calculated using nCTEQ15WZ PDFs and DSS fragmentation functions at 200\,GeV and 5\,TeV. We can see that the $z<0.2$ region does not have a substantial contribution to the cross section at 200\,GeV, and even 
the region $z<0.4$ hardly contributes above $p_T>3.0$\,GeV. 

In the lower panel we see that the higher energy (5\,TeV) shifts the $z$ bands towards lower $z$ values, with the $z<0.2$ region still contributing a non-negligible amount even at $p_T\approx 10$\,GeV. The $z<0.1$ region starts contributing below $p_T=4$\,GeV, but stays below 10\%. 
Since most fragmentation functions include data at least down to $z=0.05$ in their fits, this eliminates concerns about FF extrapolation having any significant impact on our results.

\section{Impact of SIH data on PDF fits}\label{sec:fits}
\subsection{Data selection}\label{sec:data_selection}
Before performing the fits, we need to decide which data sets to include, and which 
kinematic cuts to impose. 
Firstly, we choose not to include  the eta meson data in the current analysis 
as we only have a single FF without known uncertainties;
but, we will examine this data set in Sec.~\ref{sec:eta}.

To make sure that we can accurately describe the proton baseline 
as presented in  Fig.~\ref{fig:ppPi0},
we cut all data with $p_T<3$\,GeV.
This is a more restrictive cut than in nCTEQ15(WZ) and EPPS16~\cite{Eskola:2016oht} which both used RHIC neutral pion data with $p_T$ values down to 1.7\,GeV. 
Our $p_T$  cut is also sufficient to ensure that even at the highest $\sqrt{s_{NN}}$ of the ALICE data, the fragmentation functions are used only within their well constrained region. This cut leaves us with 77 (out of 174) ALICE and 32 (out of 77) RHIC data points. 

To account for the  fragmentation function uncertainties, 
we take our error estimate from the DSS eigenvectors and 
add them in quadrature with the systematic uncertainties of the data.

\subsection{Main PDF fits}

\begin{figure*}[pt]
	\centering    
	\includegraphics[width=\textwidth]{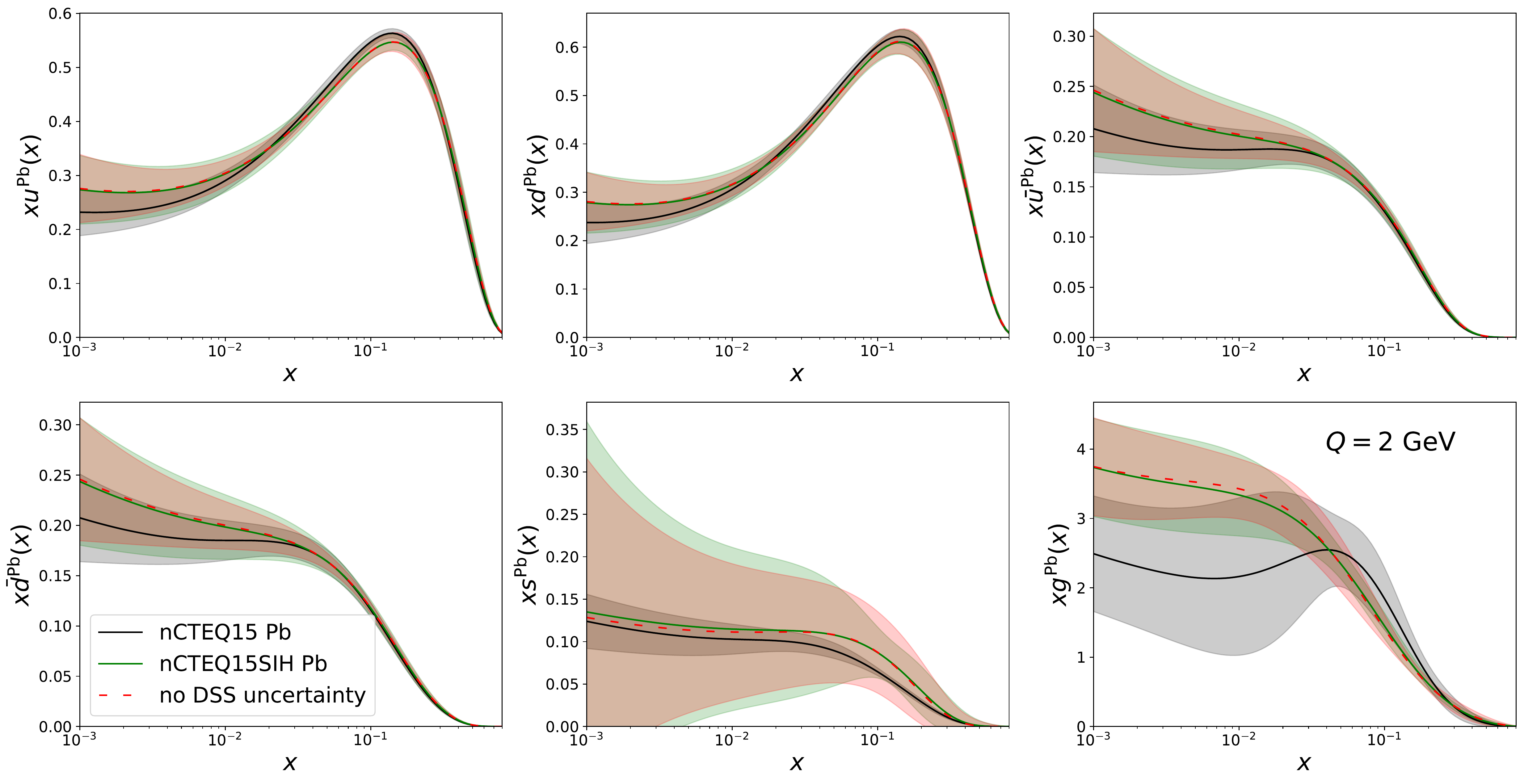}
	\caption{Lead PDFs from fits to the nCTEQ15 data + SIH data. The baseline nCTEQ15 fit is shown in black, the fit with unmodified data is shown in red and the fit where the uncertainties from the DSS FFs were added as a systematic uncertainty (nCTEQ15SIH) is shown in green.
	}
	\label{fig:main_pdfs}
\end{figure*}
\begin{figure*}[pb]
	\centering
	\includegraphics[width=\textwidth]{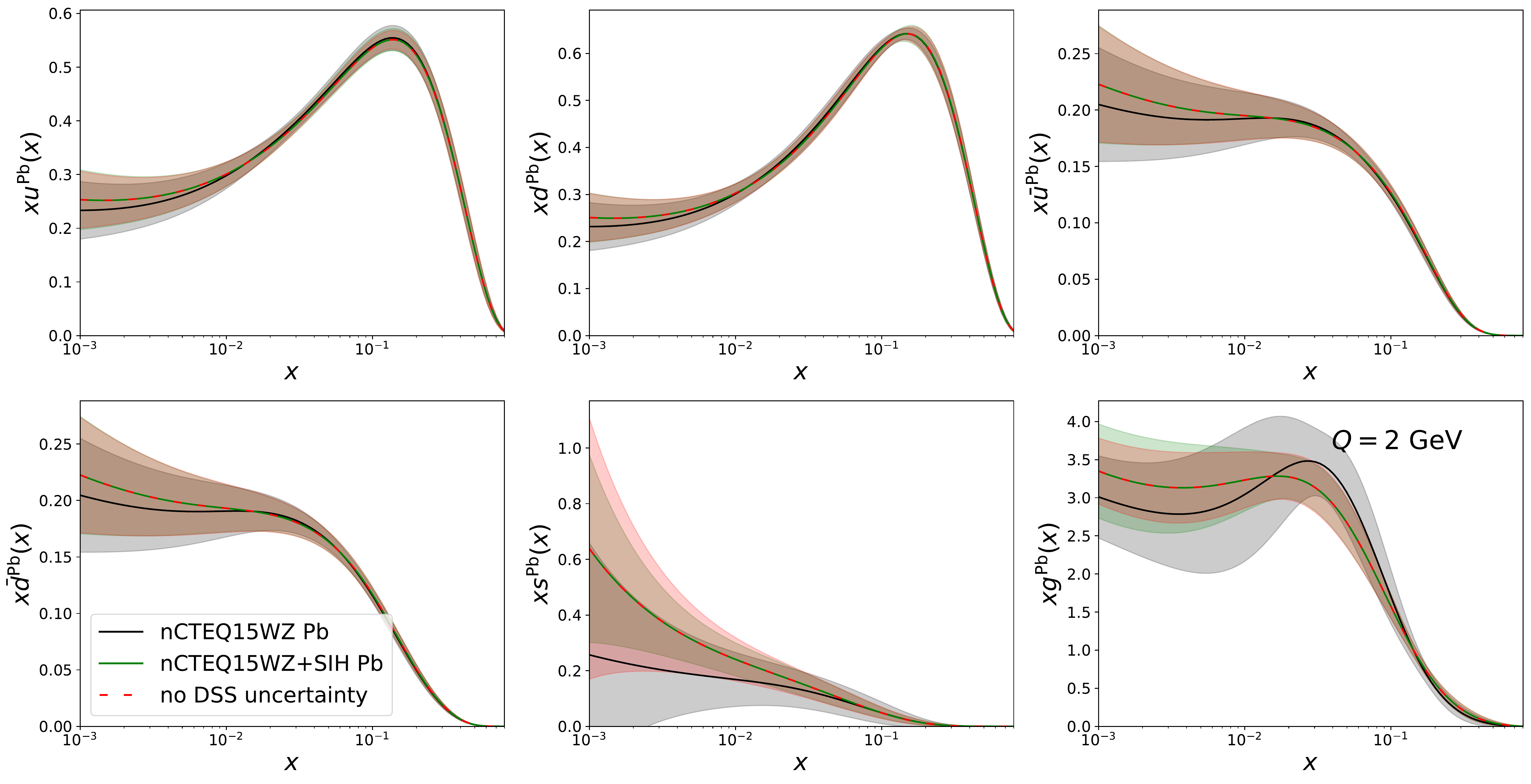}
	\caption{Lead PDFs from fits to the nCTEQ15WZ data + SIH data. The baseline nCTEQ15WZ fit is shown in black, the fit with unmodified data is shown in red and the fit where the uncertainties from the DSS FFs were added as a systematic uncertainty (nCTEQ15WZ+SIH) is shown in green.
	}
	\label{fig:main_pdf_WZ}
\end{figure*}

We now use the single inclusive hadron production data to extend both the nCTEQ15 and nCTEQ15WZ fits. 
For comparison, we will produce two baseline fits.
We produce one baseline with the BKK fragmentation functions
as this was the set used in the previous nCTEQ15 and nCTEQ15WZ analyses. 
We produce also a second baseline with the DSS fragmentation functions
as these come from a more recent analysis and include uncertainties. 
We will then compare these fits with  other available fragmentation functions
in Sec.~\ref{sec:OtherFFs}.

A short summary of the properties of the main fits are given in the following.
\begin{itemize}

    \item The included data sets are neutral pions (STAR, PHENIX, ALICE 5 and 8\,TeV), charged pions (STAR, ALICE 5\,TeV) and charged kaons (ALICE 5\,TeV).

    \item Cuts are applied below $p_T=$ 3\,GeV for all data sets.

    \item Eta mesons are not included in the current fits; we examine this data later in Sec.~\ref{sec:eta}.

    \item  PHENIX charged hadrons are excluded  by our $p_T$~cut. 

    \item Normalizations of all SIH data sets are fitted according to the prescription given in the \mbox{Appendix~\ref{sec:norm}.}

    \item Fits are performed first with  data uncertainties alone, 
    and again with uncertainties from the DSS fragmentation functions added as a systematic uncertainty to the data.

    \item Except for those items specified above, 
    all other inputs to the fit are kept equal to the baseline fit.

\end{itemize}

The resulting fits are shown in Figs.~\ref{fig:main_pdfs} and~\ref{fig:main_pdf_WZ} for the nCTEQ15 and nCTEQ15WZ baseline, respectively. 
The plots show the baseline fit in black, the fit with regular data uncertainties in red, and the fit with DSS uncertainties added to the data in green. We focus only on the lead PDF since the new data is taken on lead and gold, which is similarly heavy.

Examining Fig.~\ref{fig:main_pdfs}, the most obvious change between the baseline (black) and the new fits (red, green) is the change in the gluon which is enhanced at $x<0.05$ and suppressed at $0.05<x<0.3$. The central values of the other flavors also exhibit some slight changes as they are of course coupled to the  gluons via the DGLAP evolution. The uncertainties for up, down and strange flavors are larger in the new fits than in the baseline due to the newly opened strange parameters. This is not unexpected and was also seen in the recent nCTEQ15WZ analysis  where the same strange parameters were opened up. 
The inclusion of the DSS uncertainty does not cause any significant change in the central value but does result in an increased PDF error 
band which is most noticeable at small $x$, especially for the strange PDF. Somewhat surprisingly, the region $x\sim 0.1$ sees a slight decrease in uncertainties, likely caused by slight shifts in the Hessian basis' eigenvector directions.

In Fig.~\ref{fig:main_pdf_WZ}, the same fits are shown with $W/Z$ data included. 
Here, we see that the new fits for nCTEQ15WZ+SIH are generally (with 
the exception of the strange PDFs)
more similar to the baseline fit than was the case for the nCTEQ15SIH fits shown in Fig.~\ref{fig:main_pdfs}. 
For the gluon, we see a similar behavior as in Fig.~\ref{fig:main_pdfs},   
but slightly less pronounced due to the additional constraints from the $W/Z$ data.
A somewhat surprising feature of this fit is the enhancement of the strange quark at low $x$.  
As Fig.~\ref{fig:xsec_per_flavour} shows no particular strange sensitivity of the SIH data,
presumably this enhanced strange PDF is being driven, in part, by the 
influence of the $W/Z$ data. 
While the resulting strange PDF in the new fits is substantially larger
than the baseline at low $x$ values, it is important to recall that
the LHC heavy ion data primarily constrains the region ${x\gtrsim 0.01}$.
Including the DSS uncertainty in this case causes no visible difference in central values, but yields slightly larger uncertainty bands on the gluon. The shifted eigenvector basis results in slightly decreased strange quark uncertainties in the low-$x$ region.
\begin{table*}[tb]
\renewcommand{\arraystretch}{1.2}
    \caption{Normalization uncertainties and fitted normalizations of the SIH data sets in the nCTEQ15WZ+SIH fit.}
    \centering
    \begin{tabular}{|c|c|c|c|c|c|c|c|}
		\hline 
         & \multicolumn{2}{c|}{STAR} & PHENIX & \multicolumn{4}{c|}{ALICE}\\ 
		\hline 
		 & $\pi^0$ & $\pi^\pm$ & $\pi^0$ & 5\,TeV $\pi^0$ & 5\,TeV $\pi^\pm$ & 5\,TeV $K^\pm$ & 8\,TeV $\pi^0$ \\ 
		\hline \hline
        Normalization uncertainty & 17\% & 17\% & 10\% & 6\% & 6\% & 6\% & 3.4\%  \\
		\hline 
        Fitted normalization & 0.942 & 0.866 & 1.010 & 0.995 & 0.994 & 1.021 & 1.021 \\
		\hline 
    \end{tabular}
    \label{tab:norms}
\end{table*}
\begin{figure*}[tb]
	\centering    
	\includegraphics[width=\textwidth]{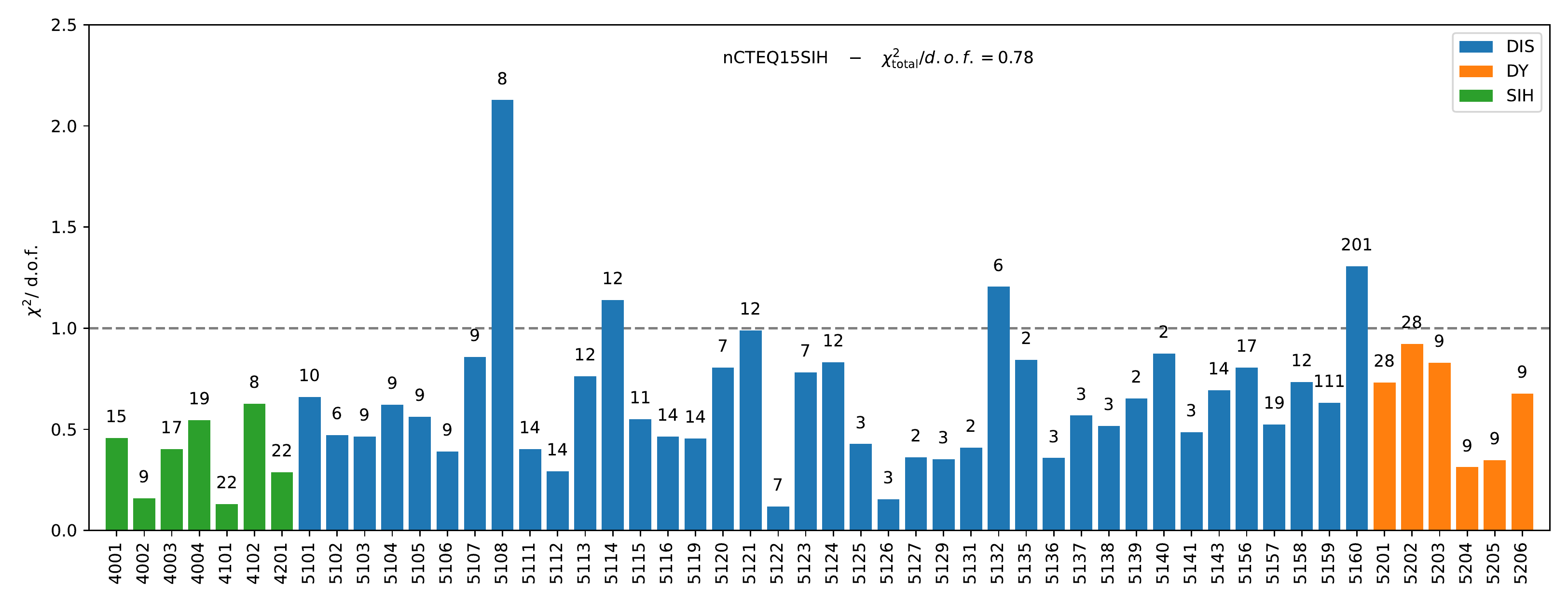}
	\centering    
	\includegraphics[width=\textwidth]{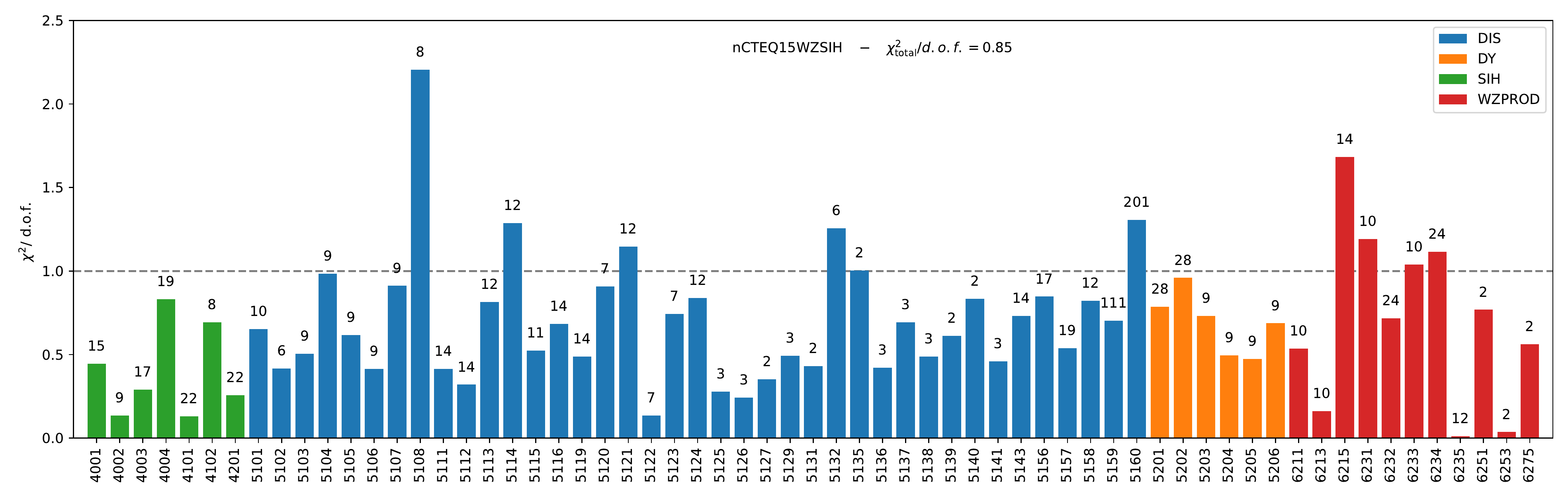}
	\caption{The $\chidof{}$ of the individual experiments for the nCTEQ15SIH fit (top panel) and the  nCTEQ15WZ+SIH fit (bottom panel). The number of data points is indicated at the top of each bar.
	The ID numbers for the SIH data are listed in Table~\ref{tab:data_table},
	and the other processes are listed in Ref.~\cite{Kusina:2020lyz}.
	\\
	}
	\label{fig:chi2perdof}
\end{figure*}
\begin{table*}[tb]	
\renewcommand{\arraystretch}{1.2}
	\caption{We present the $\chidof{}$ for the individual SIH data sets, the individual processes {DIS, DY, SIH, WZ}, and the total. The shown $\chi^2$ is the sum of regular $\chi^2$ and normalization penalty. Excluded processes are shown in parentheses. 
	Note that both  nCTEQ15 AND nCTEQ15WZ included the neutral pions from STAR and PHENIX.
	\\
	}
	\centering	
	\begin{tabular}{|c|c|c|c|c|c|c|c|c|c|c|c|c|c|}
		\hline 
		\multicolumn{13}{|c|}{$\chidof{}$ for selected experiments and processes} \\ 
		\hline
		 & \multicolumn{2}{c|}{STAR} & PHENIX & \multicolumn{4}{c|}{ALICE} & DIS  & DY & WZ & SIH & \textbf{Total}\\ 
		\hline 
		 & $\pi^0$ & $\pi^\pm$ & $\pi^0$ & 5\,TeV $\pi^0$ & 5\,TeV $\pi^\pm$ & 5\,TeV $K^\pm$ & 8\,TeV $\pi^0$ &   &  &  &  & \\ 
		\hline
		\hline
		nCTEQ15           & 0.13 & 2.68 & 0.30 & 2.53 & 0.62 & 0.71 & 1.96 & 0.86 & 0.78 & (3.74) & (1.23) & \textbf{1.28}  \\ 
		\hline 
		nCTEQ15SIH       & 0.16 & 0.69 & 0.41 & 0.48 & 0.13 & 0.29 & 0.58 & 0.87 & 0.72 & (2.32) & 0.38 & \textbf{1.00}  \\ 
		\hline 
		nCTEQ15WZ         & 0.17 & 3.24 & 0.23 & 0.67 & 0.21 & 0.41 & 1.58 & 0.90 & 0.78 & 0.90 & (0.81) & \textbf{0.90} \\ 
		\hline 
		nCTEQ15WZ+SIH     & 0.14 & 0.75 & 0.30 & 0.47 & 0.13 & 0.26 & 0.79 & 0.91 & 0.77 & 1.02 & 0.41 & \textbf{0.85}  \\
		\hline 
	\end{tabular}     
	\label{tab:chi2table_mainfits}
\end{table*}
\begin{table*}[tbh]
\renewcommand{\arraystretch}{1.2}
	\caption{The $\chidof{}$ values of the SIH  data obtained with different fragmentation functions and  PDF parameters taken from the nCTEQ15WZ+SIH fit.
	We show the DSS result  both  with (modified data) and without (unmodified data) the added systematics  arising from the fragmentation function uncertainties. 
	}
	\centering
	\vspace{0.2cm}
	\begin{tabular}{|c|c|c|c|c|c|}
		\hline 
		   DSS  & DSS & & & &  \hspace{1cm} \\[-3pt] 
		   unmodified  & modified  &  KKP & BKK & NNFF & JAM20   \\[-3pt]
		    data &  data & & & &  \\[-0pt] 
		\hline
		\hline
		 0.461  & 0.412 & 0.401 & 0.420 & 0.456 & 0.553  \\ 
		\hline 
	\end{tabular}     	
	\label{tab:chi2table_FFs}
\end{table*}

\subsection{Quality of the fits}

To judge the quality of the fits, we first take a detailed look at the resulting $\chi^2$ values. 
Figure~\ref{fig:chi2perdof} shows the $\chidof{}$ for each of the fitted data sets of the two main fits, nCTEQ15SIH and nCTEQ15WZ+SIH. 
We see that the DIS and DY data sets are still well described by the new PDF, 
and generally satisfy  $\chidof<1$, with one exception.\footnote{%
The notable exception with a large $\chidof{}$ is data set 5108 
{(Sn/D EMC-1988)}
with 8 data points. However, other analyses also found a large $\chidof{}$ 
for this data set~\cite{Eskola:2016oht, deFlorian:2011fp}. 
} %
The $W/Z$ data also remains well described when including the SIH data, 
with the exception of data set 6215 (ATLAS Run I, $Z$ production);
this behavior was also observed in the nCTEQ15WZ analysis.

More quantitative insights regarding the fit results can be obtained from Table~\ref{tab:chi2table_mainfits}, which shows a breakdown 
of the  $\chidof{}$
by experiment type and by data set.
Note that there is a small difference for the STAR and PHENIX pion results
reported here (with $p_T>3$\,GeV) and in the nCTEQ15WZ analysis which used a $p_T>1.7$\,GeV  cut.

Beginning with the nCTEQ15 fit, we see that the DIS and DY data are well described.
In contrast, the  $W/Z$ and SIH  data (which were \textbf{not} fitted) 
yield large $\chidof{}$ values.
Adding SIH data to the fit (nCTEQ15SIH) significantly improves 
the SIH data from $\chidof{}=1.23$ to $0.38$,
as well as the  $W/Z$ data  from $\chidof{}=3.74$ to $2.32$.
There is also a slight improvement in the DY data, 
with a marginal increase in the DIS $\chi^2$.

In a similar manner, the nCTEQ15WZ fit yields good $\chi^2$ values
for the DIS, DY and $W/Z$ data, but the fit to the SIH data is not optimal. 
Including the SIH data in the fit
we find an improvement from $\chidof{}=0.81$ to $0.41$.
This results in marginal shifts for the DIS and DY data, but does
increase the $W/Z$ data from  $\chidof{}=0.90$ to $1.02$.
However, the total $\chidof{}$ for the combined fit nCTEQ15WZ+SIH
is $\chidof{}=0.85$ as compared to the nCTEQ15WZ with $0.90$.

Adding either inclusive hadron data or colorless weak boson production data also improves the fit for the other data set when compared to the nCTEQ15 baseline.
This supports the consistency of our interpretation of the nuclear modification in the inclusive hadron data as cold nuclear effects. A less ambiguous signal for medium effects would therefore certainly be correlations of several particles as observed e.g., in the production of two hadrons (\cite{Acharya:2018dxw, Acharya:2019fip}).

Table~\ref{tab:norms} shows the fitted normalizations of the SIH data sets. 
All the   resulting normalization parameters are consistent with unity
within the normalization uncertainty. Therefore,  no significant normalization penalties are applied.

\subsection{Comparisons with other FFs}\label{sec:OtherFFs}

To investigate the influence of the choice of fragmentation function on the 
quality of our fit to the SIH data, 
in Table~\ref{tab:chi2table_FFs} we compute the $\chidof{}$ 
for the collection of fragmentation functions listed in Table~\ref{tab:FF_table}
using the parameters from our  nCTEQ15WZ+SIH fit.
Additionally, we show the result using  DSS both 
with and without the added systematics  arising from the fragmentation function uncertainties. 
For these two DSS results, it is clear that  including the additional 
uncertainties yields a lower $\chi^2$ value. 
The results shown with the other fragmentation functions 
are computed using the modified data including fragmentation function uncertainties. 

In Table~\ref{tab:chi2table_FFs} we find the results from KKP and BKK are quite comparable to 
the DSS result (with modified data), and the NNFF is just slightly higher. 
Nevertheless, all these results are below the DSS result with unmodified  data,
and this suggests that  the inclusion of the extra uncertainties 
taken from the DSS error bands provides our fit with a reasonable estimate
of the impact of the fragmentation function choice. 
The JAM20 fragmentation functions yield a  higher $\chidof{}$ than the others,
and this reflects the observations noted in Sec.~\ref{sec:ppPi} and Fig.~\ref{fig:ppPi0} which displayed the comparisons with the   
$p{+}p \rightarrow \pi^0{+}X$ data.

\subsection{Comparison of data and theory}

We now present a detailed comparison between our new fits 
and the SIH data in  Fig.~\ref{fig:sihfit_theopred}
which displays the nuclear ratios $R_{AA'}$ as a function of $p_T$.
Although our fits imposed   a 3\,GeV $p_T$ cut on the data,
 we extrapolate to lower $p_T$ values in the shaded regions of the figure.

Examining the nCTEQ15 PDF curves in  Fig.~\ref{fig:sihfit_theopred}, 
we notice these have a significant positive slope for most of the data sets 
as compared to the other fits. 
This observation suggests that as we add more data sets, 
the final predictions exhibit a reduced slope, and if we focus
on the fitted region with $p_T>3$\,GeV, the curves approach unity within 
approximately 10\%. 

Comparing the nCTEQ15SIH fit and its baseline (nCTEQ15), 
we see a considerable shift in the $R_{AA'}$  nuclear ratio when the SIH data is included. 
In contrast, 
the nCTEQ15WZ+SIH fit and its baseline (nCTEQ15WZ) show only subtle differences, aside from the normalization, which is not surprising  given that Table~\ref{tab:chi2table_mainfits} indicated that the $W/Z$ data pulls in the same direction as the SIH data. 
Finally,   the nCTEQ15SIH and nCTEQ15WZ+SIH fits are quite comparable, 
certainly given the uncertainty of the data.

\begin{figure}[tbh]
	\centering    
	\includegraphics[width=\linewidth]{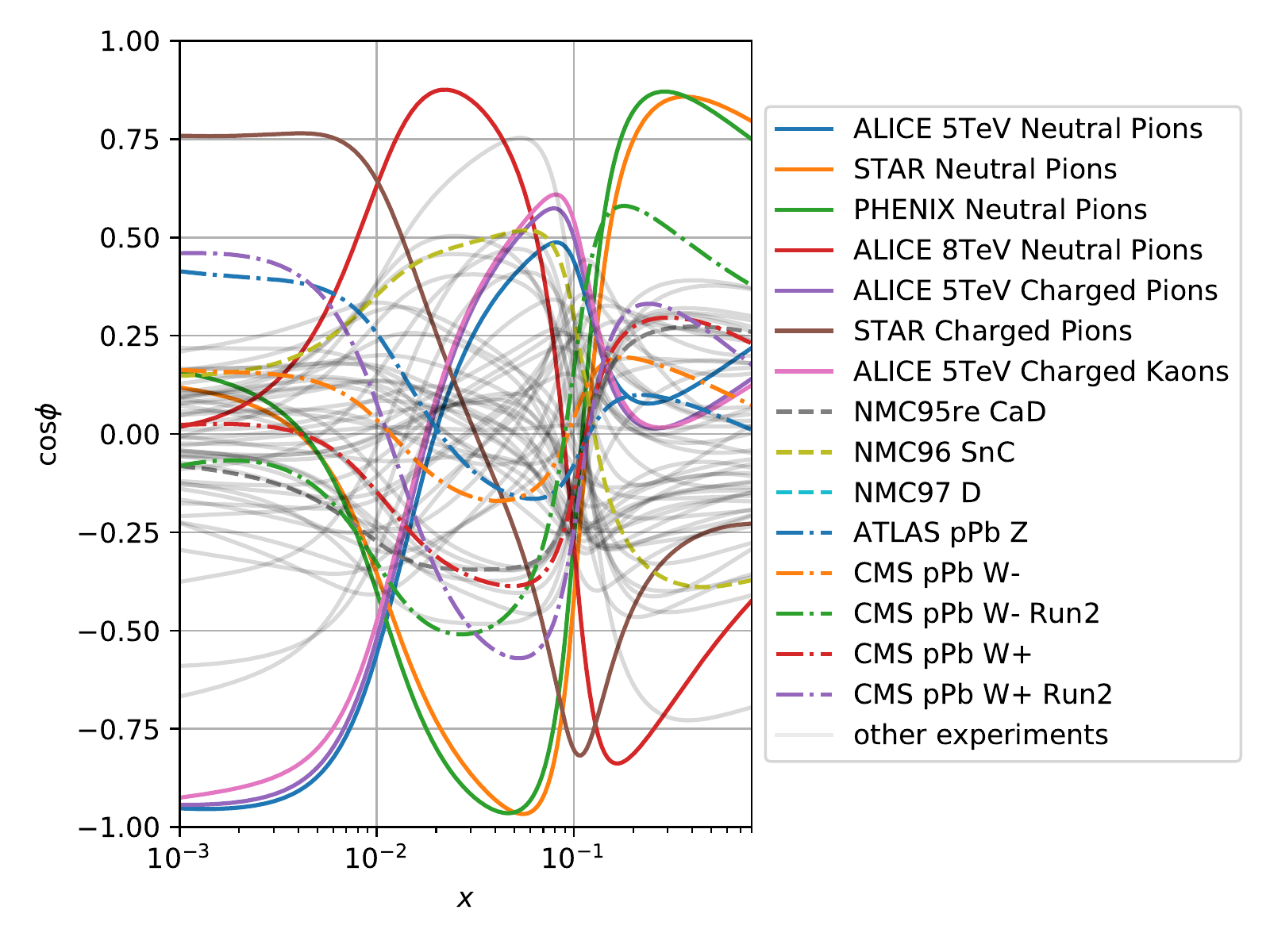}
	\caption{Cosine of the correlation angle ${\cos(\phi [g(x,Q),\chi^2(E_j)])}$
	between gluon PDF and $\chi^2$ of each experimental data set ($E_j$) for the {nCTEQ15WZ+SIH} fit at $Q=10$\,GeV.
	}
	\label{fig:cosphi}
\end{figure}

\begin{figure}[tbh]
	\centering    
	\includegraphics[width=\linewidth]{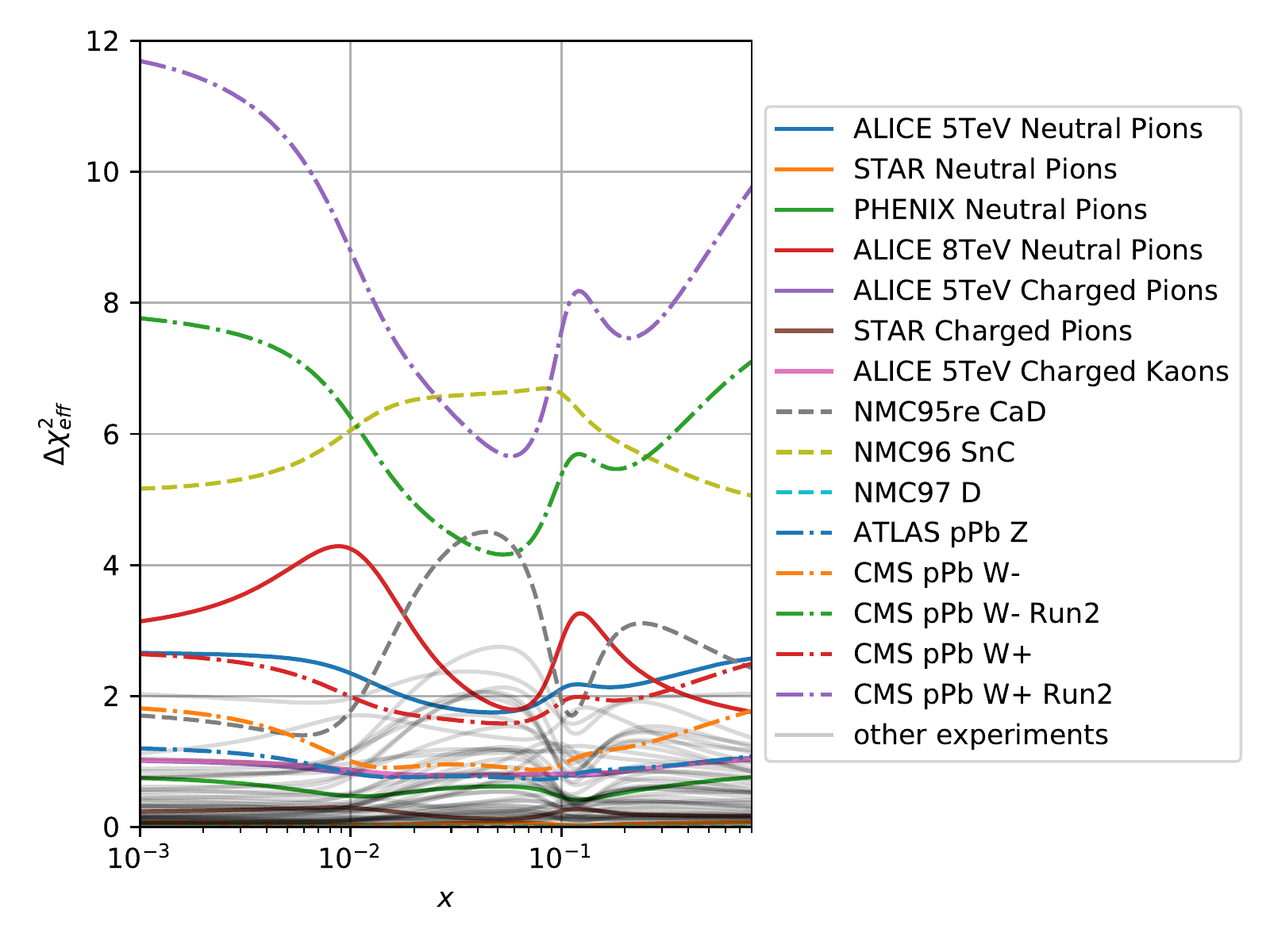}
	\caption{$\Delta\chieff{}[g(x,Q),E_j]$ for the gluon PDF with each experimental data set  ($E_j$) for the {nCTEQ15WZ+SIH} fit at $Q=10$\,GeV.}
	\label{fig:chi2eff}
\end{figure}

\begin{widetext} %

\subsection{Correlation between data  and PDFs}

By looking at the PDFs alone we cannot judge the impact of each individual new data set on the fit. Therefore we make use of two further methods to study how each data set impacts the gluon specifically. The first quantity we want to analyze is the cosine of the correlation angle between two observables $X$ and $Y$, as used in Refs.~\cite{Pumplin:2001ct, Nadolsky:2008zw}:
\def\fredVspace#1{\rule[-.3\baselineskip]{0pt}{#1 \baselineskip}}
\begin{equation}
    \cos\left( \fredVspace{1.3} \phi[X,Y]\right) = 
    \frac{\sum_{i} 
    \left( X_{i}^{(+)}-X_{i}^{(-)} \right) 
    \left( Y_{i}^{(+)}-Y_{i}^{(-)} \right) 
    }{
    \sqrt{\sum_{i'}   \left( X_{i'}^{(+)}-X_{i'}^{(-)} \right)^2 }  
    \sqrt{ \sum_{i''} \left( Y_{i''}^{(+)}-Y_{i''}^{(-)} \right)^2 } 
    }
    \quad ,
\end{equation}
where the index of each sum runs over the 19 eigenvector directions.

\begin{figure*}[p]
    	\centering
    	\includegraphics[width=0.31\textwidth]{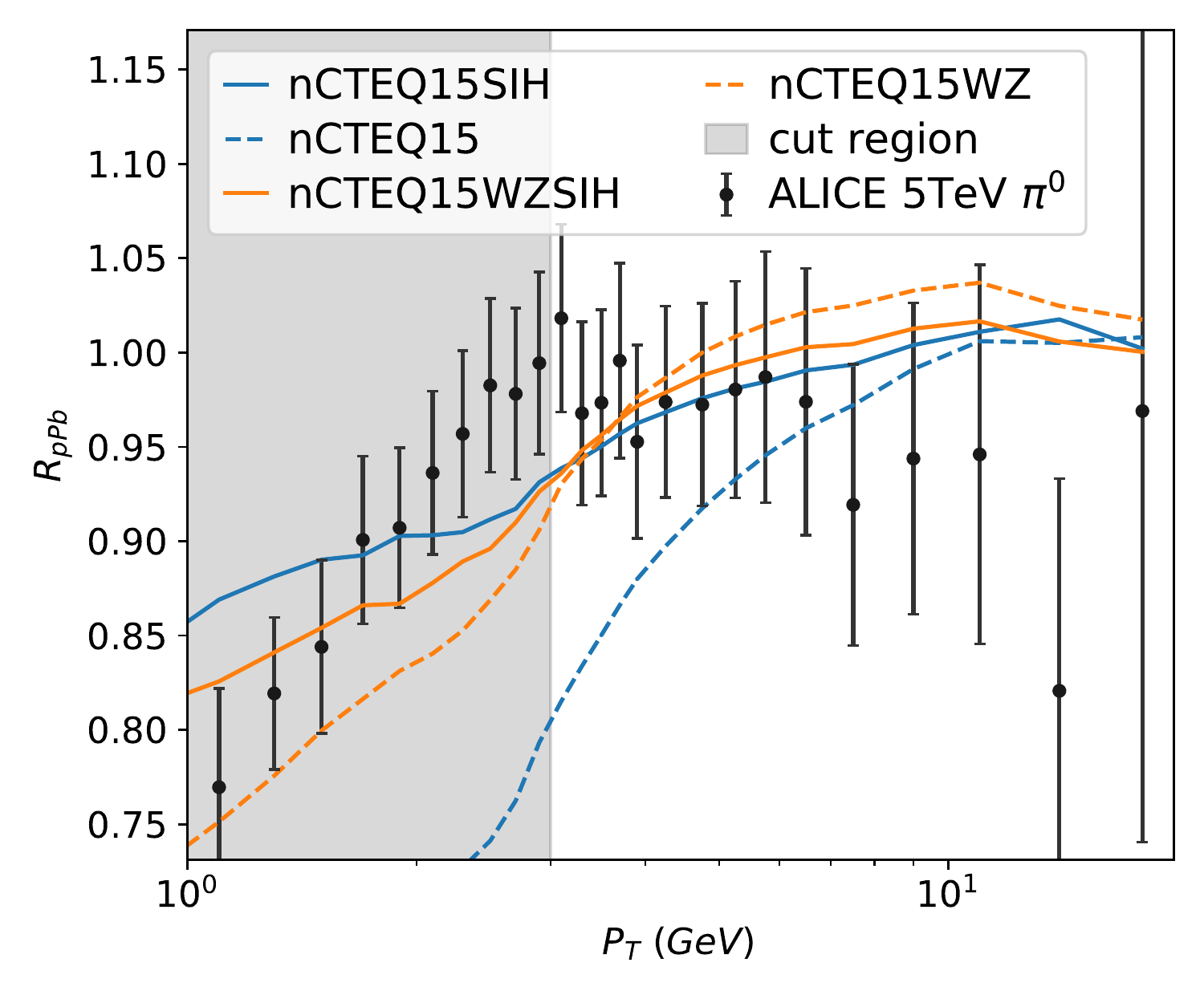}
    	\includegraphics[width=0.31\textwidth]{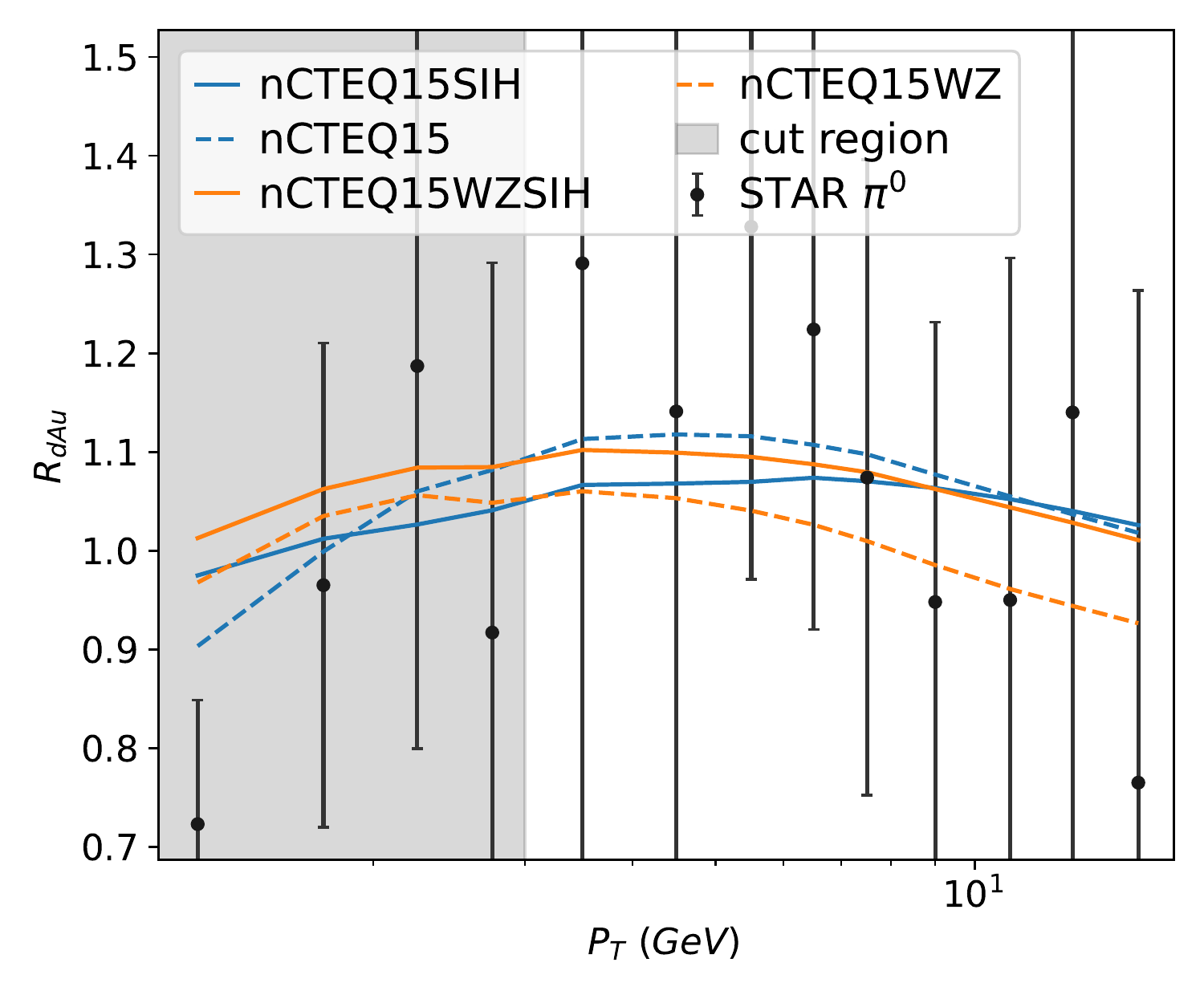}
    	\includegraphics[width=0.31\textwidth]{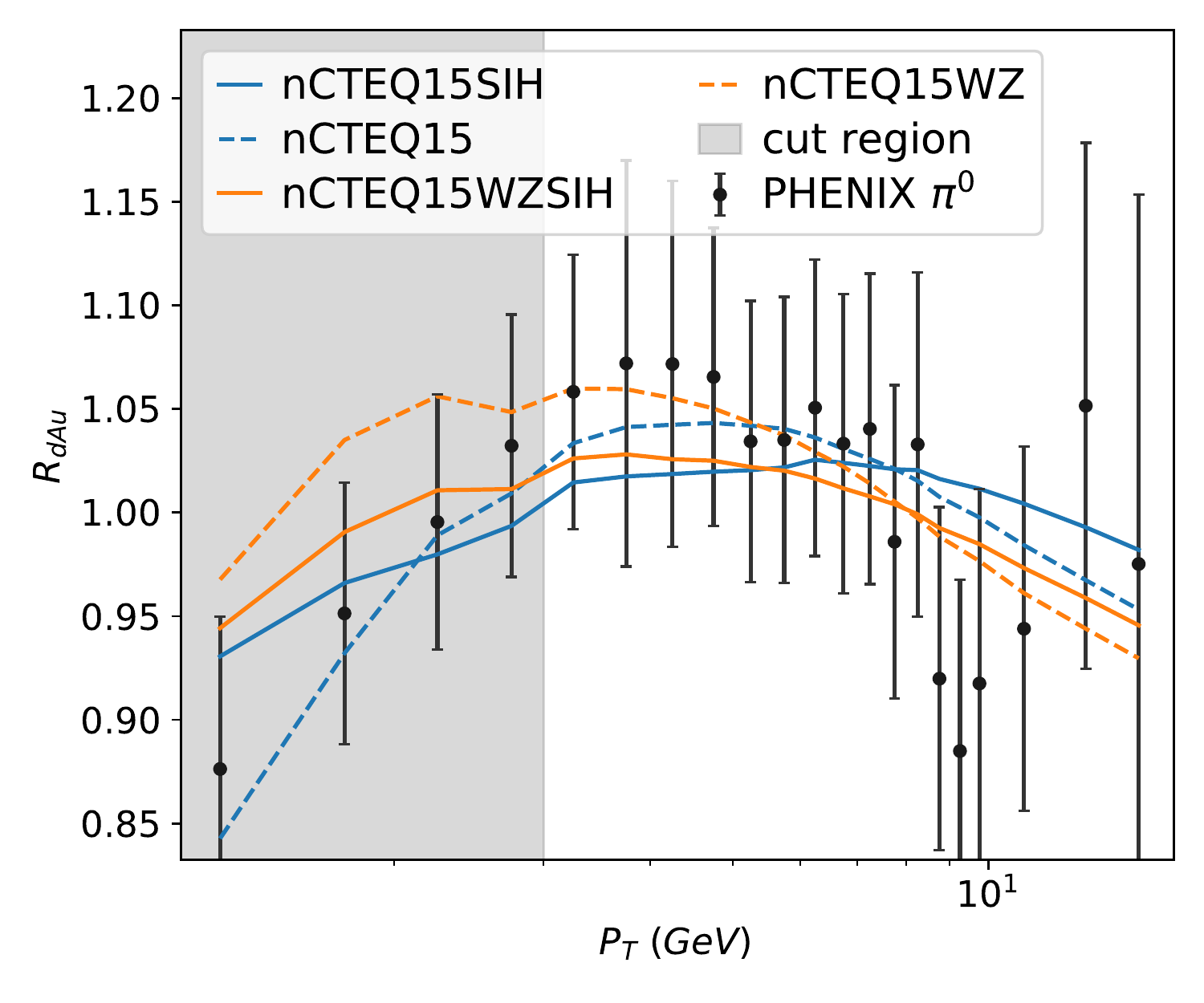}
    	\includegraphics[width=0.31\textwidth]{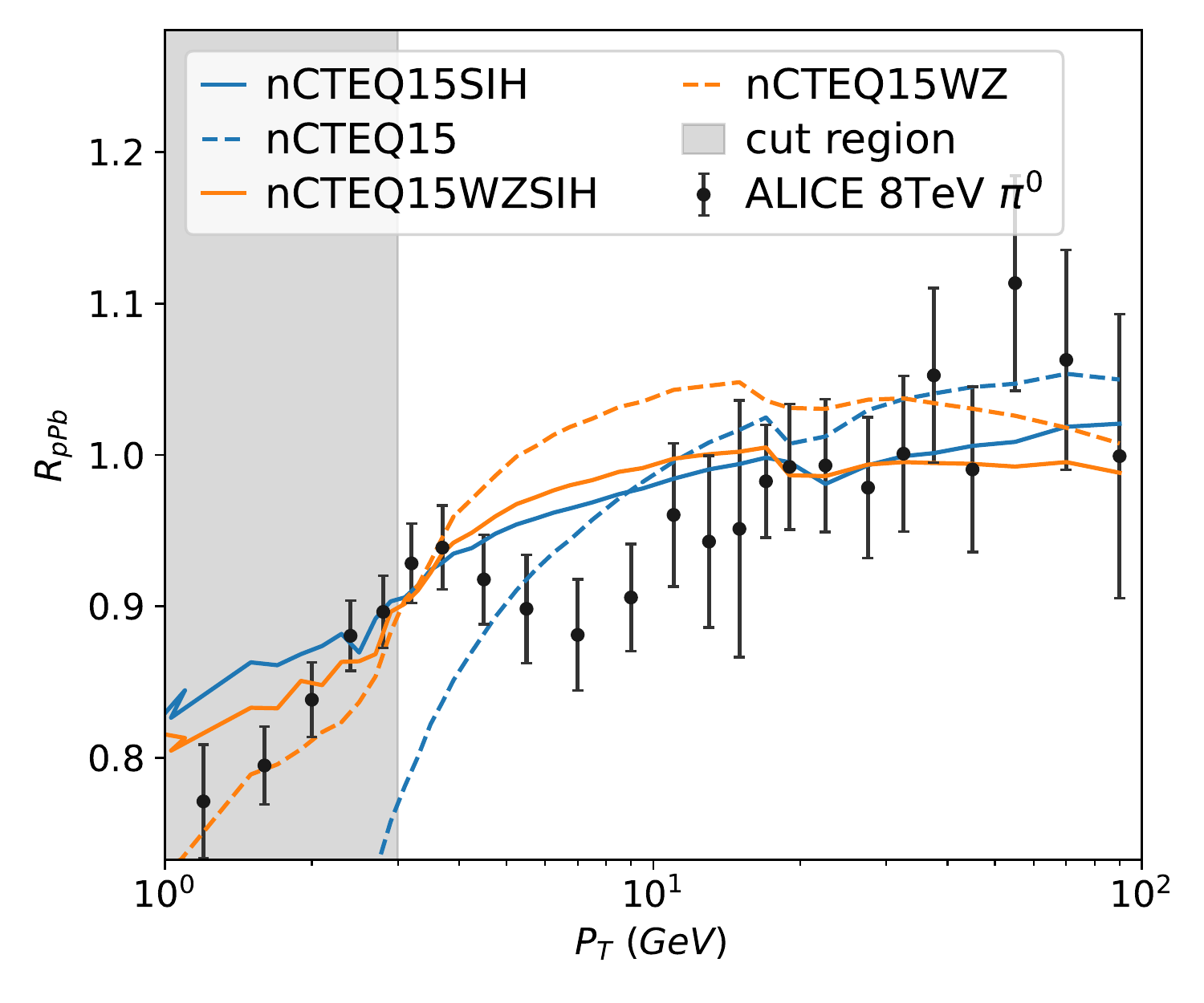}
    	\includegraphics[width=0.31\textwidth]{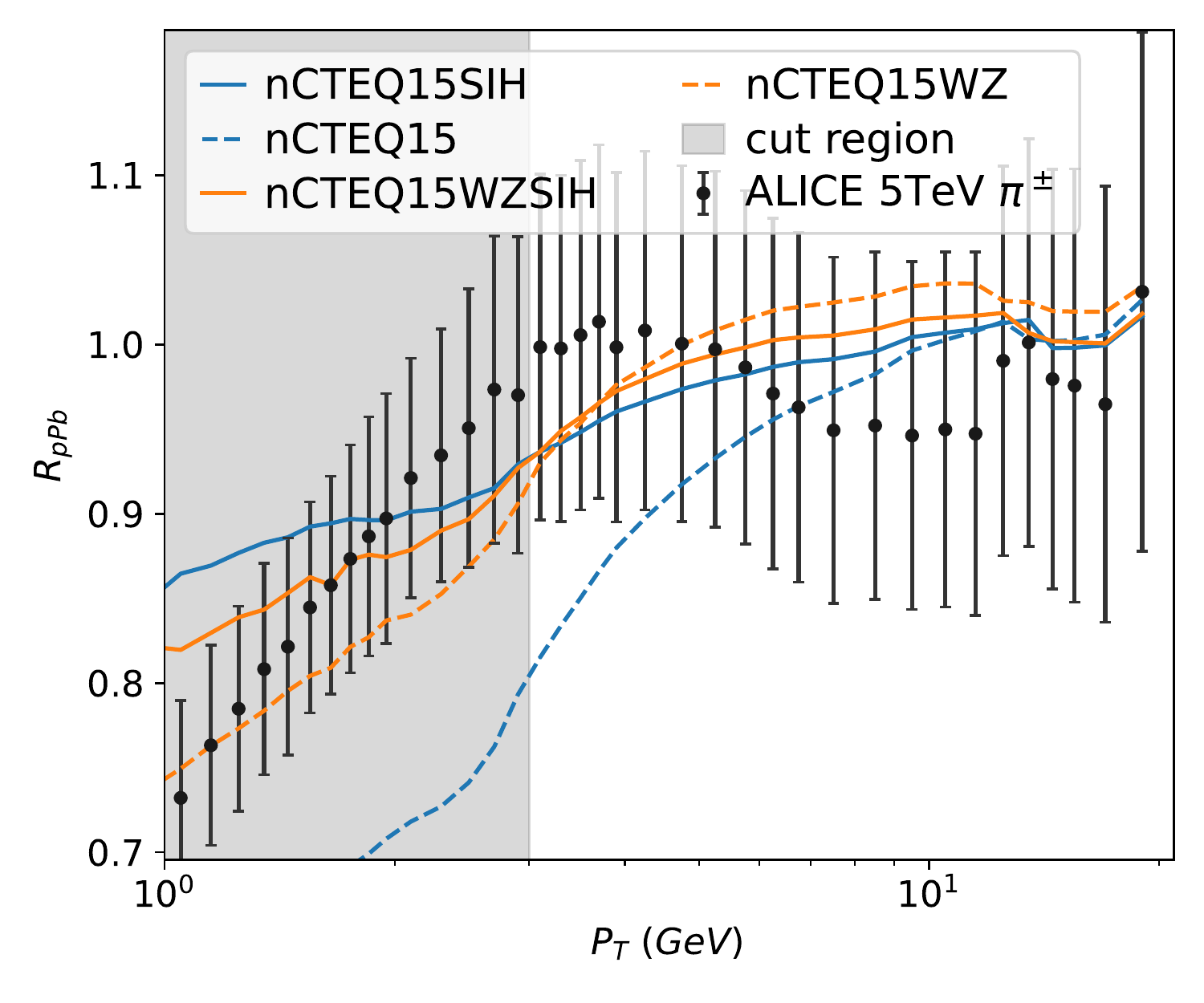}
    	\includegraphics[width=0.31\textwidth]{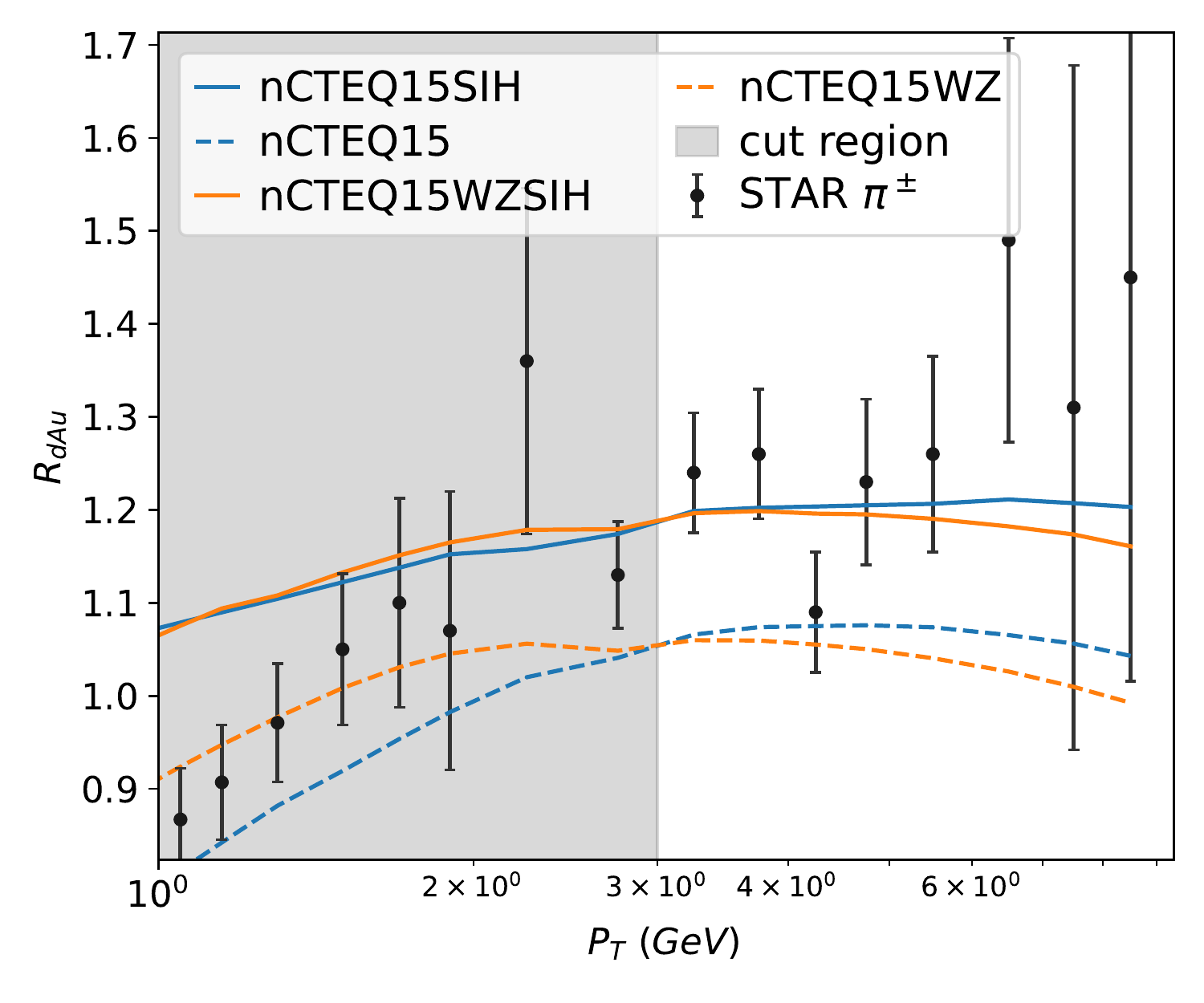}
    	\includegraphics[width=0.31\textwidth]{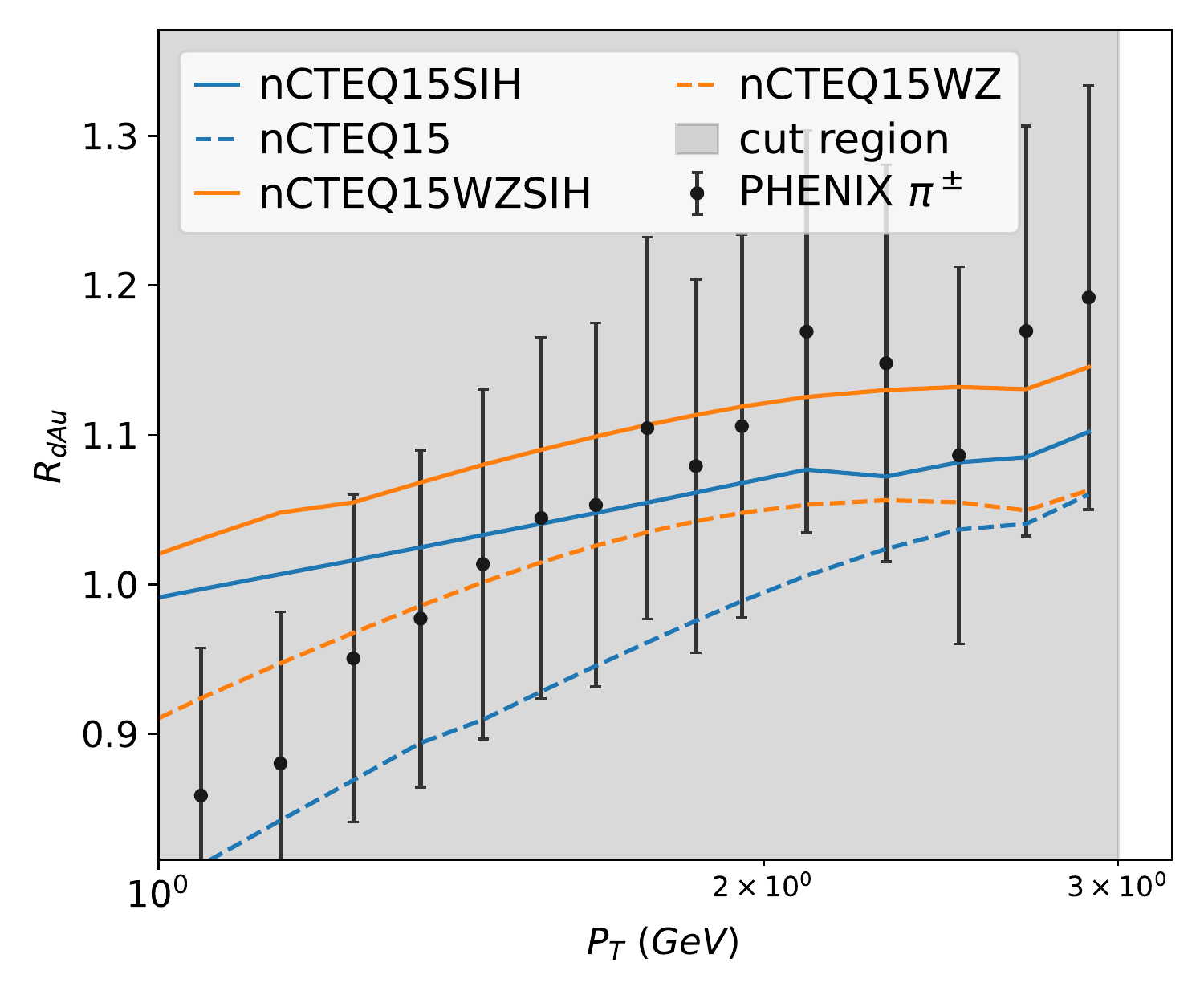}
    	\includegraphics[width=0.31\textwidth]{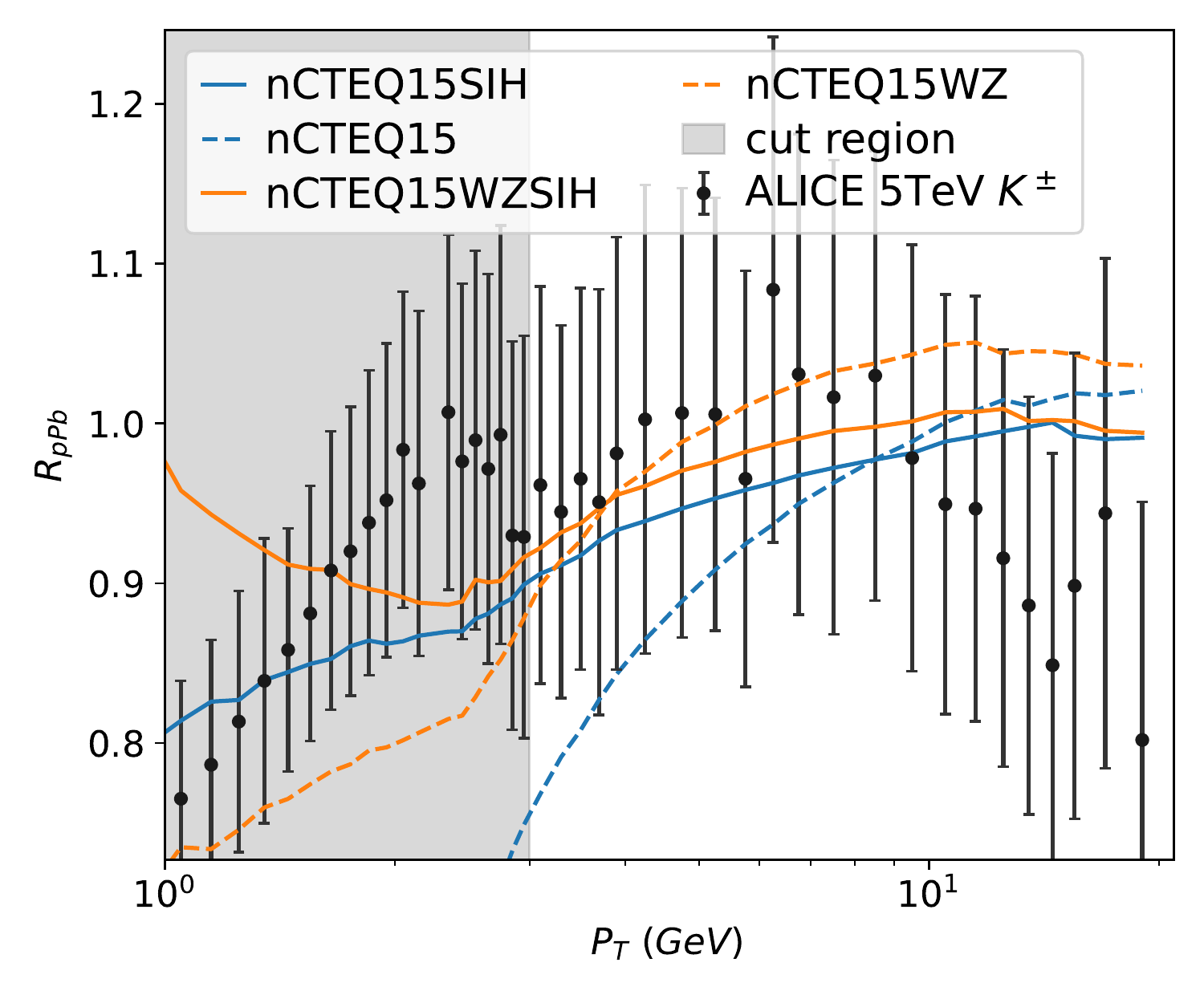}
    	\includegraphics[width=0.31\textwidth]{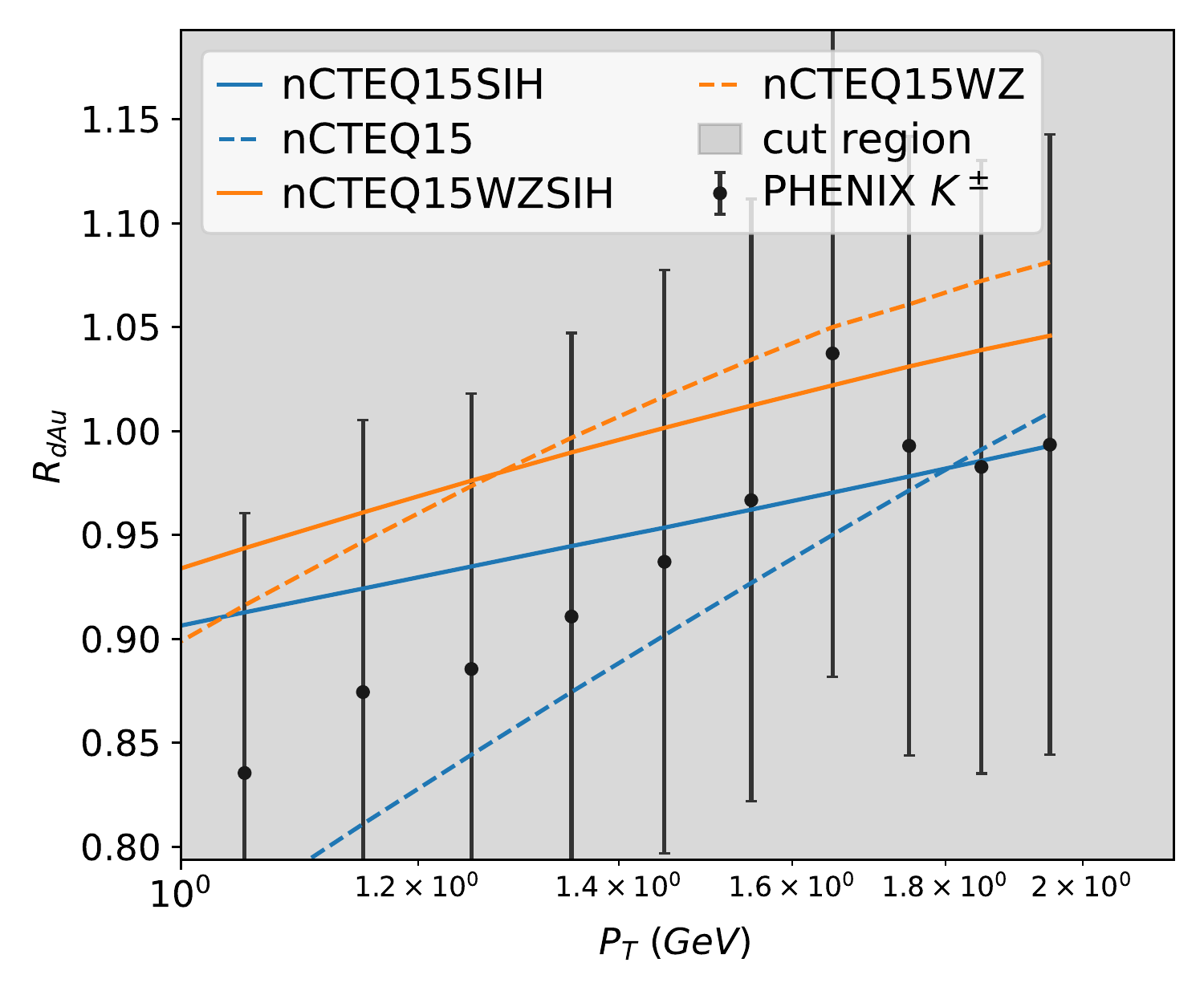}
    	\includegraphics[width=0.31\textwidth]{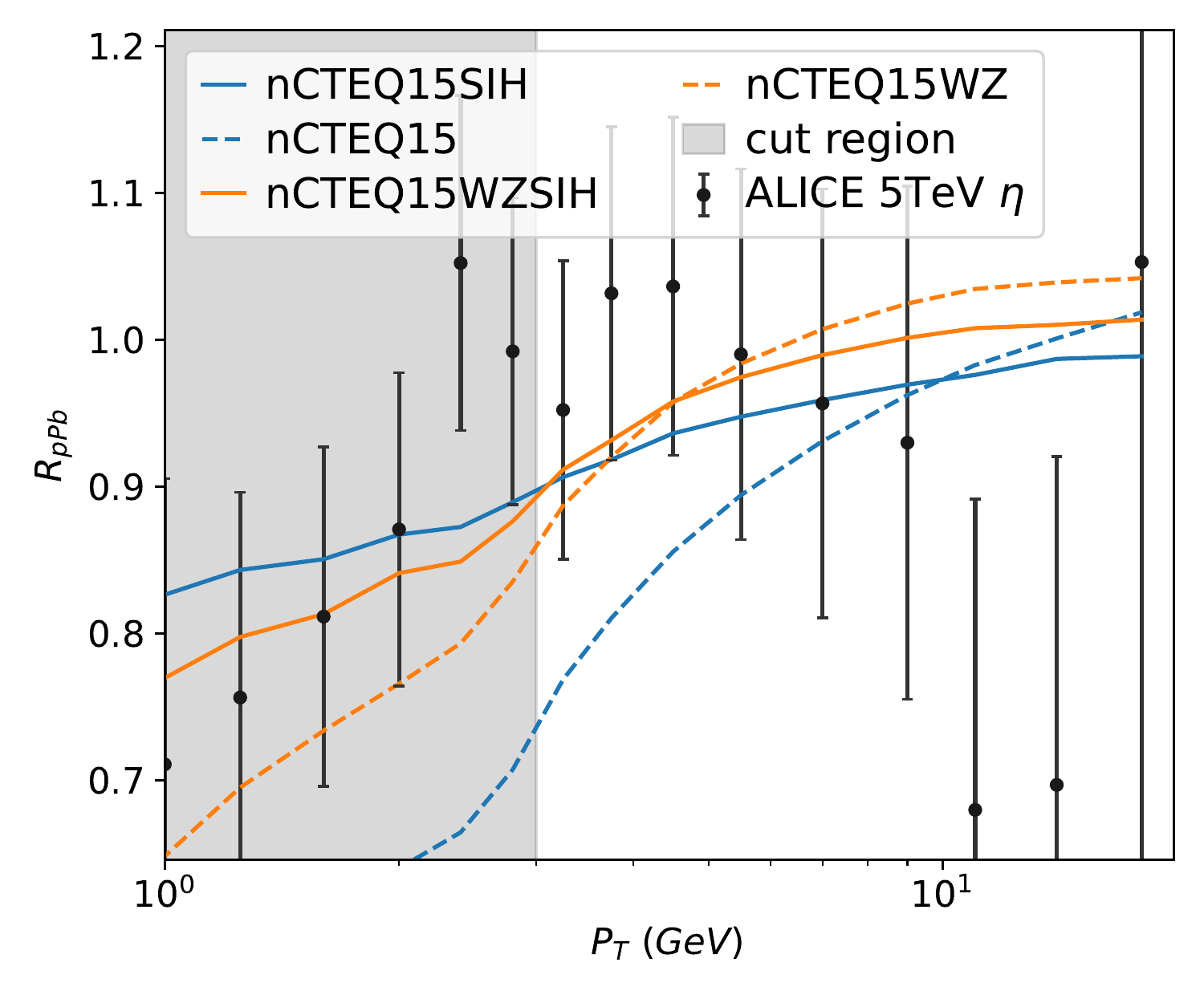}
    	\includegraphics[width=0.31\textwidth]{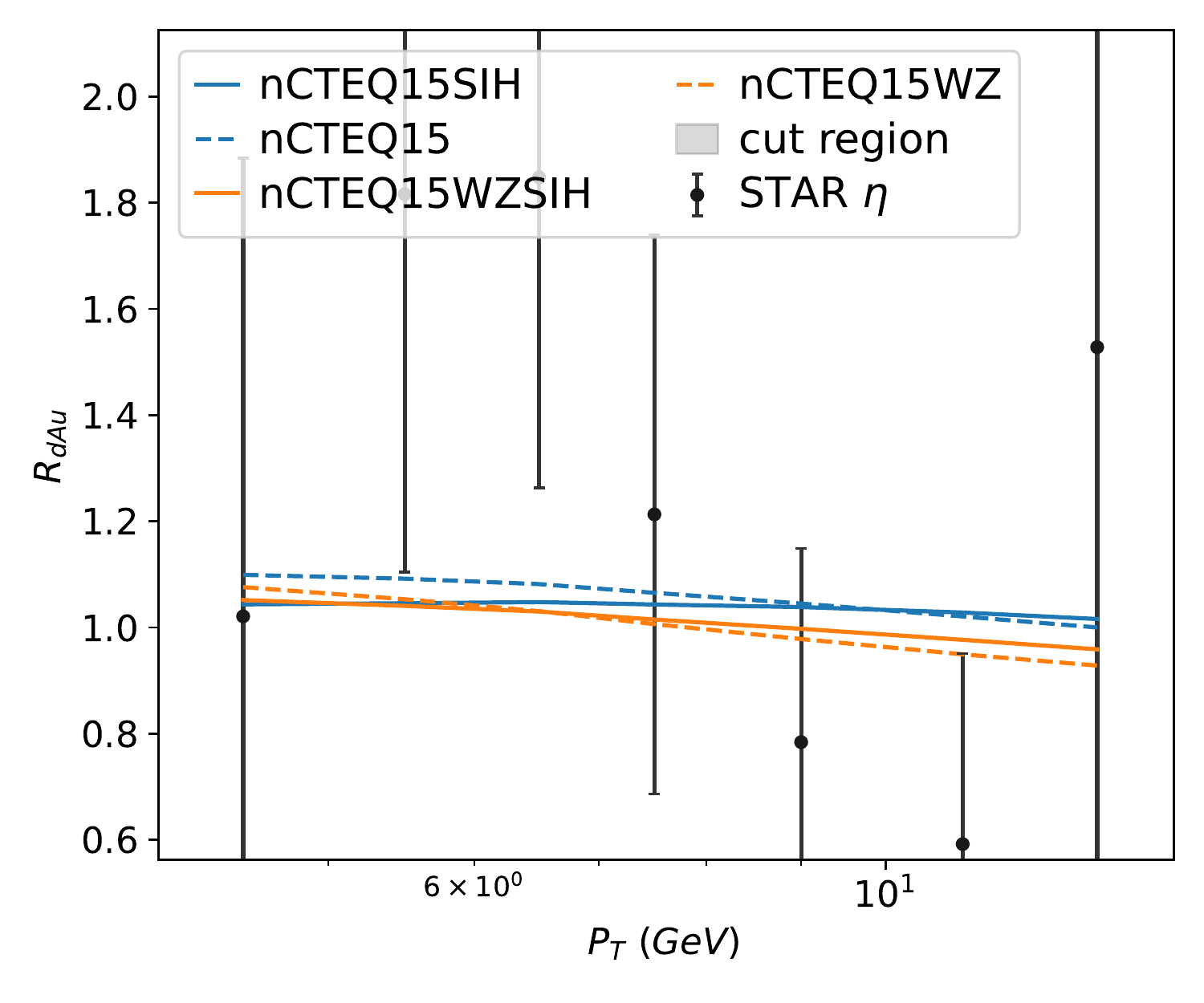}
    	\includegraphics[width=0.31\textwidth]{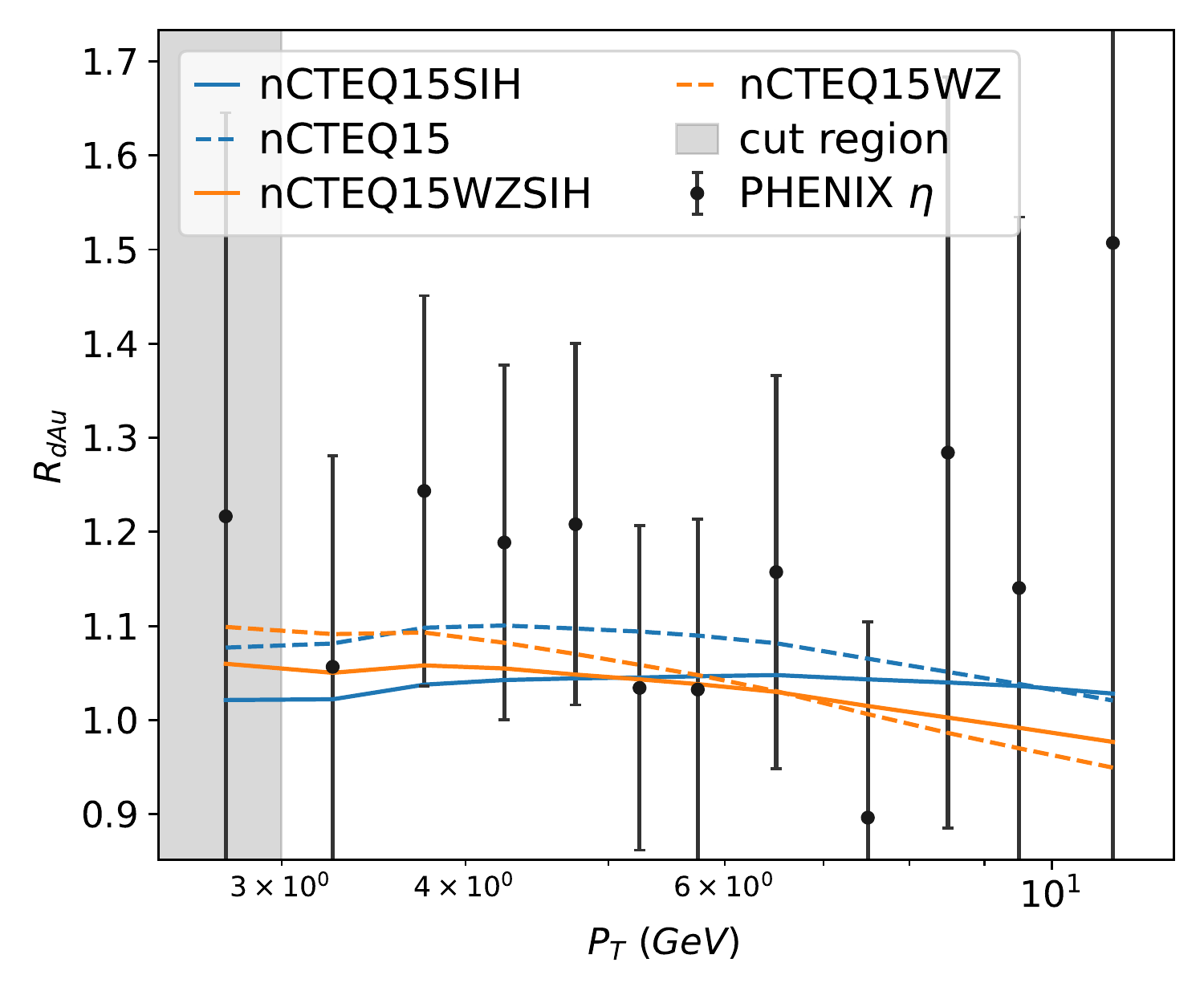}
    	\includegraphics[width=0.31\textwidth]{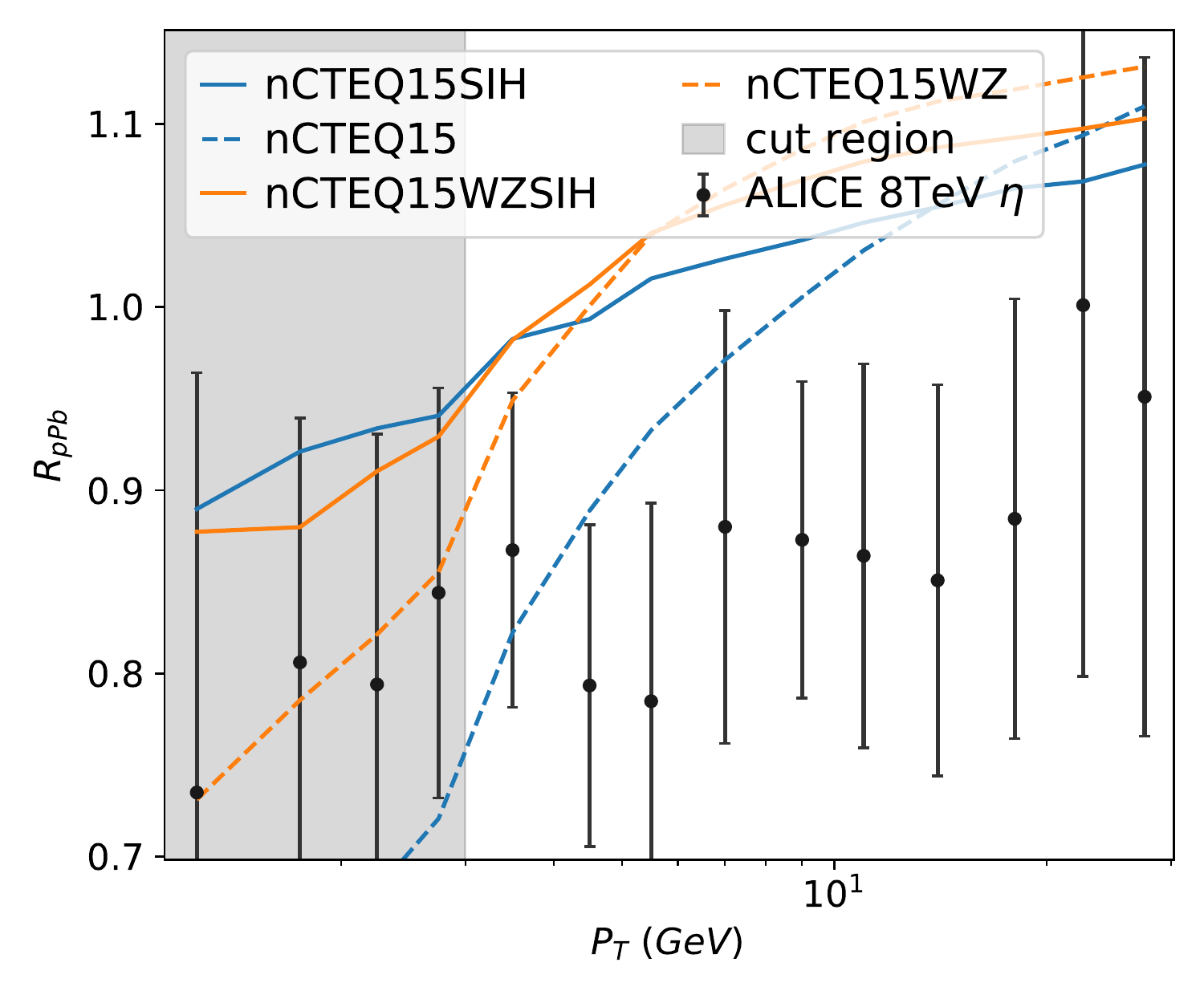}
	\caption{Theory predictions for the main fits and their respective baselines. Dashed curves indicate the baseline fits, while solid curves show the fits with SIH data included. The blue curves are based on nCTEQ15 and the orange ones on nCTEQ15WZ.}
	\label{fig:sihfit_theopred}
\end{figure*}

Another useful quantity is the effective $\Delta\chieff{}$ as introduced in Ref.~\cite{Kovarik:2015cma}. In contrast to the cosine of the correlation angle,  $\Delta\chieff{}$ is more sensitive to the number of data points and error size of the experiments  because the normalization does not cancel these factors out. 
For an experiment $E_j$ and an observable $X$, it is defined as:
\begin{equation}
        \Delta\chieff{}[X, E_j] =
        \sum_{i} \  \frac{1}{2} \ 
        \left\{  \fredVspace{2}
        \left| \chi^{2\,(+)}_{i}(E_j) - \chi^{2\,(0)}_{i}(E_j) \right|+\left| 
        \chi^{2\,(-)}_{i}(E_j) - \chi^{2\,(0)}_{i}(E_j) \right| \right\}
        \ 
        \left( \frac{X_{i}^{(+)}-X_{i}^{(-)} }{ \sqrt{\sum_{i'} \left( X_{i'}^{(+)}-X_{i'}^{(-)} \right)^2 } }\right)^2
        \ .
\end{equation}
To investigate the impact of individual experimental data sets $E_j$ on the gluon PDF $g(x,Q)$, we look at the cosine of the correlation angle 
$\cos(\phi [g(x,Q),\chi^2(E_j)])$ and the effective $\chi^2$ difference  $\Delta\chieff{}[g(x,Q),E_j]$.
Since neither of these quantities display a strong $Q$ dependence, we show them only for the value of $Q{=}10$\,GeV in Figs.~\ref{fig:cosphi} and~\ref{fig:chi2eff}. We also limit ourselves to the gluon in lead, 
as the focus of the SIH data is on the heavy elements; 
the results for gold are similar to lead.

\end{widetext}

In Fig.~\ref{fig:cosphi} we see how the 5\,TeV SIH ALICE data sets ($\pi^0,\pi^\pm,K^\pm$)
display a strong anti-correlation ($\cos\phi\sim -0.9$) with the low~$x$ gluon ($x\sim10^{-3}$) that is not seen in any of the remaining data, including the 8\,TeV ALICE neutral pions. 
This observation suggests that the 5\,TeV SIH ALICE data has significant impact 
on the resulting gluon in the small $x$ region.  
Interestingly, the correlation angle of the STAR and PHENIX neutral pion data 
are quite similar to each other, and in the region ${x\sim 5{\times} 10^{-2}}$ 
they also exhibit a strong anti-correlation \mbox{($\cos\phi\sim -0.9$),}
which then becomes strong and positive \mbox{($\cos\phi\sim +0.8$)} for larger $x$.
The  8\,TeV ALICE neutral pion data show a correlation behavior similar to the 
NMC96~SnC data set (which is the dominant DIS set due to its large size and $Q$ coverage), and somewhat opposite to the STAR and PHENIX neutral pion data.
Examining the larger $x$ region ($x>0.1$), the influence of the various data sets
is more mixed with with the STAR and PHENIX $\pi^0$ data yielding a large positive correlation and ATLAS 8\,TeV $\pi^0$ and STAR $\pi^\pm$  yielding a large negative correlation, with the result that the high $x$ gluon remains mostly unchanged in Fig.~\ref{fig:main_pdf_WZ}.

Turning to  the $\chieff{}$ in Fig.~\ref{fig:chi2eff}, 
we can see that the  CMS Run II $W^\pm$ and NMC96 SnC data  
remain the main forces determining the gluon,
with the ALICE neutral pion and NMC95re~CaD data sets also providing constraints.

Among the SIH data sets, the 8\,TeV neutral pion data has the largest $\chieff{}$, followed by the 5\,TeV neutral pion data. However, they generally do not reach values as high as the previously mentioned DIS and WZ production data. 
It is unfortunate that we must impose the $p_T>3$\,GeV cut on the SIH data due to
limitations of our perturbative theoretical calculations; this removes a large 
amount of precision SIH data from our analysis. 
Improved theoretical techniques such as resummation may allow us to 
extend our analysis to smaller $p_T$ values in the future 
so that a larger amount of the SIH data can be included in the PDF determination.

\begin{figure*}[p]
	\centering    
	\includegraphics[width=\textwidth]{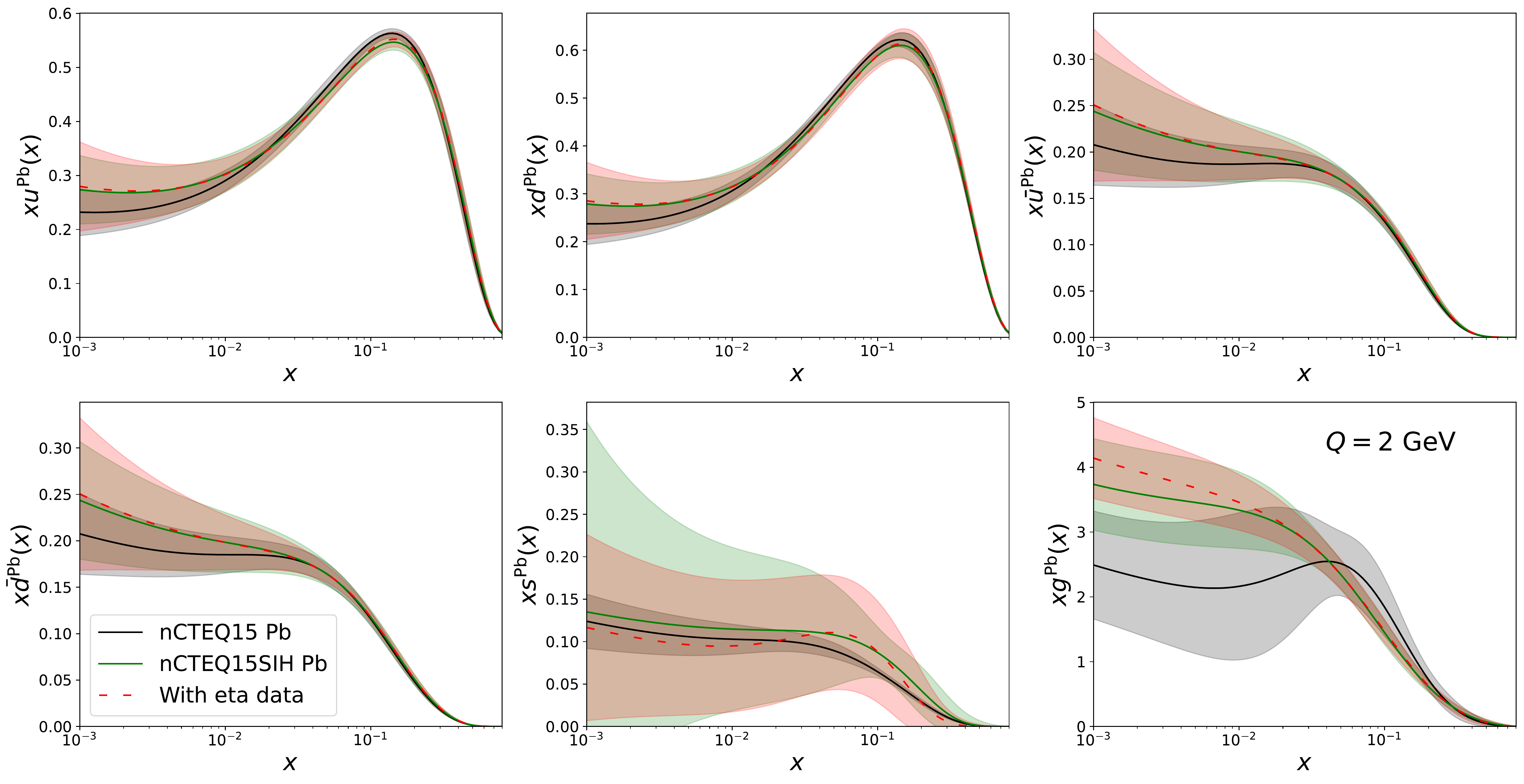}
	\caption{Lead PDFs from fits to the nCTEQ15 data including the SIH eta meson data. The baseline nCTEQ15 fit is shown in black, the fit with eta meson data is shown in red and the corresponding main fit is shown in green.
	The nCTEQ15 (black) and nCTEQ15SIH (green) are also displayed in Fig.~\ref{fig:main_pdfs}.
	}
	\label{fig:eta_pdfs}
\end{figure*}
\begin{figure*}[p]
	\centering
	\includegraphics[width=\textwidth]{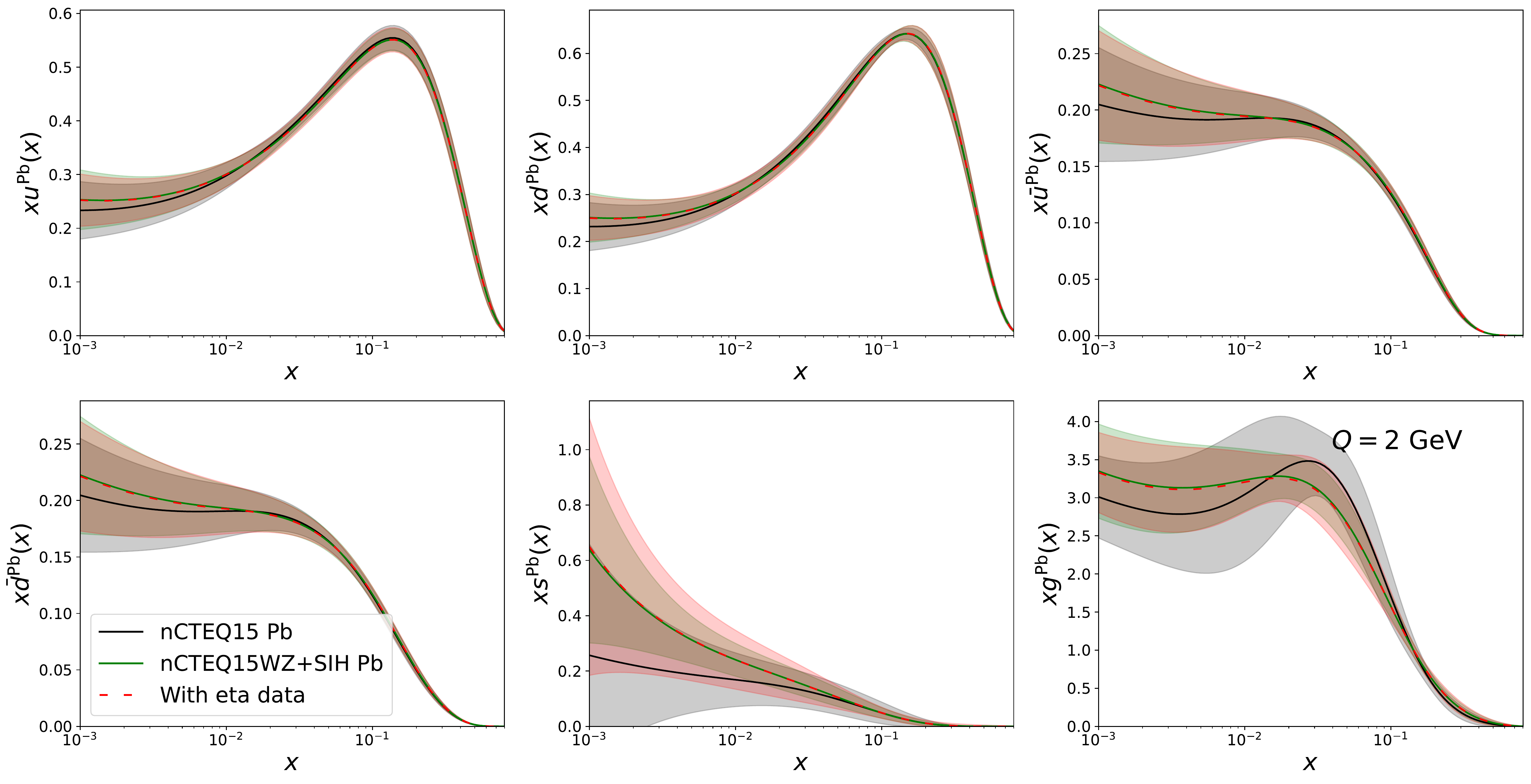}
	\caption{Lead PDFs from fits to the nCTEQ15WZ data including the SIH eta meson data. The baseline nCTEQ15WZ fit is shown in black, the fit with eta meson data is shown in red and the corresponding main fit is shown in green.
	The nCTEQ15 (black) and nCTEQ15WZ+SIH (green) are also displayed in Fig.~\ref{fig:main_pdf_WZ}.
	}
	\label{fig:eta_pdfs_WZ}
\end{figure*}

\subsection{Impact of the eta meson}\label{sec:eta}

We now investigate the  impact of  including the eta meson data. 
We will compute our fit with the DSS fragmentation functions for the pions
and kaons, and use the AESSS fragmentation functions for the eta mesons. 
The AESSS FFs do not provide any uncertainties,
so we will not include any for the eta meson data.  
Using the same $p_T\ge 3$ cut as before for the pions and kaons, 
the eta meson data now provides  an additional 18 data points from RHIC and 19 from ALICE.

The  fits including the eta meson data are shown in Figs.~\ref{fig:eta_pdfs} and~\ref{fig:eta_pdfs_WZ}, and are compared  with 
the baseline fit (nCTEQ15WZ) in black, the corresponding main fit (nCTEQ15SIH) in green,
and the fit with eta in red (nCTEQ15SIH+eta).

Examining the results of Figs.~\ref{fig:eta_pdfs},
the impact of the eta meson data yields a slight upwards shift of the gluon 
in the low $x$ region, 
and a downward modification of the  strange quark both at small $x$ and
larger $x\sim 0.2$.
The uncertainty of the gluon shrinks by a small amount,
the strange  uncertainty is reduced in the low~$x$ region while it increases at medium-$x$, 
and  the other flavors show  a slight increase at very small $x$. 

Examining the results of the second case based on the nCTEQ15WZ fit as shown in Figs.~\ref{fig:eta_pdfs_WZ}, 
we observe that the central values of the nCTEQ15WZ+SIH fit with and without the eta meson data  are virtually identical. 
Regarding the PDF uncertainties, the error bands of gluon and down-quark are reduced only by a negligible margin, while the strange quark uncertainty grows very slightly. 
Since the uncertainty of the eta fragmentation function is expected to be larger than that of pions and kaons, the net effect of including the eta meson data into the fit  would most likely be inconsequential if these additional uncertainties 
were included in the analysis.

\section{Conclusion}\label{sec:conclusions}

Using the  nCTEQ++ framework, we incorporated new data on single inclusive hadron (SIH) production from ALICE and RHIC into our PDF analysis. 
We investigated the choice of scales and fragmentation functions
(including their uncertainties), and identified a $p_T$ region  
where reliable perturbative predictions can be applied to help
constrain the PDFs. 

We obtained a good $\chidof{}$ for all the data sets, and found that
the new SIH meson data  had a noticeable impact on the gluon PDF at low to medium~$x$. 
Compared to the nCTEQ15WZ PDF set, the gluon flattens out in the region  around $x=0.05$, and the uncertainties in this region shrink. 

The necessary  $p_T > 3$\,GeV cut  limits our ability to constrain 
the PDFs in  the low~$x$ region. 
If it were possible to expand the theoretical predictions to lower 
$p_T$ values with improved calculation techniques, 
then we could use the very precise  ALICE data in this region 
to further improve the determination of the gluon PDF. 

Nevertheless, even with the current limitations on the kinematics,
the SIH data provide useful constraints on the nuclear gluon
distribution which is still one of the least constrained nPDFs. 
As such we believe the presented analysis and the obtained PDFs
provide an important step on the way to more precise knowledge
of the nuclear structure. Also on the practical level the reduced
gluon uncertainties are important for many applications. 
The PDFs of the nCTEQ15WZ+SIH fit for a selection of nuclei will 
be provided through the LHAPDF website, and those of other fits can be obtained upon request. 

\section*{Acknowledgments}
The work of P.D., M.K.\ and K.K.\ was funded by the Deutsche Forschungsgemeinschaft (DFG, German Research Foundation) – project-id 273811115 – SFB 1225. 
L.A.H., M.K., K.K.\ and K.F.M.\ also acknowledge support of the DFG through the Research Training Group GRK 2149. 
A.K. acknowledges the support of Narodowe Centrum Nauki under Grant No. 2019/34/E/ST2/00186.
The work of T.J.\ was supported by the Deutsche Forschungsgemeinschaft (DFG, German Research Foundation) under grant 396021762 - TRR 257. 
F.O.\ acknowledges support through US DOE grant  No.~DE-SC0010129,
and the National Science Foundation under Grant No.~NSF PHY-1748958.            
The work of I.S.\ was supported by the French CNRS via the IN2P3 project GLUE@NLO.

\appendix

\section{Fitting data normalizations}\label{sec:norm}
We use the $\chi^2$ prescription given in Ref.~\cite{DAgostini:1993arp} to fit the normalizations of the SIH and $W/Z$ production data. For a data set $D$ with $N$ data points and $S$ correlated systematic errors, the $\chi^2$ of the data set is given by:
\begin{align}
\begin{split}
    \chi^2_D = \sum_{i,j}^N & \left(D_i -\frac{T_i}{N_{\rm norm}} \right)(C^{-1})_{ij}\left(D_j - \frac{T_j}{N_{\rm norm}} \right) \\
    & +\left(\frac{1-N_{\rm norm}}{\sigma_{\rm norm}} \right)^2, \label{eqn:chi2norm_def}
\end{split}
\end{align}
where $\sigma_{\rm norm}$ is the normalization uncertainty and $T_i$ is the theoretical prediction for point $i$. The last term of Eq.~\eqref{eqn:chi2norm_def} is called the normalization penalty, and it vanishes when the fitted normalization is equal to unity. The penalty is scaled by the normalization uncertainty $\sigma_{\rm norm}$, which is around 0.03 for $W/Z$ production and 
ranges from 0.034 to 0.17 for SIH production.
The covariance matrix $C_{ij}$ is defined as:
\begin{align}
    C_{ij} = \sigma_i^2\delta_{ij}+\sum_\alpha^S\bar{\sigma}_{i\alpha}\bar{\sigma}_{j\alpha}
\end{align}
where $\sigma_{i}$ is the total uncorrelated uncertainty (added in quadrature) for data point $i$, and $\bar{\sigma}_{i\alpha}$ is the correlated systematic uncertainty for data point $i$ from source
$\alpha$. We use the analytical formula for the inverse of the correlation matrix found in Ref.~\cite{Stump:2001gu} to obtain:
\begin{align}
\begin{split}
    \chi^2_D = &\sum_{i} \left(\frac{D_i -T_i/N_{\rm norm}}{\sigma_{i}} \right)^2-B^TA^{-1}B\\ &+\left(\frac{1-N_{\rm norm}}{\sigma_{\rm norm}} \right)^2 
\end{split}
\end{align}
with 
\begin{align}
    A_{\alpha\gamma} = \delta_{\alpha\gamma}+\sum_i \frac{\bar{\sigma}_{i\alpha} \bar{\sigma}_{i\gamma}} {\sigma_i^2}
\end{align}
and 
\begin{align}
    B_\alpha = \sum_i\frac{\bar{\sigma}_{i\alpha}(D_i-T_i/N_{\rm norm})}{\sigma_i^2}.
\end{align}

\section{Uncertainties of other FFs}\label{sec:FFunc}
\begin{figure*}[p!]
	\centering
	\includegraphics[width=0.3\textwidth]{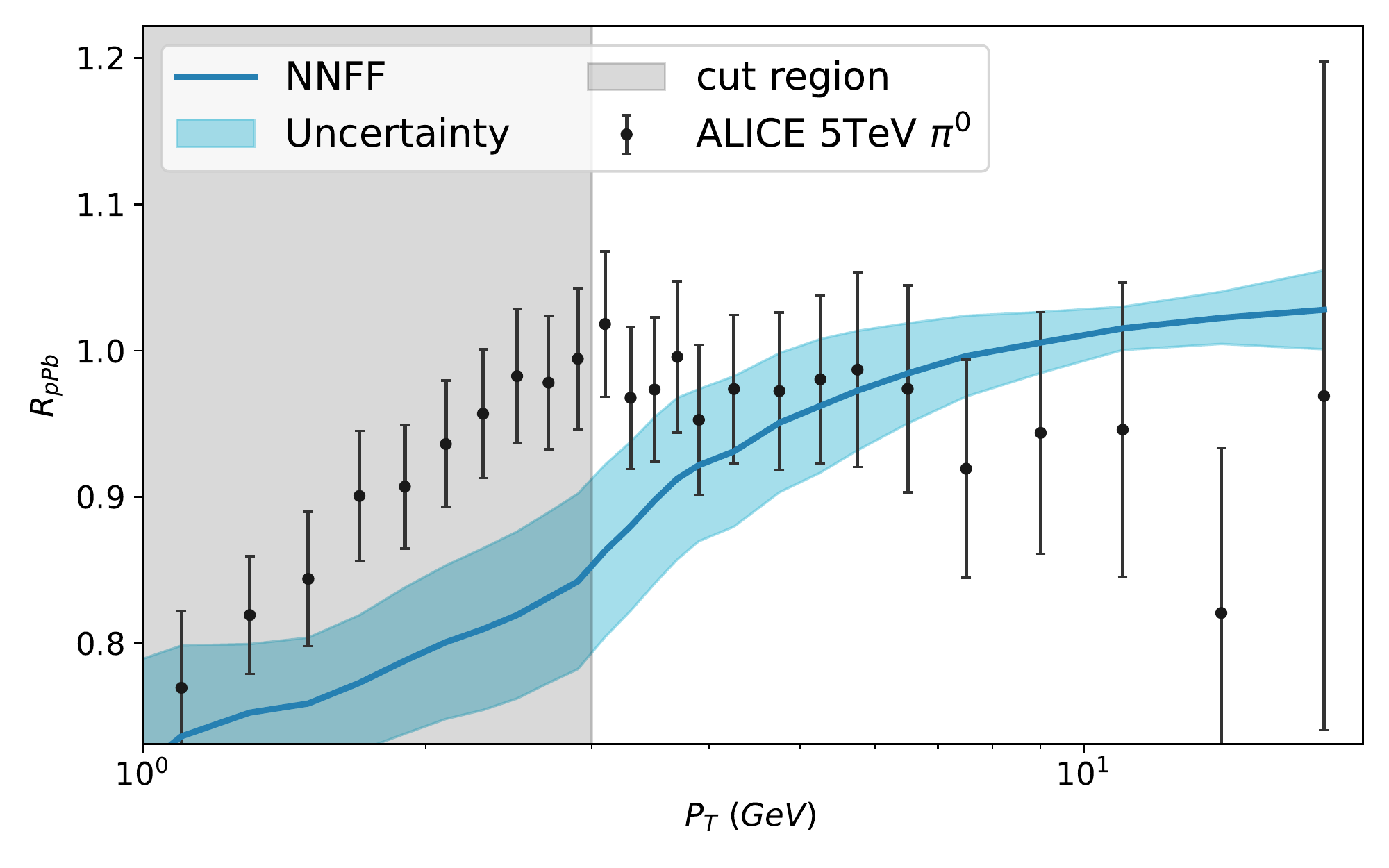}
	\includegraphics[width=0.3\textwidth]{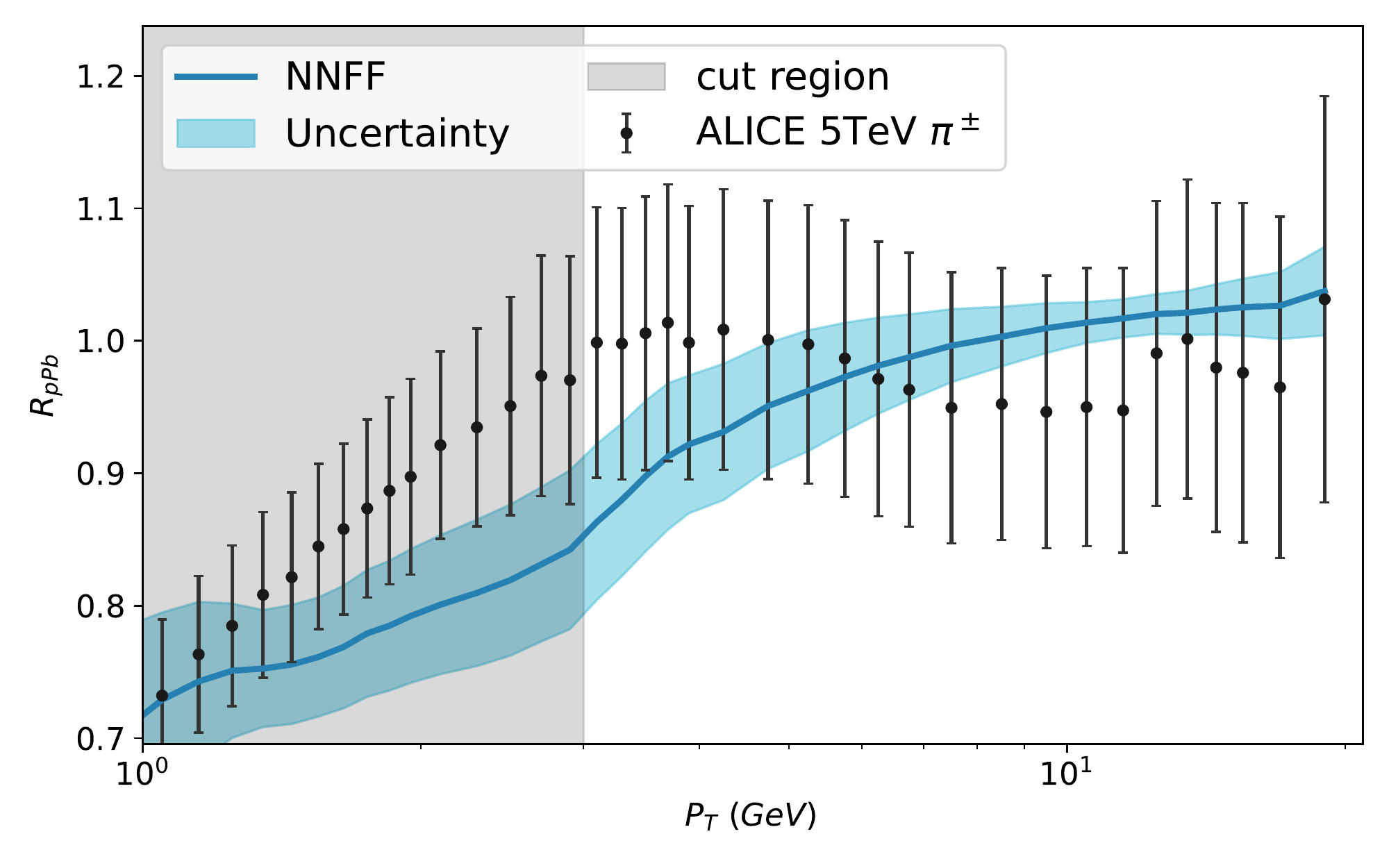}
	\includegraphics[width=0.3\textwidth]{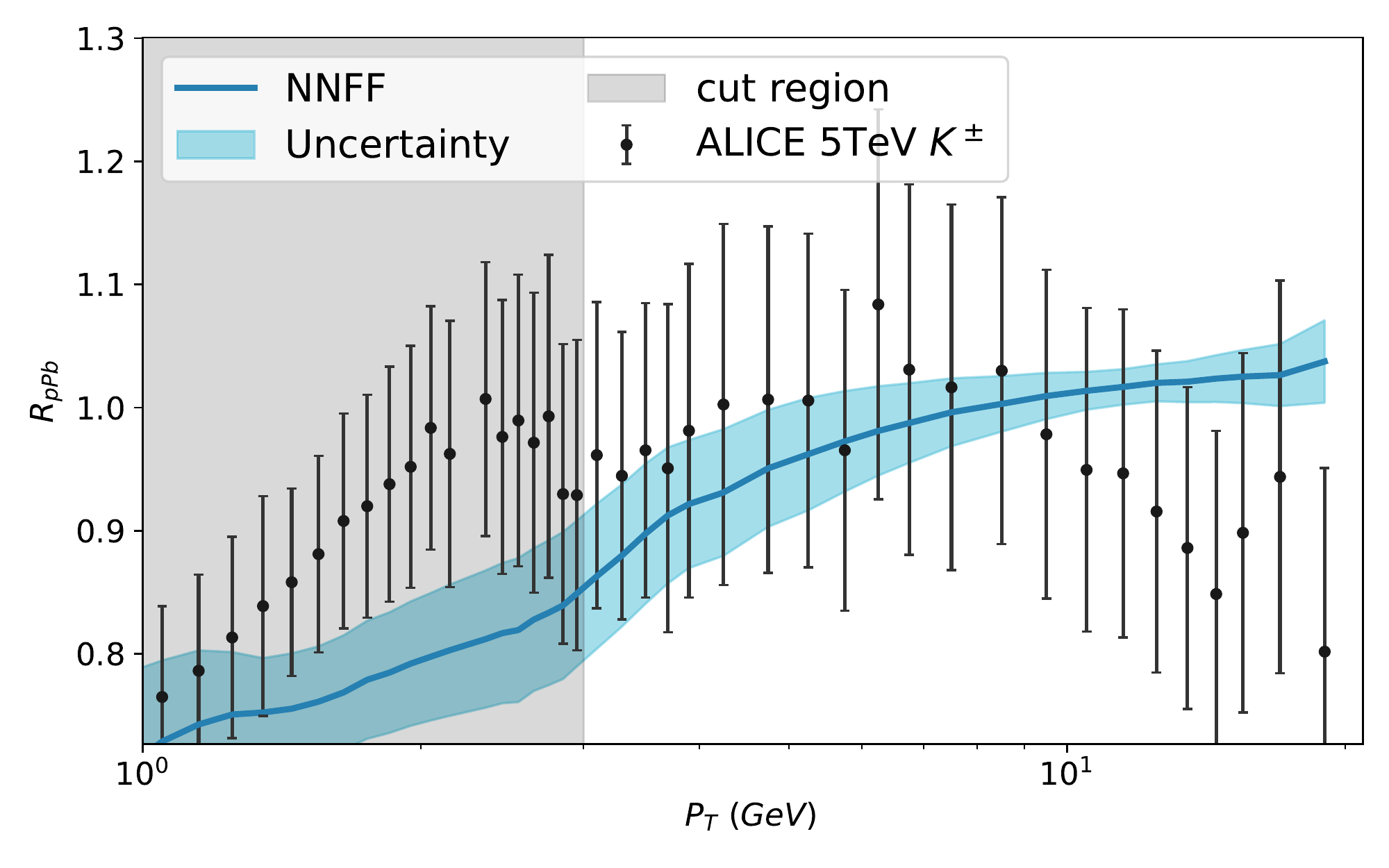}
	\includegraphics[width=0.3\textwidth]{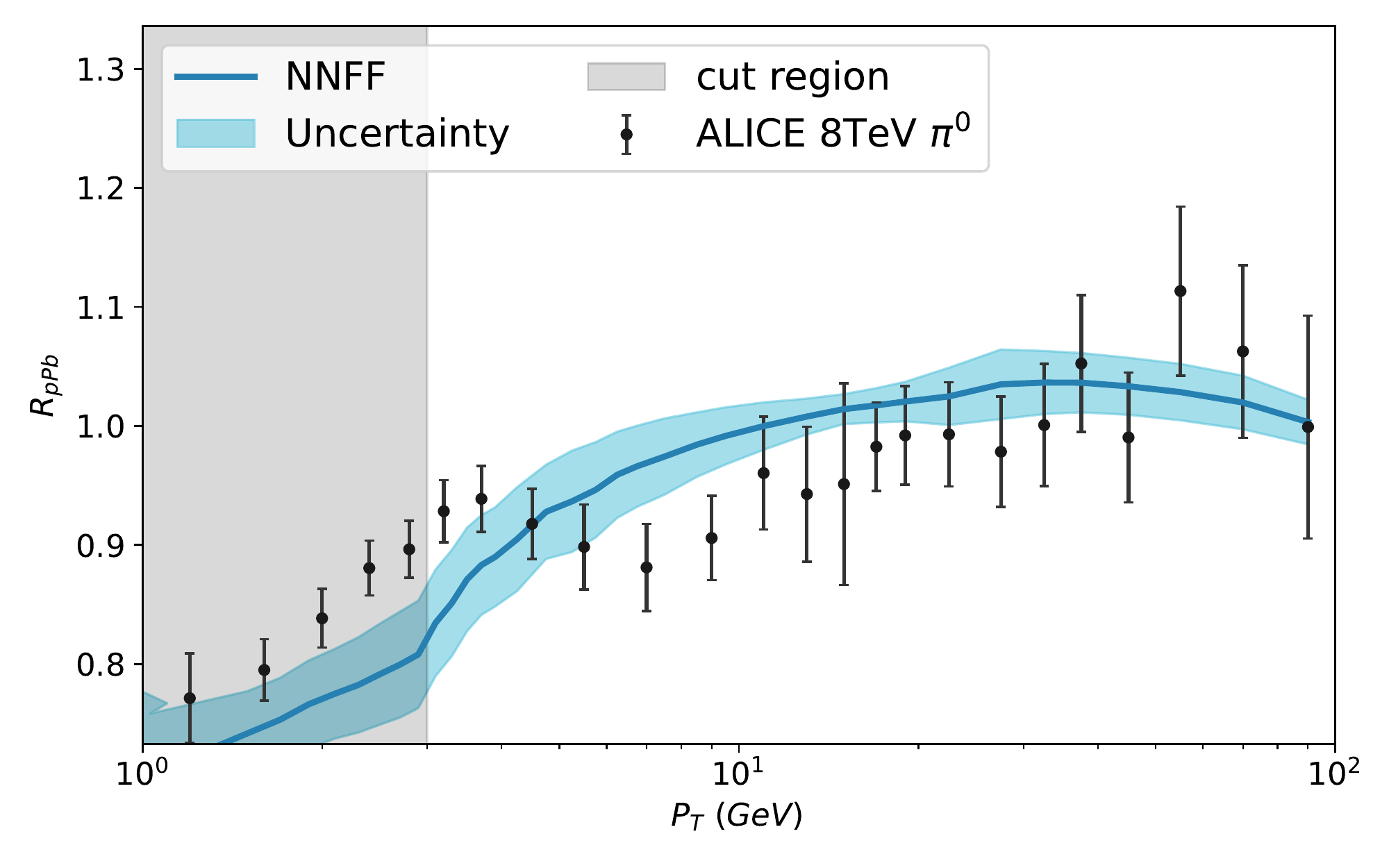}
	\includegraphics[width=0.3\textwidth]{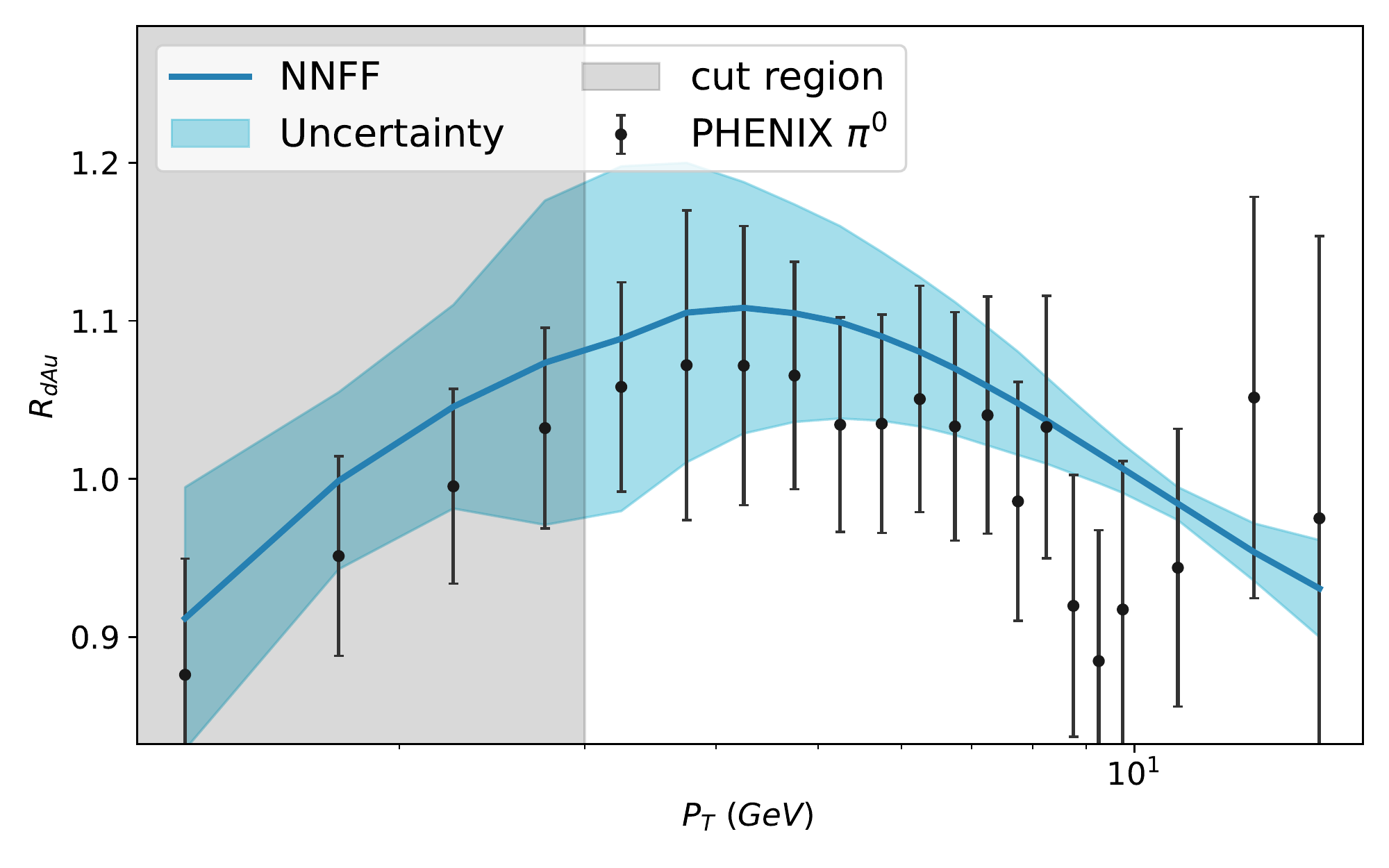}
	\includegraphics[width=0.3\textwidth]{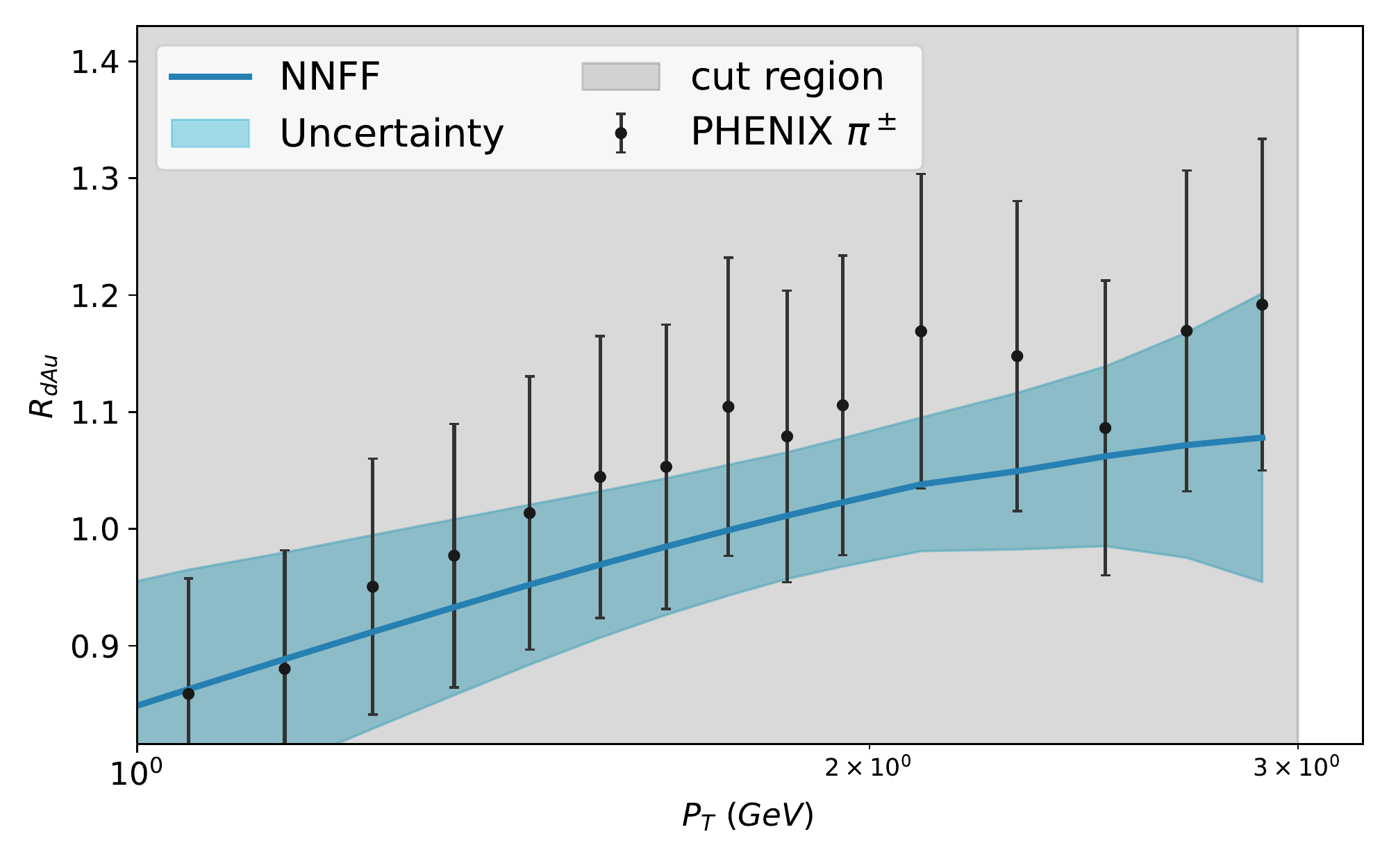}
	\includegraphics[width=0.3\textwidth]{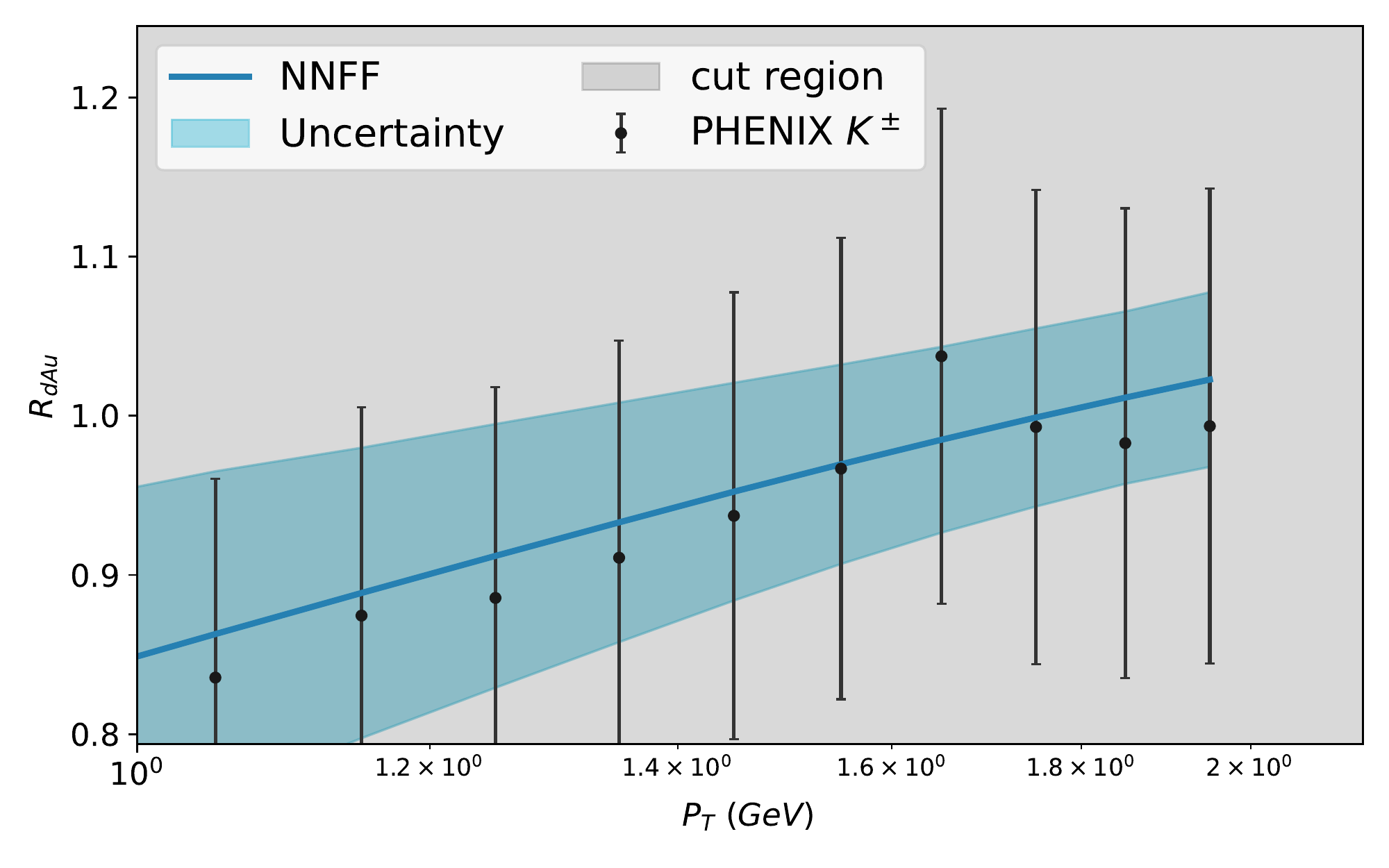}
	\includegraphics[width=0.3\textwidth]{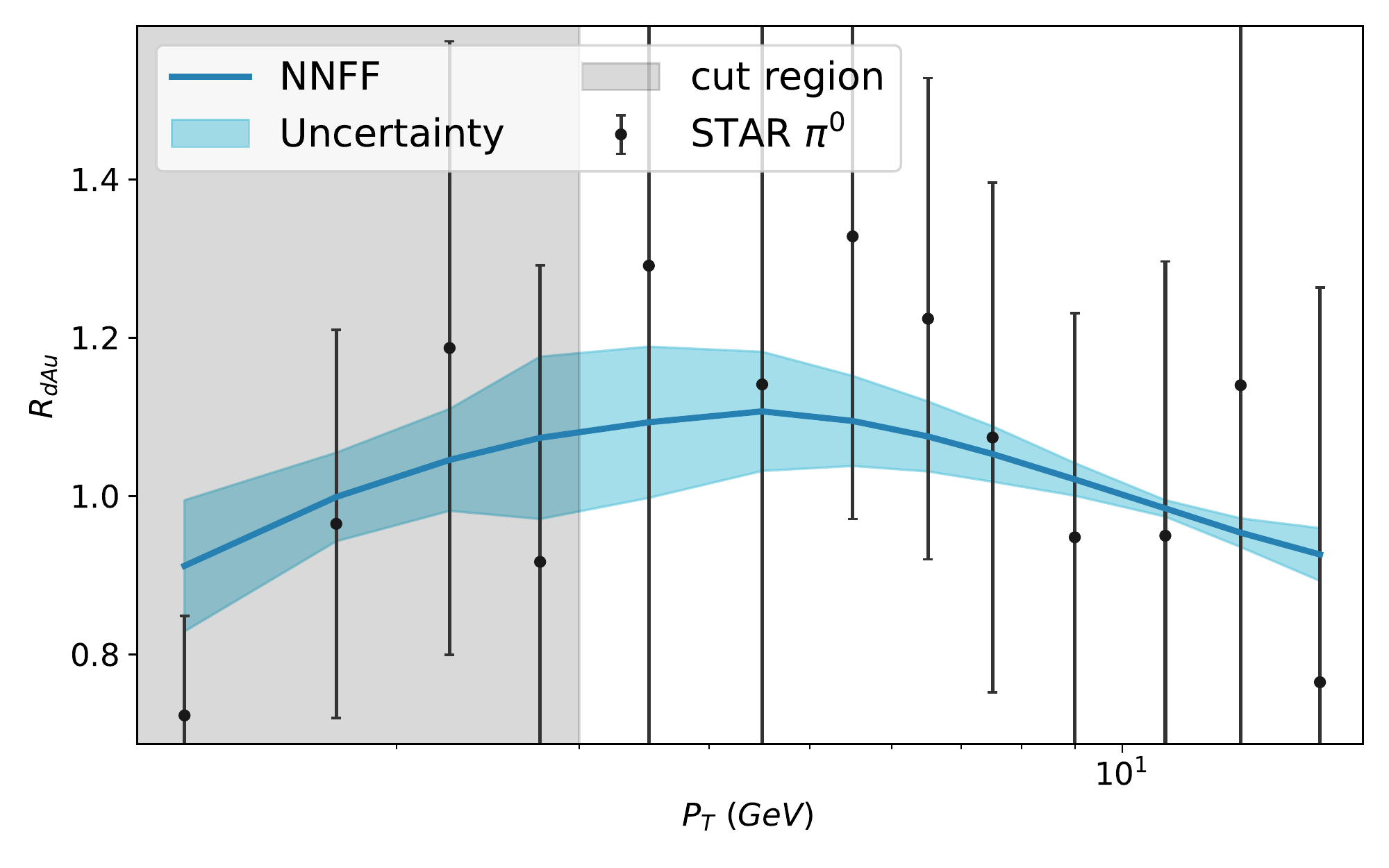}
	\includegraphics[width=0.3\textwidth]{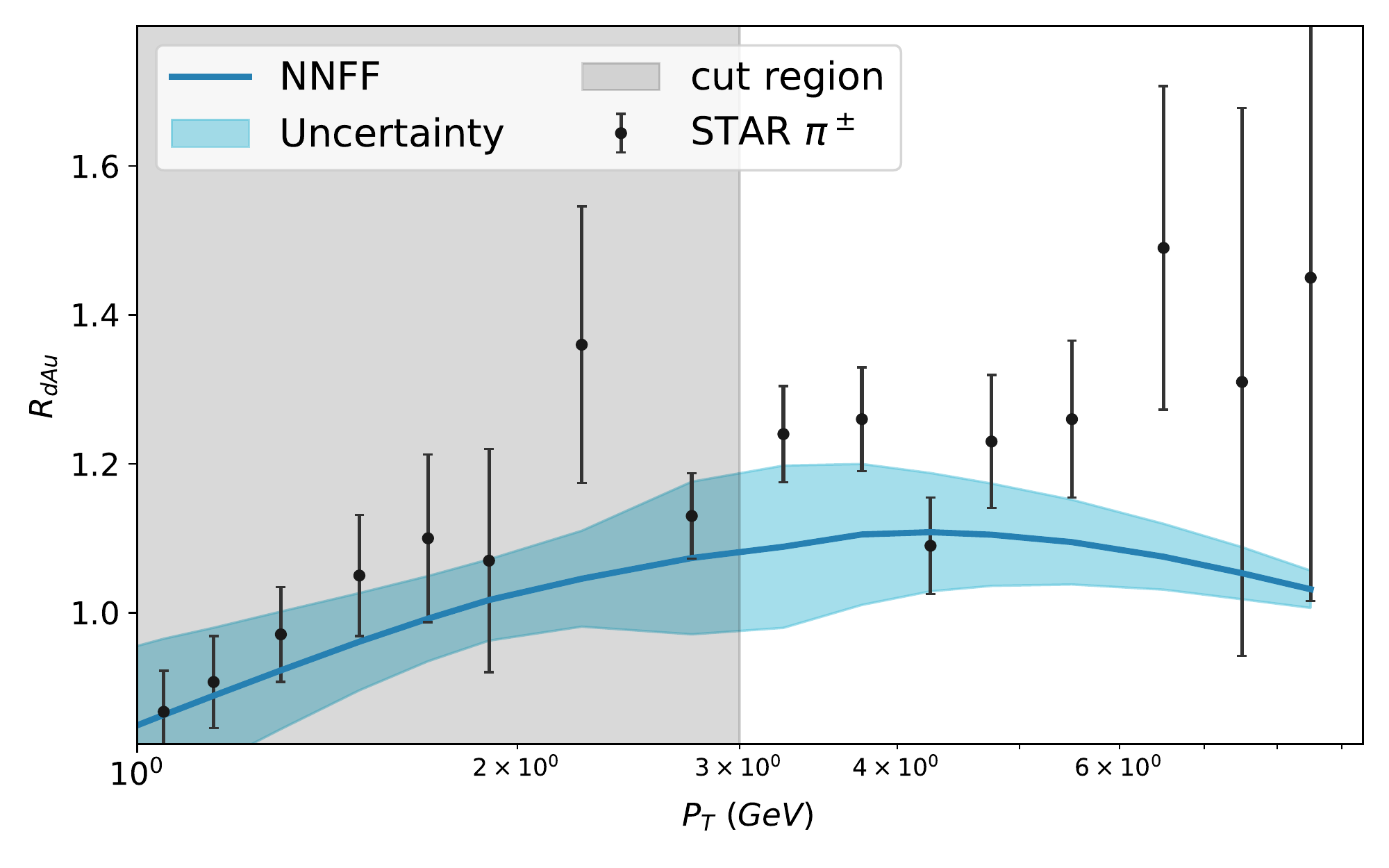}
	\caption{Uncertainties calculated from the NNFF replicas using nCTEQ15WZ PDFs. The computed uncertainties use 83 of the 101 provided replicas because the remaining 18 lead to unphysical behavior such as large jumps from one $p_T$ value to another or negative cross sections due to numerical problems.
	}
	\label{fig:NNFFunc}
\end{figure*}

\begin{figure*}[p!]
	\centering
	\includegraphics[width=0.3\textwidth]{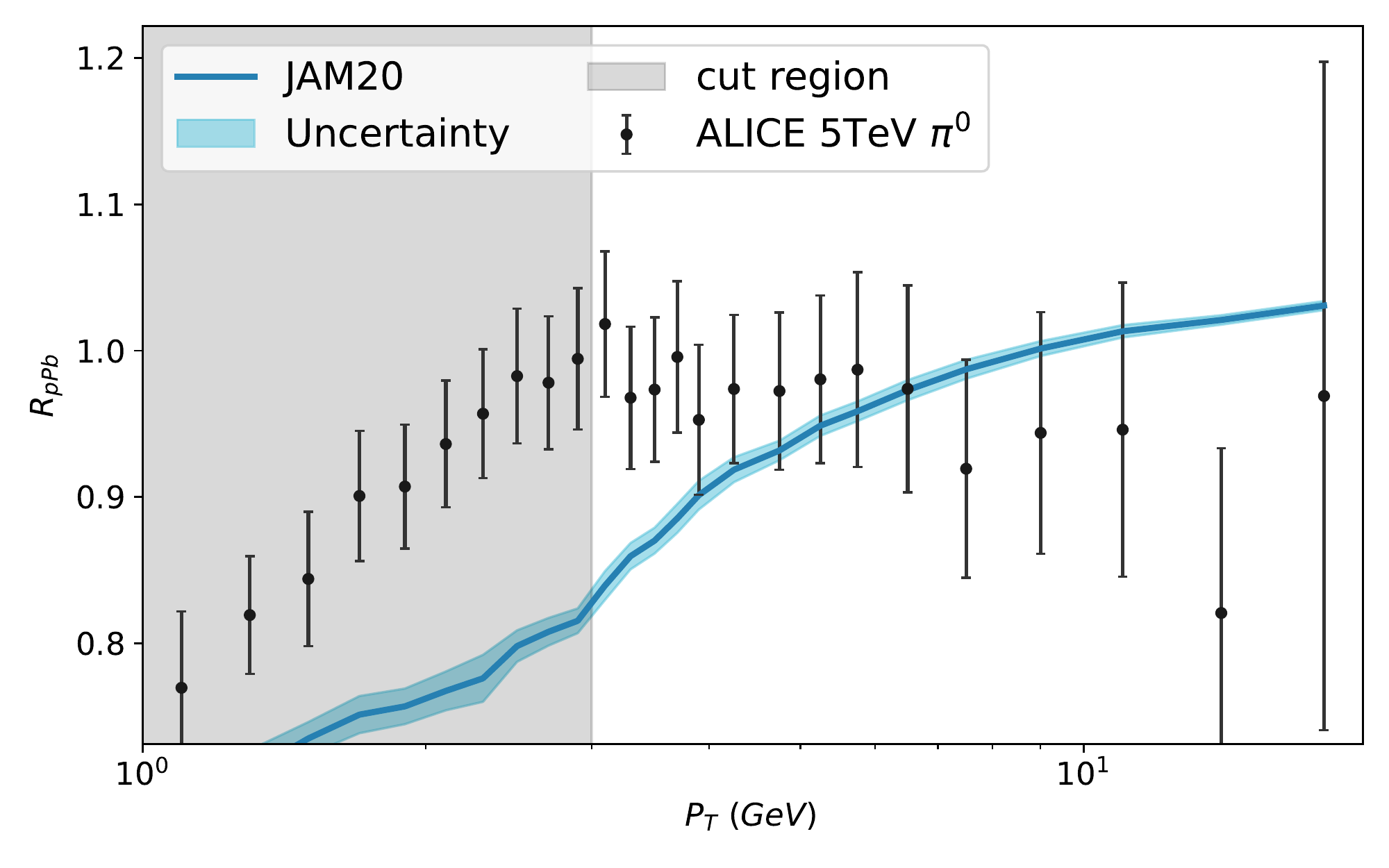}
	\includegraphics[width=0.3\textwidth]{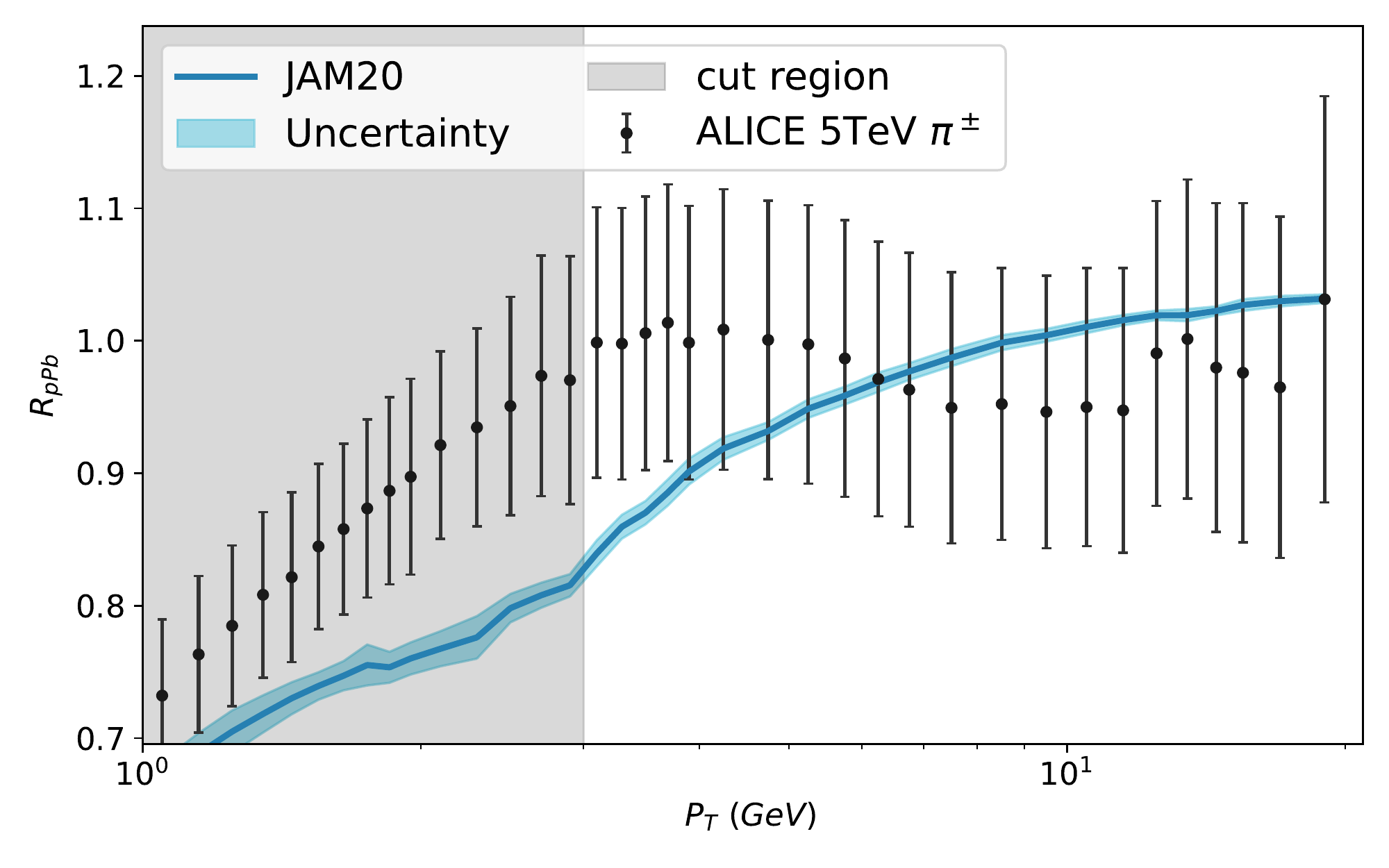}
	\includegraphics[width=0.3\textwidth]{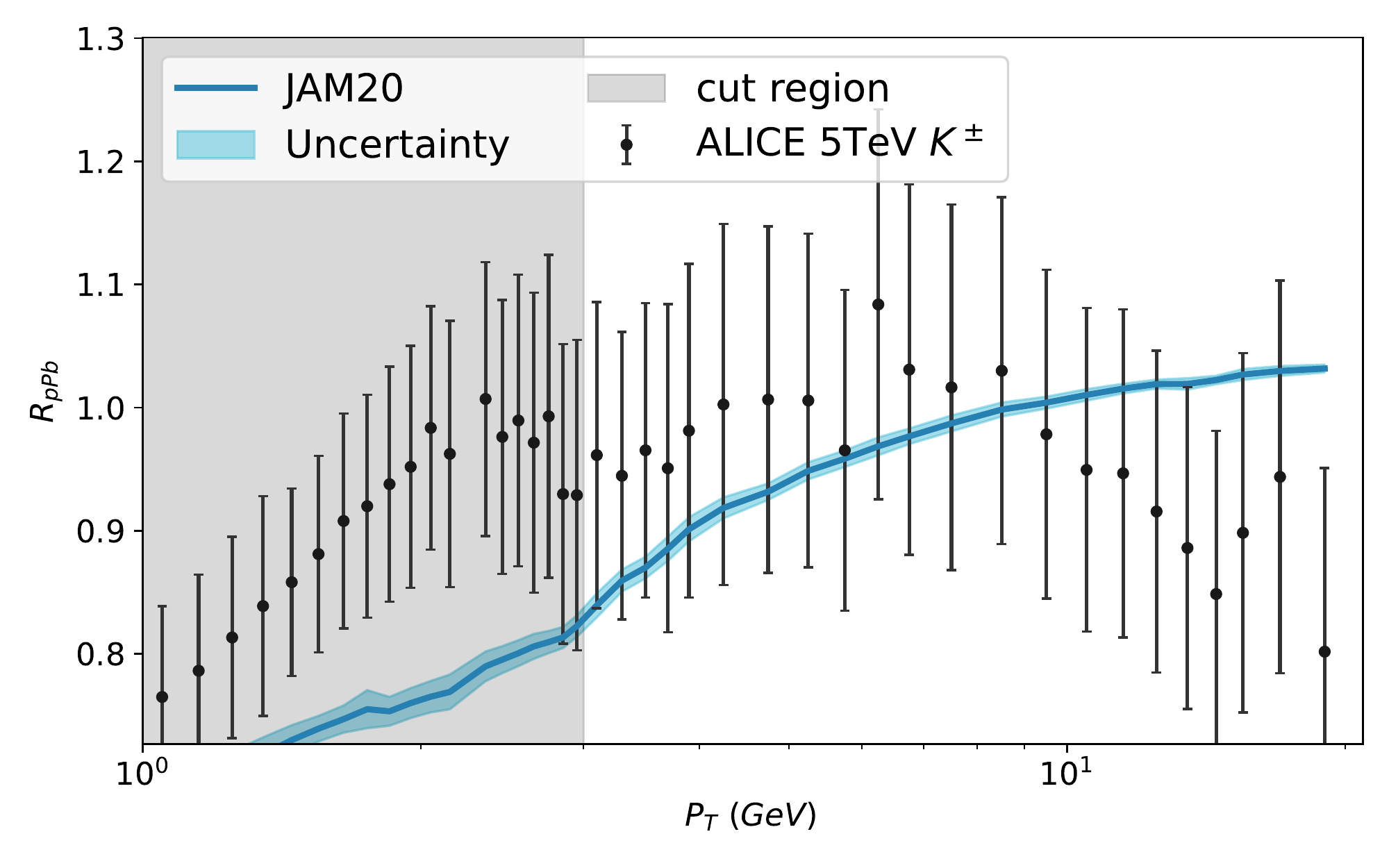}
	\includegraphics[width=0.3\textwidth]{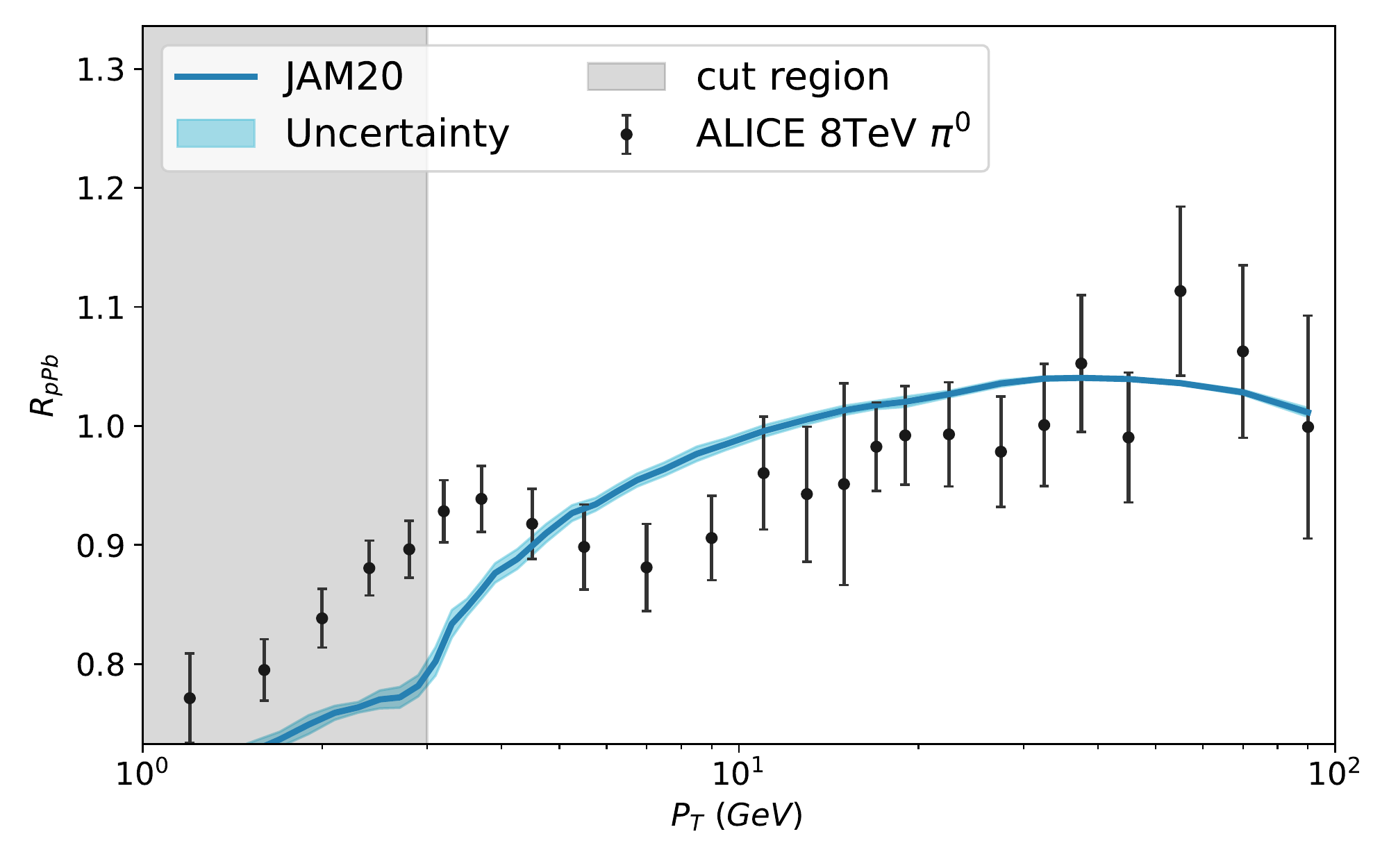}
	\includegraphics[width=0.3\textwidth]{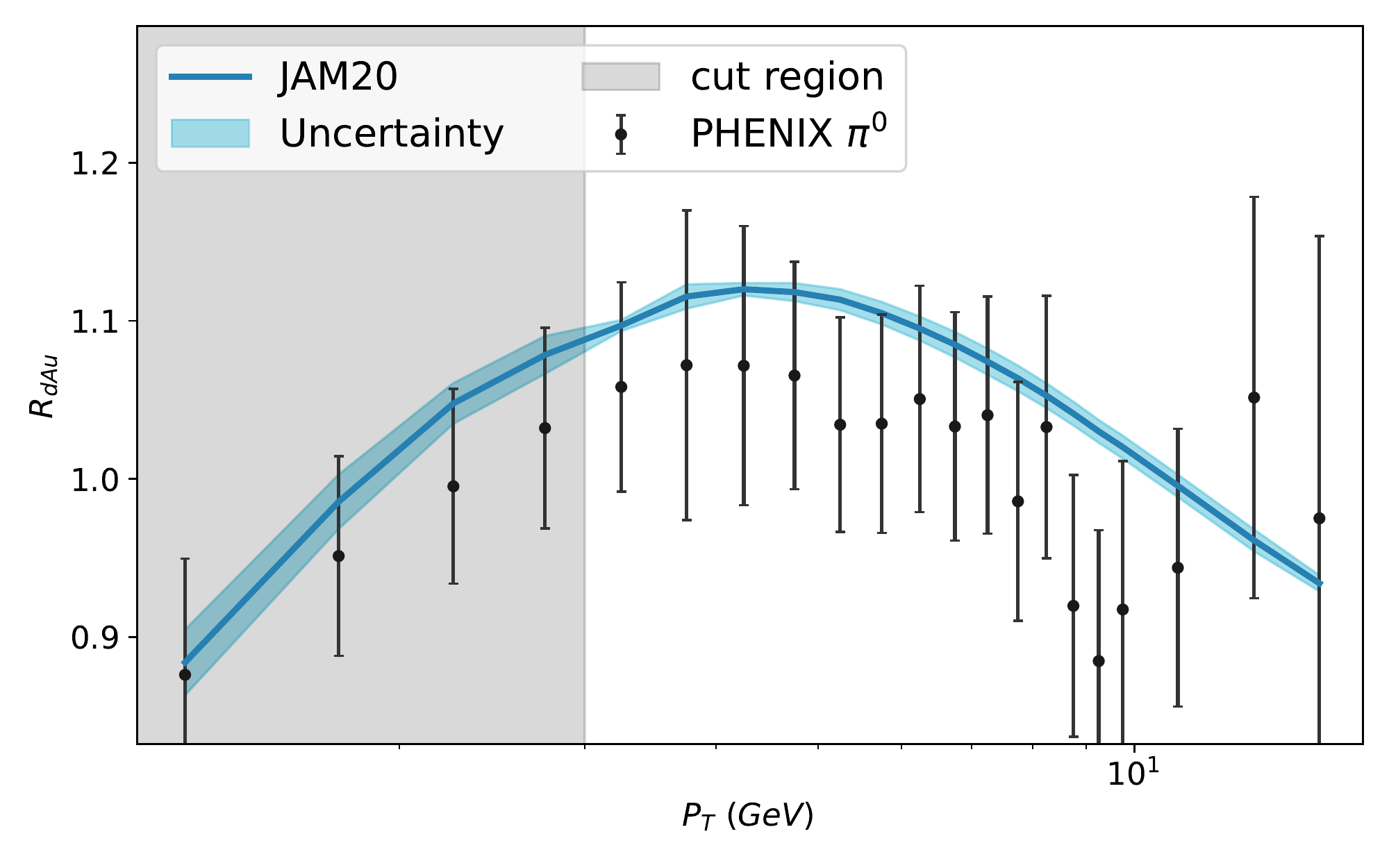}
	\includegraphics[width=0.3\textwidth]{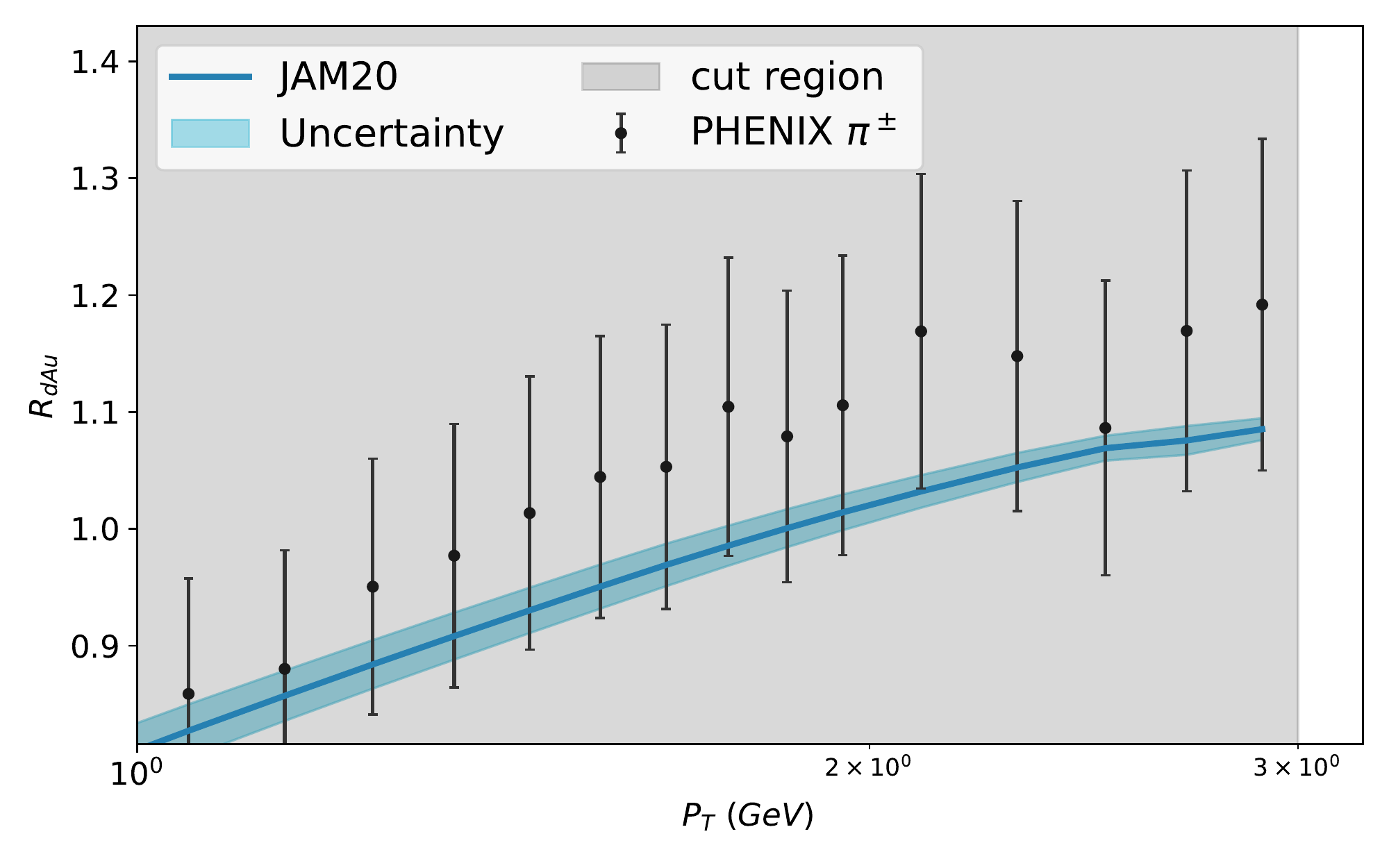}
	\includegraphics[width=0.3\textwidth]{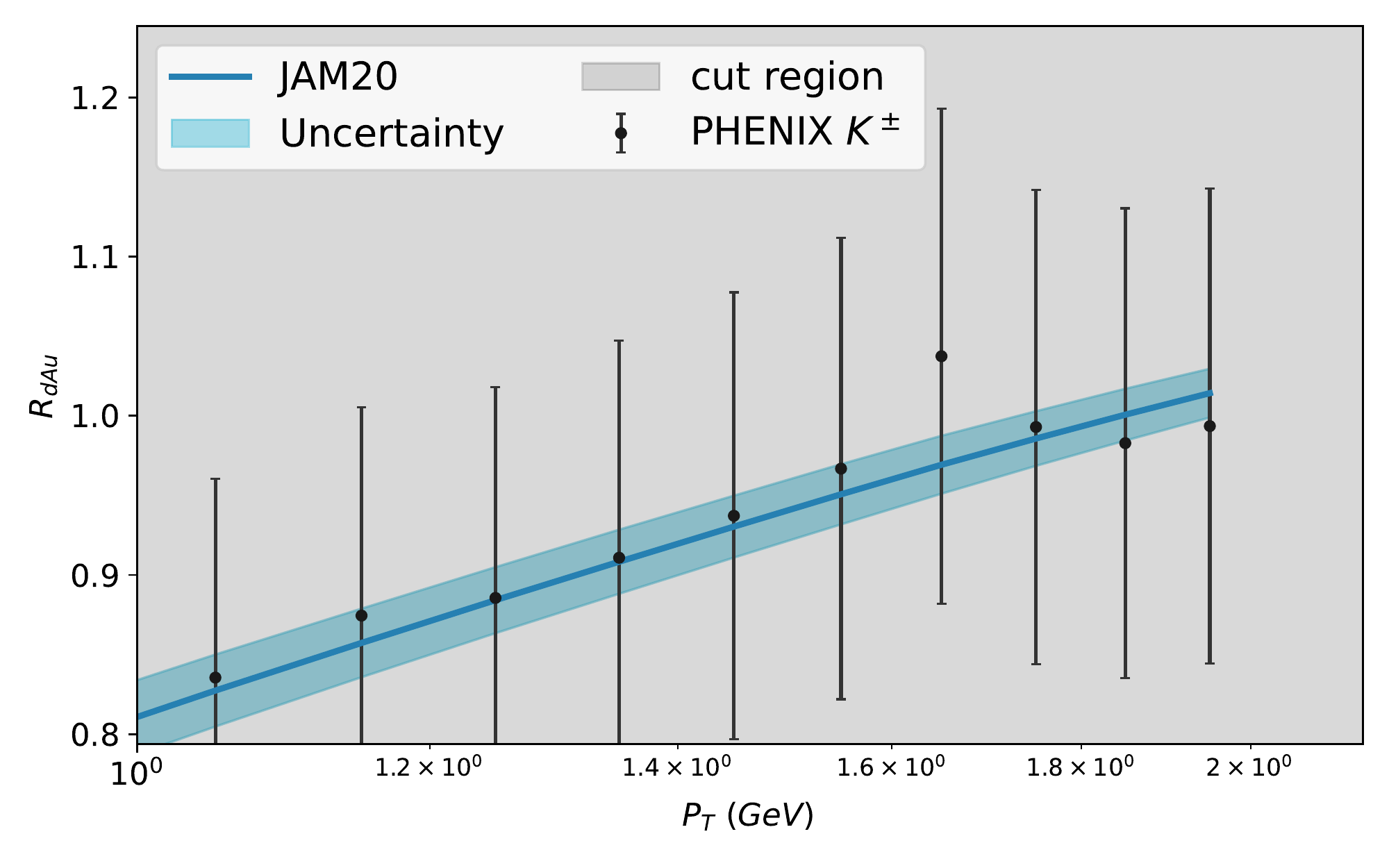}
	\includegraphics[width=0.3\textwidth]{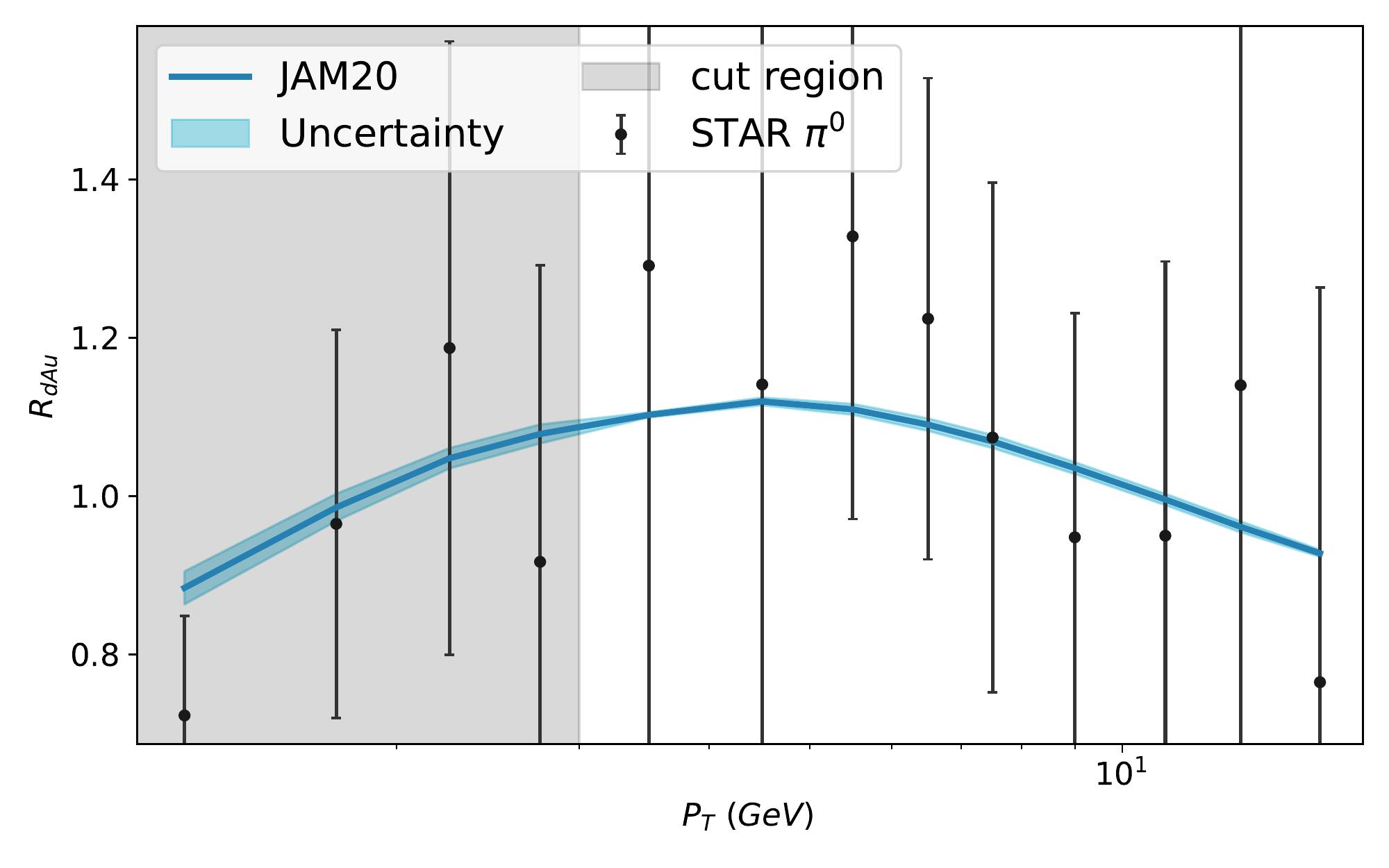}
	\includegraphics[width=0.3\textwidth]{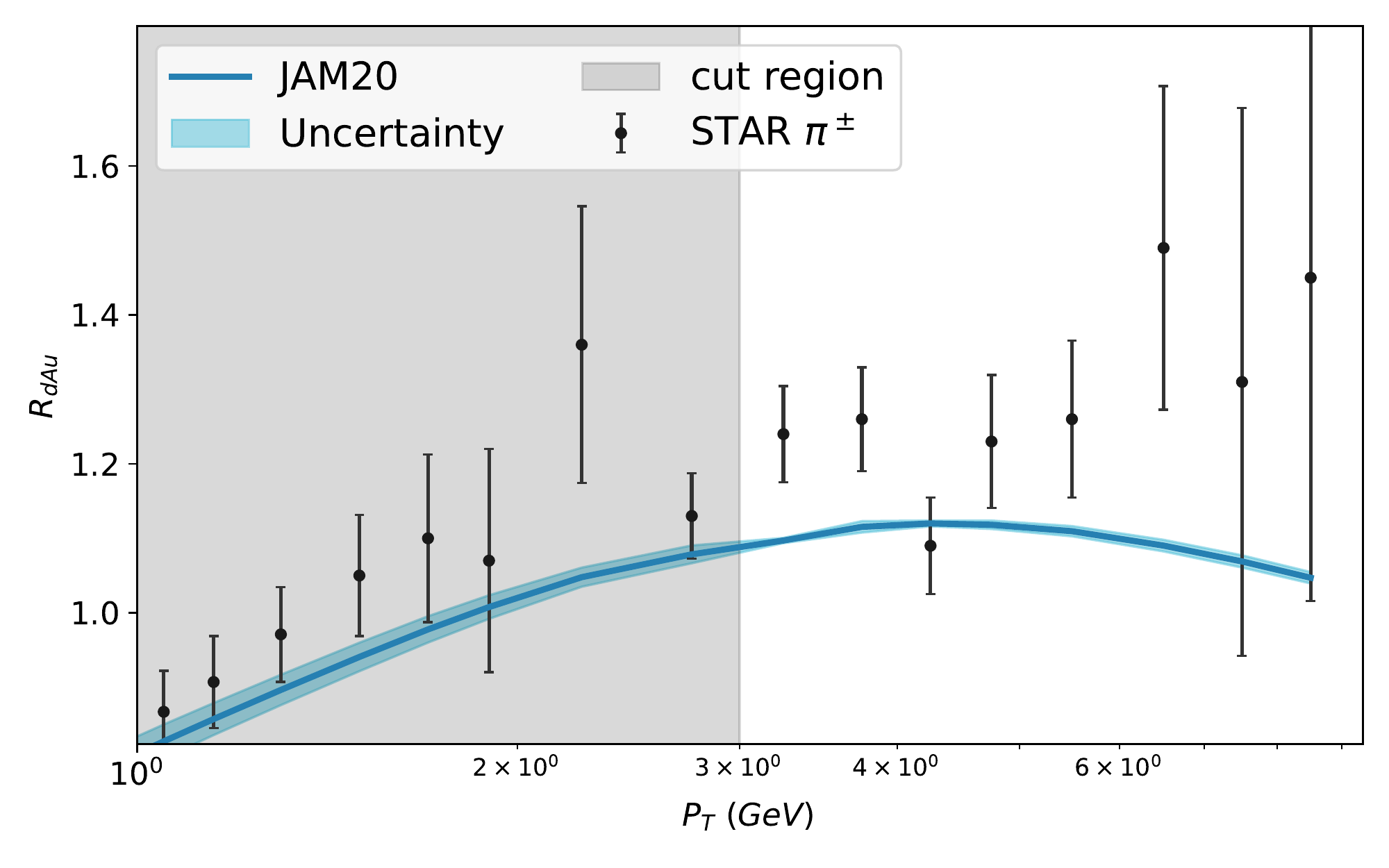}
	\caption{Uncertainties calculated from the JAM20 replicas using nCTEQ15WZ PDFs.}
	\label{fig:JAM20unc}
\end{figure*}

We compare the data with  our theoretical predictions 
with nCTEQ15WZ PDFs
using the  uncertainties taken from the  
NNFF and JAM20 fragmentation functions 
in Figs.~\ref{fig:NNFFunc}
and~\ref{fig:JAM20unc}, respectively. 

The NNFF fragmentation functions yield slightly larger uncertainties than those of DSS shown in Fig.~\ref{fig:nCTEQ15WZ_FFcomp}. 
This may be due, in part, to the use of a parameterization-free neural network instead of a ``traditional" parameterization, and a slightly smaller data set. 
The uncertainties of the JAM20 fragmentation are so small across the kinematic region with $p_T{>}1$\,GeV that they can be neglected when compared with the data uncertainty.


\newpage
\FloatBarrier

\bibliographystyle{utphys}
\bibliography{refs.bib,extra.bib}

\providecommand{\href}[2]{#2}\begingroup\raggedright\begin{thebibliography}{10}

\bibitem{Hou:2019efy}
T.-J. Hou {\em et al.}, ``{New CTEQ global analysis of quantum chromodynamics
  with high-precision data from the LHC},''
\href{http://www.arXiv.org/abs/1912.10053}{{\tt 1912.10053}}.

\bibitem{Ball:2017nwa}
{\bf NNPDF} Collaboration, R.~D. Ball {\em et al.}, ``{Parton distributions
  from high-precision collider data},'' {\em Eur. Phys. J.} {\bf C77} (2017),
  no.~10, 663,
\href{http://www.arXiv.org/abs/1706.00428}{{\tt 1706.00428}}.

\bibitem{Kovarik:2015cma}
K.~Kovarik {\em et al.}, ``{nCTEQ15 - Global analysis of nuclear parton
  distributions with uncertainties in the CTEQ framework},'' {\em Phys. Rev. D}
  {\bf 93} (2016), no.~8, 085037,
  \href{http://www.arXiv.org/abs/1509.00792}{{\tt 1509.00792}}.

\bibitem{Eskola:2016oht}
K.~J. Eskola, P.~Paakkinen, H.~Paukkunen, and C.~A. Salgado, ``{EPPS16: Nuclear
  parton distributions with LHC data},'' {\em Eur. Phys. J. C} {\bf 77} (2017),
  no.~3, 163, \href{http://www.arXiv.org/abs/1612.05741}{{\tt 1612.05741}}.

\bibitem{AbdulKhalek:2019mzd}
{\bf NNPDF} Collaboration, R.~Abdul~Khalek, J.~J. Ethier, and J.~Rojo,
  ``{Nuclear parton distributions from lepton-nucleus scattering and the impact
  of an electron-ion collider},'' {\em Eur. Phys. J.} {\bf C79} (2019), no.~6,
  471,
\href{http://www.arXiv.org/abs/1904.00018}{{\tt 1904.00018}}.

\bibitem{AbdulKhalek:2020yuc}
R.~Abdul~Khalek, J.~J. Ethier, J.~Rojo, and G.~van Weelden, ``{nNNPDF2.0: Quark
  Flavor Separation in Nuclei from LHC Data},''
  \href{http://www.arXiv.org/abs/2006.14629}{{\tt 2006.14629}}.

\bibitem{Ethier:2020way}
J.~J. Ethier and E.~R. Nocera, ``{Parton Distributions in Nucleons and
  Nuclei},'' {\em Ann. Rev. Nucl. Part. Sci.} (2020), no.~70, 1--34,
\href{http://www.arXiv.org/abs/2001.07722}{{\tt 2001.07722}}.

\bibitem{Khalek:2018mdn}
R.~Abdul~Khalek, S.~Bailey, J.~Gao, L.~Harland-Lang, and J.~Rojo, ``{Towards
  Ultimate Parton Distributions at the High-Luminosity LHC},'' {\em Eur. Phys.
  J. C} {\bf 78} (2018), no.~11, 962,
  \href{http://www.arXiv.org/abs/1810.03639}{{\tt 1810.03639}}.

\bibitem{Gao:2017yyd}
J.~Gao, L.~Harland-Lang, and J.~Rojo, ``{The Structure of the Proton in the LHC
  Precision Era},'' {\em Phys. Rept.} {\bf 742} (2018) 1--121,
  \href{http://www.arXiv.org/abs/1709.04922}{{\tt 1709.04922}}.

\bibitem{Kovarik:2019xvh}
K.~Kovarik, P.~M. Nadolsky, and D.~E. Soper, ``Hadron structure in high-energy
  collisions,'' {\em Rev.Mod.Phys.} {\bf 92} (2020) 045003,
  \href{http://www.arXiv.org/abs/1905.06957}{{\tt 1905.06957}}.

\bibitem{Alekhin:2017olj}
S.~Alekhin, J.~Blümlein, and S.~Moch, ``{Strange sea determination from
  collider data},'' {\em Phys. Lett. B} {\bf 777} (2018) 134--140,
  \href{http://www.arXiv.org/abs/1708.01067}{{\tt 1708.01067}}.

\bibitem{Nadolsky:2008zw}
P.~M. Nadolsky, H.-L. Lai, Q.-H. Cao, J.~Huston, J.~Pumplin, D.~Stump, W.-K.
  Tung, and C.~P. Yuan, ``{Implications of CTEQ global analysis for collider
  observables},'' {\em Phys. Rev. D} {\bf 78} (2008) 013004,
  \href{http://www.arXiv.org/abs/0802.0007}{{\tt 0802.0007}}.

\bibitem{Sato:2019yez}
{\bf JAM} Collaboration, N.~Sato, C.~Andres, J.~J. Ethier, and W.~Melnitchouk,
  ``{Strange quark suppression from a simultaneous Monte Carlo analysis of
  parton distributions and fragmentation functions},'' {\em Phys. Rev. D} {\bf
  101} (2020), no.~7, 074020, \href{http://www.arXiv.org/abs/1905.03788}{{\tt
  1905.03788}}.

\bibitem{Harland-Lang:2014zoa}
L.~Harland-Lang, A.~Martin, P.~Motylinski, and R.~Thorne, ``{Parton
  distributions in the LHC era: MMHT 2014 PDFs},'' {\em Eur. Phys. J. C} {\bf
  75} (2015), no.~5, 204, \href{http://www.arXiv.org/abs/1412.3989}{{\tt
  1412.3989}}.

\bibitem{Thorne:2019mpt}
R.~S. Thorne, S.~Bailey, T.~Cridge, L.~A. Harland-Lang, A.~Martin, and
  R.~Nathvani, ``{Updates of PDFs using the MMHT framework},'' {\em PoS} {\bf
  DIS2019} (2019) 036, \href{http://www.arXiv.org/abs/1907.08147}{{\tt
  1907.08147}}.

\bibitem{Ball:2009mk}
{\bf NNPDF} Collaboration, R.~D. Ball, L.~Del~Debbio, S.~Forte, A.~Guffanti,
  J.~I. Latorre, A.~Piccione, J.~Rojo, and M.~Ubiali, ``{Precision
  determination of electroweak parameters and the strange content of the proton
  from neutrino deep-inelastic scattering},'' {\em Nucl. Phys. B} {\bf 823}
  (2009) 195--233, \href{http://www.arXiv.org/abs/0906.1958}{{\tt 0906.1958}}.

\bibitem{Lin:2017snn}
H.-W. Lin {\em et al.}, ``{Parton distributions and lattice QCD calculations: a
  community white paper},'' {\em Prog. Part. Nucl. Phys.} {\bf 100} (2018)
  107--160, \href{http://www.arXiv.org/abs/1711.07916}{{\tt 1711.07916}}.

\bibitem{Lin:2020rut}
H.-W. Lin {\em et al.}, ``{Parton distributions and lattice QCD calculations:
  toward 3D structure},'' \href{http://www.arXiv.org/abs/2006.08636}{{\tt
  2006.08636}}.

\bibitem{Guzey:2019kik}
V.~Guzey and M.~Klasen, ``{Constraints on nuclear parton distributions from
  dijet photoproduction at the LHC},'' {\em Eur. Phys. J. C} {\bf 79} (2019),
  no.~5, 396, \href{http://www.arXiv.org/abs/1902.05126}{{\tt 1902.05126}}.

\bibitem{Klasen:2017kwb}
M.~Klasen, K.~Kovarik, and J.~Potthoff, ``{Nuclear parton density functions
  from jet production in DIS at an EIC},'' {\em Phys. Rev. D} {\bf 95} (2017),
  no.~9, 094013, \href{http://www.arXiv.org/abs/1703.02864}{{\tt 1703.02864}}.

\bibitem{Klasen:2018gtb}
M.~Klasen and K.~Kova\v{r}\'\i{}k, ``{Nuclear parton density functions from
  dijet photoproduction at the EIC},'' {\em Phys. Rev. D} {\bf 97} (2018),
  no.~11, 114013, \href{http://www.arXiv.org/abs/1803.10985}{{\tt 1803.10985}}.

\bibitem{Armesto:2015lrg}
N.~Armesto, H.~Paukkunen, J.~M. Pen\'\i{}n, C.~A. Salgado, and P.~Zurita, ``{An
  analysis of the impact of LHC Run I proton\textendash{}lead data on nuclear
  parton densities},'' {\em Eur. Phys. J. C} {\bf 76} (2016), no.~4, 218,
  \href{http://www.arXiv.org/abs/1512.01528}{{\tt 1512.01528}}.

\bibitem{Armesto:2006ph}
N.~Armesto, ``{Nuclear shadowing},'' {\em J. Phys. G} {\bf 32} (2006)
  R367--R394, \href{http://www.arXiv.org/abs/hep-ph/0604108}{{\tt
  hep-ph/0604108}}.

\bibitem{Frankfurt:2011cs}
L.~Frankfurt, V.~Guzey, and M.~Strikman, ``{Leading Twist Nuclear Shadowing
  Phenomena in Hard Processes with Nuclei},'' {\em Phys. Rept.} {\bf 512}
  (2012) 255--393, \href{http://www.arXiv.org/abs/1106.2091}{{\tt 1106.2091}}.

\bibitem{Kopeliovich:2012kw}
B.~Z. Kopeliovich, J.~G. Morfin, and I.~Schmidt, ``{Nuclear Shadowing in
  Electro-Weak Interactions},'' {\em Prog. Part. Nucl. Phys.} {\bf 68} (2013)
  314--372, \href{http://www.arXiv.org/abs/1208.6541}{{\tt 1208.6541}}.

\bibitem{Kulagin:2004ie}
S.~A. Kulagin and R.~Petti, ``{Global study of nuclear structure functions},''
  {\em Nucl. Phys. A} {\bf 765} (2006) 126--187,
  \href{http://www.arXiv.org/abs/hep-ph/0412425}{{\tt hep-ph/0412425}}.

\bibitem{Brodsky:1989qz}
S.~J. Brodsky and H.~J. Lu, ``{Shadowing and Antishadowing of Nuclear Structure
  Functions},'' {\em Phys. Rev. Lett.} {\bf 64} (1990) 1342.

\bibitem{Brodsky:2004qa}
S.~J. Brodsky, I.~Schmidt, and J.-J. Yang, ``{Nuclear antishadowing in neutrino
  deep inelastic scattering},'' {\em Phys. Rev. D} {\bf 70} (2004) 116003,
  \href{http://www.arXiv.org/abs/hep-ph/0409279}{{\tt hep-ph/0409279}}.

\bibitem{Geesaman:1995yd}
D.~F. Geesaman, K.~Saito, and A.~W. Thomas, ``{The nuclear EMC effect},'' {\em
  Ann. Rev. Nucl. Part. Sci.} {\bf 45} (1995) 337--390.

\bibitem{Norton:2003cb}
P.~R. Norton, ``{The EMC effect},'' {\em Rept. Prog. Phys.} {\bf 66} (2003)
  1253--1297.

\bibitem{Hen:2013oha}
O.~Hen, D.~W. Higinbotham, G.~A. Miller, E.~Piasetzky, and L.~B. Weinstein,
  ``{The EMC Effect and High Momentum Nucleons in Nuclei},'' {\em Int. J. Mod.
  Phys. E} {\bf 22} (2013) 1330017,
  \href{http://www.arXiv.org/abs/1304.2813}{{\tt 1304.2813}}.

\bibitem{Malace:2014uea}
S.~Malace, D.~Gaskell, D.~W. Higinbotham, and I.~Cloet, ``{The Challenge of the
  EMC Effect: existing data and future directions},'' {\em Int. J. Mod. Phys.
  E} {\bf 23} (2014), no.~08, 1430013,
  \href{http://www.arXiv.org/abs/1405.1270}{{\tt 1405.1270}}.

\bibitem{Hen:2016kwk}
O.~Hen, G.~A. Miller, E.~Piasetzky, and L.~B. Weinstein, ``{Nucleon-Nucleon
  Correlations, Short-lived Excitations, and the Quarks Within},'' {\em Rev.
  Mod. Phys.} {\bf 89} (2017), no.~4, 045002,
  \href{http://www.arXiv.org/abs/1611.09748}{{\tt 1611.09748}}.

\bibitem{Iancu:2000hn}
E.~Iancu, A.~Leonidov, and L.~D. McLerran, ``{Nonlinear gluon evolution in the
  color glass condensate. 1.},'' {\em Nucl. Phys. A} {\bf 692} (2001) 583--645,
  \href{http://www.arXiv.org/abs/hep-ph/0011241}{{\tt hep-ph/0011241}}.

\bibitem{Gelis:2010nm}
F.~Gelis, E.~Iancu, J.~Jalilian-Marian, and R.~Venugopalan, ``{The Color Glass
  Condensate},'' {\em Ann. Rev. Nucl. Part. Sci.} {\bf 60} (2010) 463--489,
  \href{http://www.arXiv.org/abs/1002.0333}{{\tt 1002.0333}}.

\bibitem{Andronic:2017pug}
A.~Andronic, P.~Braun-Munzinger, K.~Redlich, and J.~Stachel, ``{Decoding the
  phase structure of QCD via particle production at high energy},'' {\em
  Nature} {\bf 561} (2018), no.~7723, 321--330,
  \href{http://www.arXiv.org/abs/1710.09425}{{\tt 1710.09425}}.

\bibitem{ALICE:2016fzo}
{\bf ALICE} Collaboration, J.~Adam {\em et al.}, ``{Enhanced production of
  multi-strange hadrons in high-multiplicity proton-proton collisions},'' {\em
  Nature Phys.} {\bf 13} (2017) 535--539,
  \href{http://www.arXiv.org/abs/1606.07424}{{\tt 1606.07424}}.

\bibitem{Ortiz:2019osu}
{\bf ALICE, ATLAS, CMS, LHCb} Collaboration, A.~Ortiz, ``{Particle production
  and flow-like effects in small systems},'' {\em PoS} {\bf LHCP2019} (2019)
  091, \href{http://www.arXiv.org/abs/1909.03937}{{\tt 1909.03937}}.

\bibitem{Kusina:2020lyz}
A.~Kusina {\em et al.}, ``{Impact of LHC vector boson production in heavy ion
  collisions on strange PDFs},'' {\em Eur. Phys. J. C} {\bf 80} (2020), no.~10,
  968, \href{http://www.arXiv.org/abs/2007.09100}{{\tt 2007.09100}}.

\bibitem{deFlorian:2014xna}
D.~de~Florian, R.~Sassot, M.~Epele, R.~J. Hern\'andez-Pinto, and M.~Stratmann,
  ``{Parton-to-Pion Fragmentation Reloaded},'' {\em Phys. Rev. D} {\bf 91}
  (2015), no.~1, 014035, \href{http://www.arXiv.org/abs/1410.6027}{{\tt
  1410.6027}}.

\bibitem{Olness:2003wz}
F.~Olness, J.~Pumplin, D.~Stump, J.~Huston, P.~M. Nadolsky, H.~L. Lai,
  S.~Kretzer, J.~F. Owens, and W.~K. Tung, ``{Neutrino dimuon production and
  the strangeness asymmetry of the nucleon},'' {\em Eur. Phys. J. C} {\bf 40}
  (2005) 145--156, \href{http://www.arXiv.org/abs/hep-ph/0312323}{{\tt
  hep-ph/0312323}}.

\bibitem{Segarra:2020gtj}
E.~P. Segarra {\em et al.}, ``{nCTEQ15HIX -- Extending nPDF Analyses into the
  High-$x$, Low $Q^2$ Region},''
  \href{http://www.arXiv.org/abs/2012.11566}{{\tt 2012.11566}}.

\bibitem{Salam:2008qg}
G.~P. Salam and J.~Rojo, ``{A Higher Order Perturbative Parton Evolution
  Toolkit (HOPPET)},'' {\em Comput. Phys. Commun.} {\bf 180} (2009) 120--156,
  \href{http://www.arXiv.org/abs/0804.3755}{{\tt 0804.3755}}.

\bibitem{Carli:2010rw}
T.~Carli, D.~Clements, A.~Cooper-Sarkar, C.~Gwenlan, G.~P. Salam, F.~Siegert,
  P.~Starovoitov, and M.~Sutton, ``{A posteriori inclusion of parton density
  functions in NLO QCD final-state calculations at hadron colliders: The
  APPLGRID Project},'' {\em Eur. Phys. J. C} {\bf 66} (2010) 503--524,
  \href{http://www.arXiv.org/abs/0911.2985}{{\tt 0911.2985}}.

\bibitem{INCNLO}
M.~Werlen, ``{INCNLO}-direct photon and inclusive hadron production code
  website.'' {Version 1.4. http://lapth.cnrs.fr/PHOX\_FAMILY}.

\bibitem{Collins:1989gx}
J.~C. Collins, D.~E. Soper, and G.~F. Sterman, ``{Factorization of Hard
  Processes in QCD},'' {\em Adv. Ser. Direct. High Energy Phys.} {\bf 5} (1989)
  1--91, \href{http://www.arXiv.org/abs/hep-ph/0409313}{{\tt hep-ph/0409313}}.

\bibitem{Albino:2008gy}
S.~Albino, ``{The Hadronization of partons},'' {\em Rev. Mod. Phys.} {\bf 82}
  (2010) 2489--2556, \href{http://www.arXiv.org/abs/0810.4255}{{\tt
  0810.4255}}.

\bibitem{Schienbein:2007gr}
I.~Schienbein {\em et al.}, ``{A Review of Target Mass Corrections},'' {\em J.
  Phys. G} {\bf 35} (2008) 053101,
  \href{http://www.arXiv.org/abs/0709.1775}{{\tt 0709.1775}}.

\bibitem{tmc}
{\bf nCTEQ} Collaboration, I.~Schienbein {\em et al.}, ``Target mass
  corrections in lepton-nucleus {DIS} revisited.'' SMU-HEP-21-01 (In
  preparation).

\bibitem{Aurenche:1999nz}
P.~Aurenche, M.~Fontannaz, J.~P. Guillet, B.~A. Kniehl, and M.~Werlen, ``{Large
  $p_T$ inclusive $\pi^0$ cross-sections and next-to-leading-order QCD
  predictions},'' {\em Eur. Phys. J. C} {\bf 13} (2000) 347--355,
  \href{http://www.arXiv.org/abs/hep-ph/9910252}{{\tt hep-ph/9910252}}.

\bibitem{Adler:2006wg}
{\bf PHENIX} Collaboration, S.~S. Adler {\em et al.}, ``{Centrality dependence
  of $\pi^0$ and $\eta$ production at large transverse momentum in
  $\sqrt{s_{NN}}$ = 200 GeV d{+}Au collisions},'' {\em Phys. Rev. Lett.} {\bf
  98} (2007) 172302, \href{http://www.arXiv.org/abs/nucl-ex/0610036}{{\tt
  nucl-ex/0610036}}.

\bibitem{Adare:2013esx}
{\bf PHENIX} Collaboration, A.~Adare {\em et al.}, ``{Spectra and ratios of
  identified particles in Au+Au and $d$+Au collisions at
  $\sqrt{s_{\mathrm{N}\mathrm{N}}}=200$ GeV},'' {\em Phys. Rev. C} {\bf 88}
  (2013), no.~2, 024906, \href{http://www.arXiv.org/abs/1304.3410}{{\tt
  1304.3410}}.

\bibitem{Abelev:2009hx}
{\bf STAR} Collaboration, B.~I. Abelev {\em et al.}, ``{Inclusive $\pi^0$,
  $\eta$, and direct photon production at high transverse momentum in $p+p$ and
  $d+$Au collisions at $\sqrt{s_{\mathrm{N}\mathrm{N}}}=200$ GeV},'' {\em Phys.
  Rev. C} {\bf 81} (2010) 064904,
  \href{http://www.arXiv.org/abs/0912.3838}{{\tt 0912.3838}}.

\bibitem{Adams:2006nd}
{\bf STAR} Collaboration, J.~Adams {\em et al.}, ``{Identified hadron spectra
  at large transverse momentum in p{+}p and d{+}Au collisions at
  $\sqrt{s_{NN}}$ = 200 GeV},'' {\em Phys. Lett. B} {\bf 637} (2006) 161--169,
  \href{http://www.arXiv.org/abs/nucl-ex/0601033}{{\tt nucl-ex/0601033}}.

\bibitem{Acharya:2018hzf}
{\bf ALICE} Collaboration, S.~Acharya {\em et al.}, ``{Neutral pion and $\eta$
  meson production in p-Pb collisions at $\sqrt{s_\mathrm{NN}} = 5.02$ TeV},''
  {\em Eur. Phys. J. C} {\bf 78} (2018), no.~8, 624,
  \href{http://www.arXiv.org/abs/1801.07051}{{\tt 1801.07051}}.

\bibitem{Adam:2016dau}
{\bf ALICE} Collaboration, J.~Adam {\em et al.}, ``{Multiplicity dependence of
  charged pion, kaon, and (anti)proton production at large transverse momentum
  in p-Pb collisions at $\sqrt{s_{\mathrm{N}\mathrm{N}}}$ = 5.02 TeV},'' {\em
  Phys. Lett. B} {\bf 760} (2016) 720--735,
  \href{http://www.arXiv.org/abs/1601.03658}{{\tt 1601.03658}}.

\bibitem{Acharya:2021yrj}
{\bf ALICE} Collaboration, S.~Acharya {\em et al.}, ``{Nuclear modification
  factor of light neutral-meson spectra up to high transverse momentum in p-Pb
  collisions at $\sqrt{s_{\mathrm{N}\mathrm{N}}}$ = 8.16 TeV},''
  \href{http://www.arXiv.org/abs/2104.03116}{{\tt 2104.03116}}.

\bibitem{DAgostini:1993arp}
G.~D'Agostini, ``{On the use of the covariance matrix to fit correlated
  data},'' {\em Nucl. Instrum. Meth. A} {\bf 346} (1994) 306--311.

\bibitem{Binnewies:1994ju}
J.~Binnewies, B.~A. Kniehl, and G.~Kramer, ``{Next-to-leading order
  fragmentation functions for pions and kaons},'' {\em Z. Phys. C} {\bf 65}
  (1995) 471--480, \href{http://www.arXiv.org/abs/hep-ph/9407347}{{\tt
  hep-ph/9407347}}.

\bibitem{Kniehl:2000fe}
B.~A. Kniehl, G.~Kramer, and B.~Potter, ``{Fragmentation functions for pions,
  kaons, and protons at next-to-leading order},'' {\em Nucl. Phys. B} {\bf 582}
  (2000) 514--536, \href{http://www.arXiv.org/abs/hep-ph/0010289}{{\tt
  hep-ph/0010289}}.

\bibitem{Kretzer:2000yf}
S.~Kretzer, ``{Fragmentation functions from flavor inclusive and flavor tagged
  $e^+ e^-$ annihilations},'' {\em Phys. Rev. D} {\bf 62} (2000) 054001,
  \href{http://www.arXiv.org/abs/hep-ph/0003177}{{\tt hep-ph/0003177}}.

\bibitem{Hirai:2007cx}
M.~Hirai, S.~Kumano, T.~H. Nagai, and K.~Sudoh, ``{Determination of
  fragmentation functions and their uncertainties},'' {\em Phys. Rev. D} {\bf
  75} (2007) 094009, \href{http://www.arXiv.org/abs/hep-ph/0702250}{{\tt
  hep-ph/0702250}}.

\bibitem{Albino:2008fy}
S.~Albino, B.~A. Kniehl, and G.~Kramer, ``{AKK Update: Improvements from New
  Theoretical Input and Experimental Data},'' {\em Nucl. Phys. B} {\bf 803}
  (2008) 42--104, \href{http://www.arXiv.org/abs/0803.2768}{{\tt 0803.2768}}.

\bibitem{Bertone:2017tyb}
{\bf NNPDF} Collaboration, V.~Bertone, S.~Carrazza, N.~P. Hartland, E.~R.
  Nocera, and J.~Rojo, ``{A determination of the fragmentation functions of
  pions, kaons, and protons with faithful uncertainties},'' {\em Eur. Phys. J.
  C} {\bf 77} (2017), no.~8, 516,
  \href{http://www.arXiv.org/abs/1706.07049}{{\tt 1706.07049}}.

\bibitem{Moffat:2021dji}
E.~Moffat, W.~Melnitchouk, T.~Rogers, and N.~Sato, ``{Simultaneous Monte Carlo
  analysis of parton densities and fragmentation functions},''
  \href{http://www.arXiv.org/abs/2101.04664}{{\tt 2101.04664}}.

\bibitem{deFlorian:2017lwf}
D.~de~Florian, M.~Epele, R.~J. Hernandez-Pinto, R.~Sassot, and M.~Stratmann,
  ``{Parton-to-Kaon Fragmentation Revisited},'' {\em Phys. Rev. D} {\bf 95}
  (2017), no.~9, 094019, \href{http://www.arXiv.org/abs/1702.06353}{{\tt
  1702.06353}}.

\bibitem{Aidala:2010bn}
C.~A. Aidala, F.~Ellinghaus, R.~Sassot, J.~P. Seele, and M.~Stratmann,
  ``{Global Analysis of Fragmentation Functions for Eta Mesons},'' {\em Phys.
  Rev. D} {\bf 83} (2011) 034002,
  \href{http://www.arXiv.org/abs/1009.6145}{{\tt 1009.6145}}.

\bibitem{dEnterria:2013sgr}
D.~d'Enterria, K.~J. Eskola, I.~Helenius, and H.~Paukkunen, ``{Confronting
  current NLO parton fragmentation functions with inclusive charged-particle
  spectra at hadron colliders},'' {\em Nucl. Phys. B} {\bf 883} (2014)
  615--628, \href{http://www.arXiv.org/abs/1311.1415}{{\tt 1311.1415}}.

\bibitem{Metz:2016swz}
A.~Metz and A.~Vossen, ``{Parton Fragmentation Functions},'' {\em Prog. Part.
  Nucl. Phys.} {\bf 91} (2016) 136--202,
  \href{http://www.arXiv.org/abs/1607.02521}{{\tt 1607.02521}}.

\bibitem{Hirai:2016loo}
M.~Hirai, H.~Kawamura, S.~Kumano, and K.~Saito, ``{Impacts of B-factory
  measurements on determination of fragmentation functions from
  electron-positron annihilation data},'' {\em PTEP} {\bf 2016} (2016), no.~11,
  113B04, \href{http://www.arXiv.org/abs/1608.04067}{{\tt 1608.04067}}.

\bibitem{Soleymaninia:2018uiv}
M.~Soleymaninia, M.~Goharipour, and H.~Khanpour, ``{First QCD analysis of
  charged hadron fragmentation functions and their uncertainties at
  next-to-next-to-leading order},'' {\em Phys. Rev. D} {\bf 98} (2018), no.~7,
  074002, \href{http://www.arXiv.org/abs/1805.04847}{{\tt 1805.04847}}.

\bibitem{Bertone:2018ecm}
{\bf NNPDF} Collaboration, V.~Bertone, N.~P. Hartland, E.~R. Nocera, J.~Rojo,
  and L.~Rottoli, ``{Charged hadron fragmentation functions from collider
  data},'' {\em Eur. Phys. J. C} {\bf 78} (2018), no.~8, 651,
  \href{http://www.arXiv.org/abs/1807.03310}{{\tt 1807.03310}}.

\bibitem{Sassot:2009sh}
R.~Sassot, M.~Stratmann, and P.~Zurita, ``{Fragmentation Functions in Nuclear
  Media},'' {\em Phys. Rev. D} {\bf 81} (2010) 054001,
  \href{http://www.arXiv.org/abs/0912.1311}{{\tt 0912.1311}}.

\bibitem{Sievert:2019cwq}
M.~D. Sievert, I.~Vitev, and B.~Yoon, ``{A complete set of in-medium splitting
  functions to any order in opacity},'' {\em Phys. Lett. B} {\bf 795} (2019)
  502--510, \href{http://www.arXiv.org/abs/1903.06170}{{\tt 1903.06170}}.

\bibitem{Adler:2003pb}
{\bf PHENIX} Collaboration, S.~S. Adler {\em et al.}, ``{Mid-rapidity neutral
  pion production in proton proton collisions at $\sqrt{s}$ = 200-GeV},'' {\em
  Phys. Rev. Lett.} {\bf 91} (2003) 241803,
  \href{http://www.arXiv.org/abs/hep-ex/0304038}{{\tt hep-ex/0304038}}.

\bibitem{Abelev:2012cn}
{\bf ALICE} Collaboration, B.~Abelev {\em et al.}, ``{Neutral pion and $\eta$
  meson production in proton-proton collisions at $\sqrt{s}=0.9$ TeV and
  $\sqrt{s}=7$ TeV},'' {\em Phys. Lett. B} {\bf 717} (2012) 162--172,
  \href{http://www.arXiv.org/abs/1205.5724}{{\tt 1205.5724}}.

\bibitem{Martin:2009iq}
A.~D. Martin, W.~J. Stirling, R.~S. Thorne, and G.~Watt, ``{Parton
  distributions for the LHC},'' {\em Eur. Phys. J. C} {\bf 63} (2009) 189--285,
  \href{http://www.arXiv.org/abs/0901.0002}{{\tt 0901.0002}}.

\bibitem{Aad:2016zif}
{\bf ATLAS} Collaboration, G.~Aad {\em et al.}, ``{Transverse momentum,
  rapidity, and centrality dependence of inclusive charged-particle production
  in $\sqrt{s_{\mathrm{N}\mathrm{N}}}=5.02$ TeV $p$ + Pb collisions measured by
  the ATLAS experiment},'' {\em Phys. Lett. B} {\bf 763} (2016) 313--336,
  \href{http://www.arXiv.org/abs/1605.06436}{{\tt 1605.06436}}.

\bibitem{Khachatryan:2016odn}
{\bf CMS} Collaboration, V.~Khachatryan {\em et al.}, ``{Charged-particle
  nuclear modification factors in PbPb and pPb collisions at $
  \sqrt{s_{\mathrm{N}\mathrm{N}}}=5.02 $ TeV},'' {\em JHEP} {\bf 04} (2017)
  039, \href{http://www.arXiv.org/abs/1611.01664}{{\tt 1611.01664}}.

\bibitem{Acharya:2018qsh}
{\bf ALICE} Collaboration, S.~Acharya {\em et al.}, ``{Transverse momentum
  spectra and nuclear modification factors of charged particles in pp, p-Pb and
  Pb-Pb collisions at the LHC},'' {\em JHEP} {\bf 11} (2018) 013,
  \href{http://www.arXiv.org/abs/1802.09145}{{\tt 1802.09145}}.

\bibitem{deFlorian:2011fp}
D.~de~Florian, R.~Sassot, P.~Zurita, and M.~Stratmann, ``{Global Analysis of
  Nuclear Parton Distributions},'' {\em Phys. Rev. D} {\bf 85} (2012) 074028,
  \href{http://www.arXiv.org/abs/1112.6324}{{\tt 1112.6324}}.

\bibitem{Acharya:2018dxw}
{\bf ALICE} Collaboration, S.~Acharya {\em et al.}, ``{Jet fragmentation
  transverse momentum measurements from di-hadron correlations in $\sqrt{s}$ =
  7 TeV pp and $\sqrt{s_{\rm{NN}}}$ = 5.02 TeV p-Pb collisions},'' {\em JHEP}
  {\bf 03} (2019) 169, \href{http://www.arXiv.org/abs/1811.09742}{{\tt
  1811.09742}}.

\bibitem{Acharya:2019fip}
{\bf ALICE} Collaboration, S.~Acharya {\em et al.}, ``{One-dimensional charged
  kaon femtoscopy in p-Pb collisions at $\sqrt{s_{\rm NN}}$ = 5.02 TeV},'' {\em
  Phys. Rev. C} {\bf 100} (2019), no.~2, 024002,
  \href{http://www.arXiv.org/abs/1903.12310}{{\tt 1903.12310}}.

\bibitem{Pumplin:2001ct}
J.~Pumplin, D.~Stump, R.~Brock, D.~Casey, J.~Huston, J.~Kalk, H.~L. Lai, and
  W.~K. Tung, ``{Uncertainties of predictions from parton distribution
  functions. 2. The Hessian method},'' {\em Phys. Rev. D} {\bf 65} (2001)
  014013, \href{http://www.arXiv.org/abs/hep-ph/0101032}{{\tt hep-ph/0101032}}.

\bibitem{Stump:2001gu}
D.~Stump, J.~Pumplin, R.~Brock, D.~Casey, J.~Huston, J.~Kalk, H.~L. Lai, and
  W.~K. Tung, ``{Uncertainties of predictions from parton distribution
  functions. 1. The Lagrange multiplier method},'' {\em Phys. Rev. D} {\bf 65}
  (2001) 014012, \href{http://www.arXiv.org/abs/hep-ph/0101051}{{\tt
  hep-ph/0101051}}.

\end{thebibliography}\endgroup

\end{document}